\documentclass[useAMS,usenatbib]{mn2e}
\usepackage{graphicx,subfigure,color,afterpage}
\usepackage[total={17.8cm,24.0cm},centering]{geometry}
\usepackage{times}

\def\hi{H{\small I}}
\def\sauron{{\tt SAURON}}

\def\atlas{{{ATLAS}}$^{\rm 3D}$}
\def\kms{km s$^{-1}$}

\def\arcsec{$^{\prime \prime}$}
\definecolor{Mygrey}{gray}{0.75}

\newcommand{\ltsimeq}{\raisebox{-0.6ex}{$\,\stackrel{\raisebox{-.2ex}{$\textstyle <$}}{\sim}\,$}}
\newcommand{\gtsimeq}{\raisebox{-0.6ex}{$\,\stackrel{\raisebox{-.2ex}{$\textstyle >$}}{\sim}\,$}}
\newcommand{\farc}{\mbox{\ensuremath{.\!\!^{\prime\prime}}}}

\usepackage[compact]{titlesec}
\titlespacing{\section}{0pt}{*2}{*1}
\title[ IFU observations of the AGN driven outflow in NGC\,1266]{Gemini GMOS and WHT SAURON integral-field spectrograph observations of the AGN driven outflow in NGC\,1266}

\author[Timothy A. Davis et al.]{\parbox{\textwidth}{Timothy A. Davis,$^{1}$\thanks{E-mail:\texttt{tdavis@eso.org}}
Davor Krajnovi\'c,$^{1}$
Richard M. McDermid,$^{2}$
Martin Bureau,$^{3}$
Marc Sarzi,$^{4}$
Kristina Nyland,$^{5}$
Katherine Alatalo,$^{6}$
Estelle Bayet,$^{2}$
Leo Blitz,$^{6}$
Maxime Bois,$^{7}$
Fr\'ed\'eric Bournaud,$^{8}$
Michele Cappellari,$^{2}$
Alison Crocker,$^{9}$
Roger L. Davies,$^{2}$
P. T. de Zeeuw,$^{1,10}$
Pierre-Alain Duc,$^{8}$
Eric Emsellem,$^{1,11}$
Sadegh Khochfar,$^{12}$
Harald Kuntschner,$^{13}$
Pierre-Yves Lablanche,$^{1,11}$
Raffaella Morganti,$^{14,15}$
Thorsten Naab,$^{16}$
Tom Oosterloo,$^{14,15}$
Nicholas Scott,$^{17}$
Paolo Serra,$^{14}$
Anne-Marie Weijmans,$^{18}$\thanks{Dunlap Fellow} 
and Lisa M. Young$^{5}$\thanks{Adjunct Astronomer with NRAO}}\vspace{0.4cm}\\ 
\parbox{\textwidth}{
$^{1}$European Southern Observatory, Karl-Schwarzschild-Str. 2, 85748 Garching, Germany\\
$^{2}$Gemini Observatory, Northern Operations Centre, 670 N. A`ohoku Place, Hilo, HI 96720, USA\\
$^{3}$Sub-Dept. of Astrophysics, Dept. of Physics, University of Oxford, Denys Wilkinson Building, Keble Road, Oxford, OX1 3RH, UK\\
$^{4}$Centre for Astrophysics Research, University of Hertfordshire, Hatfield, Herts AL1 9AB, UK\\
$^{5}$Physics Department, New Mexico Institute of Mining and Technology, Socorro, NM 87801, USA\\
$^{6}$Department of Astronomy, Campbell Hall, University of California, Berkeley, CA 94720, USA\\
$^{7}$Observatoire de Paris, LERMA and CNRS, 61 Av. de l`Observatoire, F-75014 Paris, France\\
$^{8}$Laboratoire AIM Paris-Saclay, CEA/IRFU/SAp -- CNRS -- Universit\'e Paris Diderot, 91191 Gif-sur-Yvette Cedex, France\\
$^{9}$University of Massachussets, Amherst, USA\\
$^{10}$Sterrewacht Leiden, Leiden University, Postbus 9513, 2300 RA Leiden, the Netherlands\\
$^{11}$Universit\'e Lyon 1, Observatoire de Lyon, Centre de Recherche Astrophysique de Lyon and Ecole Normale Sup\'erieure de Lyon, 9 avenue Charles Andr\'e, F-69230 Saint-Genis Laval, France\\
$^{12}$Max-Planck Institut f\"ur extraterrestrische Physik, PO Box 1312, D-85478 Garching, Germany\\
$^{13}$Space Telescope European Coordinating Facility, European Southern Observatory, Karl-Schwarzschild-Str. 2, 85748 Garching, Germany\\
$^{14}$Netherlands Institute for Radio Astronomy (ASTRON), Postbus 2, 7990 AA Dwingeloo, The Netherlands\\
$^{15}$Kapteyn Astronomical Institute, University of Groningen, Postbus 800, 9700 AV Groningen, The Netherlands\\
$^{16}$Max-Planck-Institut f\"ur Astrophysik, Karl-Schwarzschild-Str. 1, 85741 Garching, Germany\\
$^{17}$Centre for Astrophysics \& Supercomputing, Swinburne University of Technology, PO Box 218, Hawthorn, VIC 3122, Australia\\
$^{18}$Dunlap Institute for Astronomy \& Astrophysics, University of Toronto, 50 St. George Street, Toronto, ON M5S 3H4, Canada
}
}

\begin{document}

\date{Accepted 2012 July 23.  Received 2012 July 5; in original form 2012 June 13}

\pagerange{\pageref{firstpage}--\pageref{lastpage}} \pubyear{2012}

\maketitle

\label{firstpage}
\clearpage
\begin{abstract}

We use the SAURON and GMOS integral field spectrographs to observe the active galactic nucleus (AGN) powered outflow in NGC\,1266. This unusual galaxy is relatively nearby (D=30 Mpc), allowing us to investigate the process of AGN feedback in action. 
We present maps of the kinematics and line strengths of the ionised gas emission lines H$\alpha$, H$\beta$, [OIII], [OI], [NII] and [SII], and report on the detection of Sodium D absorption. We use these tracers to explore the structure of the source, derive the ionised and atomic gas kinematics and investigate the gas excitation and physical conditions.
NGC\,1266 contains two ionised gas components along most lines of sight, tracing the ongoing outflow and a component closer to the galaxy systemic, the origin of which is unclear.  This gas appears to be disturbed by a nascent AGN jet. 
 We confirm that the outflow in NGC\,1266 is truly multiphase, containing radio plasma, atomic, molecular and ionised gas and X-ray emitting plasma. The outflow has velocities up to $\pm$900 \kms\ away from the systemic velocity, and is very likely to be removing significant amounts of cold gas from the galaxy.
The LINER-like line-emission in NGC\,1266 is extended, and likely arises from fast shocks caused by the interaction of the radio jet with the ISM.  These shocks have velocities of up to 800 \kms, which match well with the observed velocity of the outflow. Sodium D equivalent width profiles are used to set constraints on the size and orientation of the outflow. The ionised gas morphology correlates with the nascent radio jets observed in 1.4 Ghz and 5 Ghz continuum emission, supporting the suggestion that an AGN jet is providing the energy required to drive the outflow.

\end{abstract}


\section{Introduction}

In recent years the idea feedback from an active galactic nucleus (AGN; e.g. \citealt{2005ApJ...620L..79S,2006MNRAS.365...11C}) could be responsible for the quenching of star formation has grown in popularity. Such quenching seems to be required to create the red-sequence galaxies we observe today \citep[e.g.][]{2004ApJ...600..681B}. There is circumstantial evidence
to support AGN-driven quenching, such as the study by \cite{2007MNRAS.382.1415S} suggesting that AGN are predominantly
found in green valley galaxies, but direct evidence for removal/heating of cold star-forming gas is rare.  

The physical mechanism by which an AGN could drive molecular gas out of a galaxy is still debated. Radiation pressure is thought to be important in star formation-driven outflows (e.g. \citealt{2005ApJ...618..569M}), and is potentially implicated in AGN-powered `quasar mode' outflows (e.g. \citealt{1999ApJ...516...27A,2009MNRAS.397.1791K}). Kinetic feedback from an AGN jet can provide sufficient power to directly push through the ISM of a galaxy and entrain or destroy it (e.g. \citealt{2010ApJ...716..131R,McNul2012}), but it is unclear if the geometry of a bipolar jet, which often emerges perpendicular to the nuclear disk, will allow the jet to remove the ISM from an entire galactic disk. Broad-line winds can deposit significant momentum into gas surrounding an AGN, which could also lead to large outflows (e.g. \citealt{2010ApJ...722..642O}). Alternatively, heating by X-rays and cosmic rays could destroy/alter the molecular clouds close to an AGN, removing the need to expel them from the galaxy (e.g. \citealt{1991ApJ...382..416B,Ferland}). These processes should be distinguishable if we can identify and study local galaxies where AGN feedback is ongoing. 

\begin{figure} 
\begin{center} 
\includegraphics[width=0.48\textwidth,clip,trim=10cm 3.0cm 11cm 1.5cm]{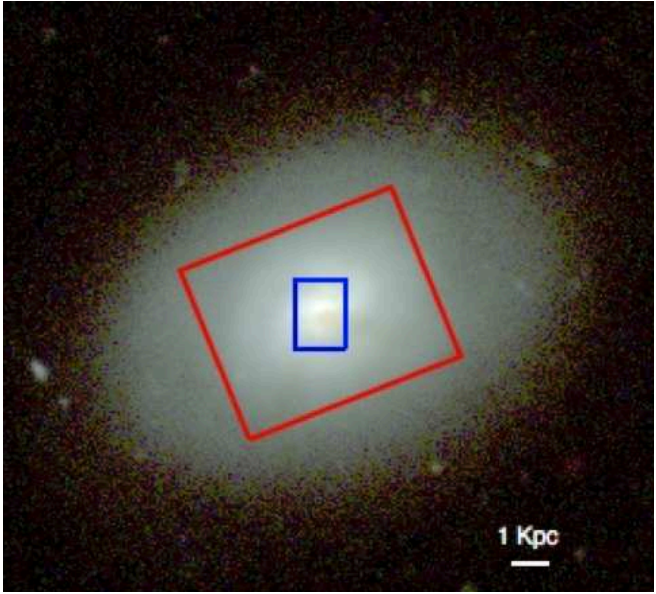}
\end{center}
\caption{\small SINGs \citep{SINGS} $B$,$V$ and $R$ band composite three colour image of S0 galaxy NGC\,1266. The white bar shows a linear scale of 1 Kpc (6\farc94 at an adopted distance of 29.9 Mpc; \atlas\ Paper I). Overlaid are the total field of view of our \sauron\ IFU (red) and GMOS IFU (blue) observations. }
\label{ngc1266color}
\end{figure}

Our recent Combined Array for Research in Millimetre-wave Astronomy (CARMA) and Sub-Millimetre Array (SMA) observations of the nearby lenticular galaxy NGC\,1266 suggest that it harbors a massive
AGN-driven molecular outflow, providing an excellent local laboratory for studying AGN-driven quenching \citep[][hearafter A2011]{2011ApJ...735...88A}.
{NGC\,1266} is a nearby (D= 29.9 Mpc; derived from recession velocity in \citealt{2011MNRAS.413..813C}; hereafter \atlas\ Paper I), early-type galaxy (ETG) in the southern sky ($\delta=-$2$^{\circ}$), which was studied as part of the \atlas\
project. A three colour image of this galaxy (from \citealt{SINGS}) is presented in Figure \ref{ngc1266color}. While typical CO spectra from early-type galaxies reveal the double-horned profile characteristic
of gas in a relaxed disk with a flat rotation curve, the spectrum of NGC\,1266 shows a narrow
central peak (FWHM $\approx$120 \kms) with non-Gaussian wings out to at least $\pm$400 \kms\ with respect to the systemic velocity \citep{2011MNRAS.414..940Y}.
Imaging of the high-velocity components using the SMA revealed that the wings resolve into 
redshifted and blueshifted lobes (A2011), coincident with H$\alpha$ emission \citep{2003PASP..115..928K}, 1.4 GHz continuum \citep{2006A&A...449..559B}, and thermal bremsstrahlung emission (detected
with Chandra; A2011; Fig. 3). 
Molecular gas observations suggest that 3$\times$10$^8$ M$_{\odot}$ of molecular
gas is contained within the central 100 pc of NGC\,1266, and that at least 5$\times$10$^7$ M$_{\odot}$ of this gas is involved in a molecular outflow (A2011).
 This is thus the first observed large-scale expulsion of molecular gas from a non-starbursting ETG in the local
universe, and this presents a unique opportunity to study this powerful process in action.

In this paper we present SAURON (Spectrographic Areal Unit for Research on Optical Nebulae) and Gemini  Multi-Object Spectrograph (GMOS) integral-field unit (IFU) observations of the ionised gas in NGC\,1266. {By investigating the ionised gas kinematics and line ratios we hope to constrain the outflow parameters and ionisation mechanisms and thus shed light on the mechanism driving gas from the galaxy}. In Section \ref{dataredux} we present the data, and describe our reduction processes. We then present the derived maps of the gas kinematics and line fluxes. In Section \ref{results} we discuss the kinematic structure of the system, gas excitation mechanisms and the driving force behind the outflow. Finally we conclude and discuss prospects for the future in Section \ref{conclude}.

\section{Data Reduction and Results}
\label{dataredux}
\subsection{SAURON data}

SAURON is an integral-field spectrograph built at Lyon Observatory 
and mounted at the Cassegrain focus of the William Herschel Telescope (WHT). It is based on the TIGER concept 
\citep{Bacon:1995p3377}, using a microlens array to sample the field of 
view. Details of the instrument can be found in \cite{Bacon:2001p1477}. 
The SAURON data of NGC\,1266 was taken at the William Herschel Telescope (WHT), on the night of 10-11 January 2008, as part of the \atlas\ observing campaign (\atlas\ Paper I). The galaxy was observed with the low-resolution mode of SAURON, covering a field of view of about 33\arcsec $\times$ 41\arcsec\ with 0$^{\prime\prime}_{.}$94 $\times$ 0$^{\prime\prime}_{.}$94 lenslets. The field of view (FOV) of our observations is shown in red on Figure \ref{ngc1266color}. \sauron\ covers the wavelength range from 4810-5350 \AA\ with a spectral resolution of 105 \kms.

The basic reduction of the SAURON observation was accomplished using the standard \atlas\ pipeline. Details of this process, including extraction of the stellar kinematics {are} presented in \atlas\ Paper II \citep{A3DII}. In brief, the two observed datacubes were merged and processed as described in \cite{Emsellem:2004p1497}, 
using the Voronoi binning scheme developed by \cite{Cappellari:2003p3284}. {This binning scheme maximises the scientific potential of the data by ensuring a minimum signal-to-noise ratio of 40 per spatial and spectral pixel. This does however result in an non-uniform spatial resolution, here varying from 0\farc8 $\times$ 0\farc8 for unbinned spaxels in the central regions, to 10\arcsec $\times$ 7\arcsec in the largest outer bin.}

\begin{figure}
\begin{center} 
\includegraphics[width=0.5\textwidth,clip,trim=0cm 0.0cm 0cm 0cm]{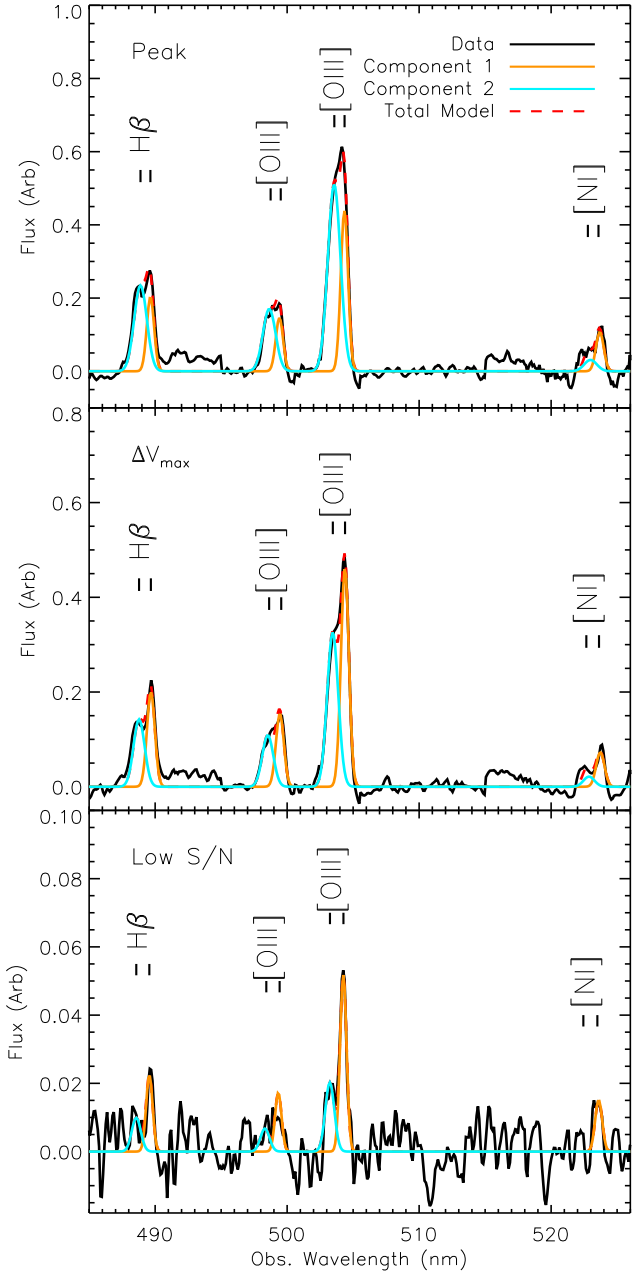}
\end{center}
\caption{\small Stellar emission subtracted SAURON spectrum from single bins (black solid line) with line identifications. The top panel shows the spaxel with the largest line flux (x=0\arcsec, y=-4\arcsec), the middle panel shows the spaxel with the biggest difference in the fitted velocities (x=0\farc8 y=-5\farc17) and the bottom panel shows the lowest flux region where a two component fit can be constrained (x=-4\farc0 y=-5\farc17). Overlaid is the two component fit produced by out fitting routine, as described in Section \ref{sauronfit}. Component one (nearest the galaxy systemic) is shown in orange, while component two is shown in blue. The red dashed line shows the sum of component one and two, which closely matches the observed data. }
\label{sauronspec}
\end{figure}

The SAURON stellar kinematics were derived using a penalized 
pixel fitting routine \citep{Cappellari:2004p3283}, providing
parametric estimates of the line-of-sight velocity distribution for each bin. During the extraction of the stellar kinematics, the GANDALF code \citep{Sarzi:2006p1474} was used to simultaneously extract the ionised gas line fluxes and kinematics.  The standard GANDALF reduction completed in the pipeline (using a single gaussian for the lines) is insufficient in this source, due to the complex structure of the ionised gas outflow (see Figure \ref{sauronspec}). We have reanalyzed the datacube using a multi-gaussian technique (as described below) after the subtraction of the stellar continuum.

\subsubsection{Emission line fitting}
\label{sauronfit}
The SAURON spectra include the H$\beta$, [OIII] and [NI] ionised gas spectral lines. As can be seen in Figure \ref{sauronspec} some of the binned spaxels show clear signs of having two ionised gas components along the line of sight with different velocities. In order to fit these profiles we created an IDL procedure based on the Non-Linear Least Squares Fitting code \textit{mpfit} \citep{MarkwardtIDL}. 
In this procedure we perform two fits, and compare the chi-square to determine if two components are needed at each position. 

In the first fit, we assume a single ionised gas component is present, and fit the H$\beta$, [OIII] and [NI] lines with single gaussians. These gaussians are constrained to have the same kinematics (velocity and velocity dispersion). Additionally, we confine the velocity of the lines to be within 1000 \kms\ of the galaxy systemic (2170 \kms; \atlas Paper I), and to have a velocity dispersion greater than the instrumental resolution, and less than a convolved velocity dispersion of $\approx$200 \kms. 
Initial guesses at the ionised gas kinematics were made by assuming the ionised gas co-rotates with the stars, with a velocity dispersion of 120 \kms. We constrain the fitting by forcing each gaussian to have a peak at least 3 times larger than the noise in the continuum, or to be zero. Any flux which had a 1$\sigma$ error bar that included zero was set to zero. Initial guesses of the line fluxes were estimated by taking the maximum flux within the allowed velocity range of each line. The two [OIII] lines in our spectral range have a fixed line ratio determined by the energy structure of the atom, and we fixed the line ratio assumed in our fit to agree with the observed line ratio (F$_{5007}$=2.99$\times$F$_{4959}$: \citealt{2000MNRAS.312..813S,2007MNRAS.374.1181D}). 

In the second fit we assume two, independent ionised gas components are present in each bin, and fit each component with its own set of independent linked gaussians. As before, it is assumed that the lines in each component trace the same kinematics (velocity and velocity dispersion). Once again we confine the velocity of each of the components to be within 1000 \kms\ of the galaxy systemic,  and to have a velocity dispersion greater than the instrumental resolution. We use different upper bounds for the first and second components of the gas distribution. Component one is forced to have a velocity dispersion less than 200 \kms, as before. In general the second component is needed where the outflow is present, and is thus allowed to have a higher velocity dispersion. We allowed the lines in the second component to have a maximum velocity dispersion of 360 \kms. In practice however good fits were found with velocity dispersions $<$300\kms. The same limits were used on the line fluxes as described above. 

Once the two fits described above were complete for each bin, we tested (using an F-test, as implemented in the mpfit package \citealt{MarkwardtIDL}) if adding the additional free parameters to our model of the emission lines produced a significantly better fit, over and above the improvement expected when one adds free parameters.
The F-test can be used as an indicator of where fitting two components produces better models, but the best threshold to take should be determined by visually inspecting the fits obtained (as the values tested for are at the extreme edge of the possible distribution). 
In this work we visually chose a threshold that corresponds to an improvement in the chi-square of 60\% when adding in the additional parameters. When a spaxel did not satisfy this criterion then the values from the single gaussian fit were used, and the second component set to zero. Where two components were found to be necessary we denoted the component closest to the galaxy systemic as component one, and the faster component as component two. 

In an attempt to ensure that the fits were robust, and spatially continuous, we implemented an iterative fitting regime where the fitting processes described above were performed for each spaxel in turn. Then the resulting two dimensional flux and velocity maps were smoothed using a gaussian kernel, and then these smoothed values used as the initial guesses for the next iteration of the fitting procedure. Using this procedure we found that the parameters usually converged within three iterations, with very little variance between fitting attempts. Figure \ref{sauronspec} shows SAURON spectra from a single bin, overlaid with the two component fit. {The top panel shows the spaxel with the largest line flux, the middle panel shows the spaxel with the biggest difference in the fitted velocities and the bottom panel shows the lowest flux region where a two component fit can be constrained. Clearly in the low flux regions of the cube the fitted velocities are driven by the [OIII]$_{5007}$ line, and the parameters have correspondingly higher uncertainties. }

\begin{figure*} 
\begin{center} 
\subfigure[]{\includegraphics[width=0.42\textwidth,clip,trim=0cm 0.5cm 0cm 1cm]{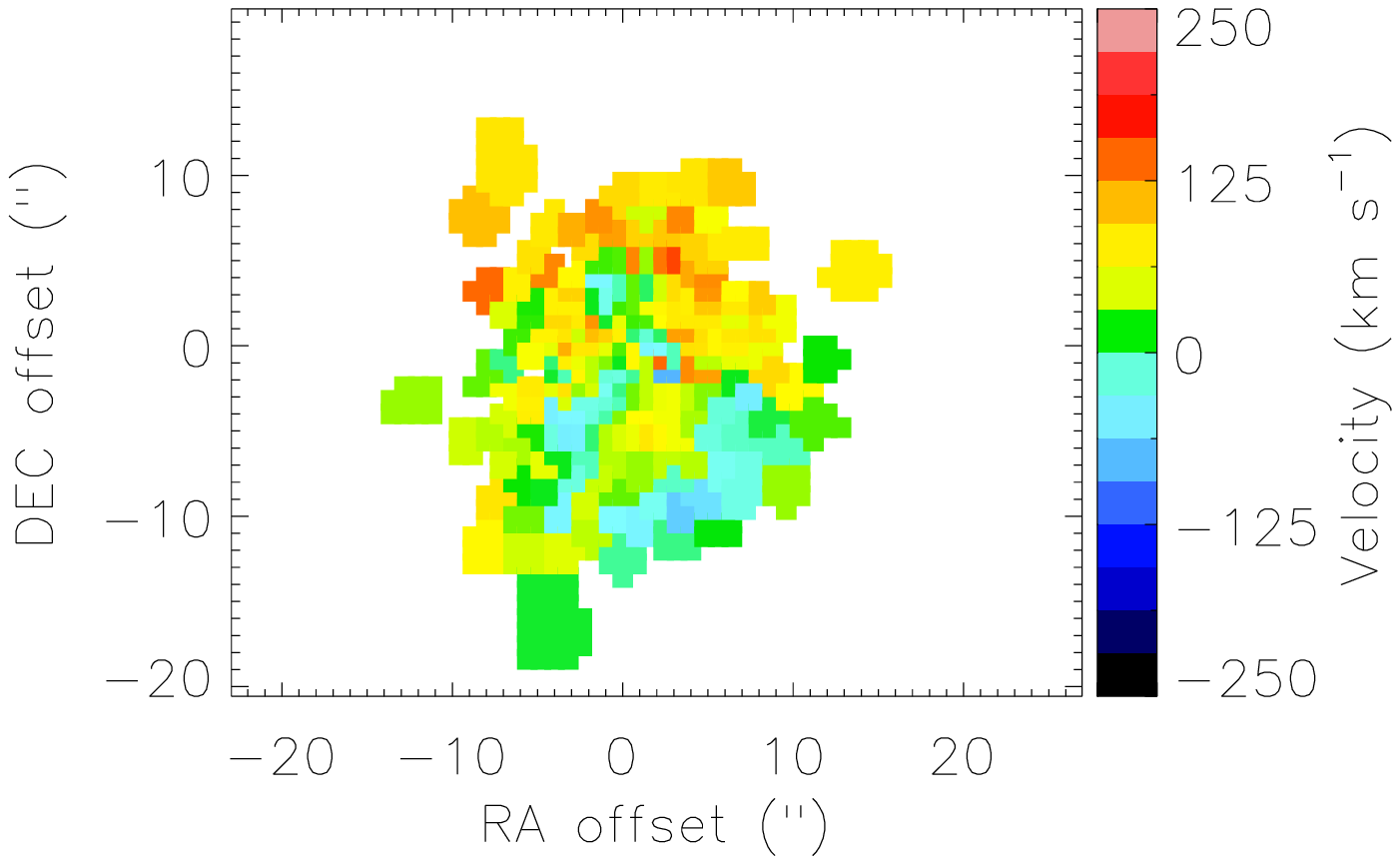}}
\subfigure[]{\includegraphics[width=0.42\textwidth,clip,trim=0cm 0.5cm 0cm 1cm]{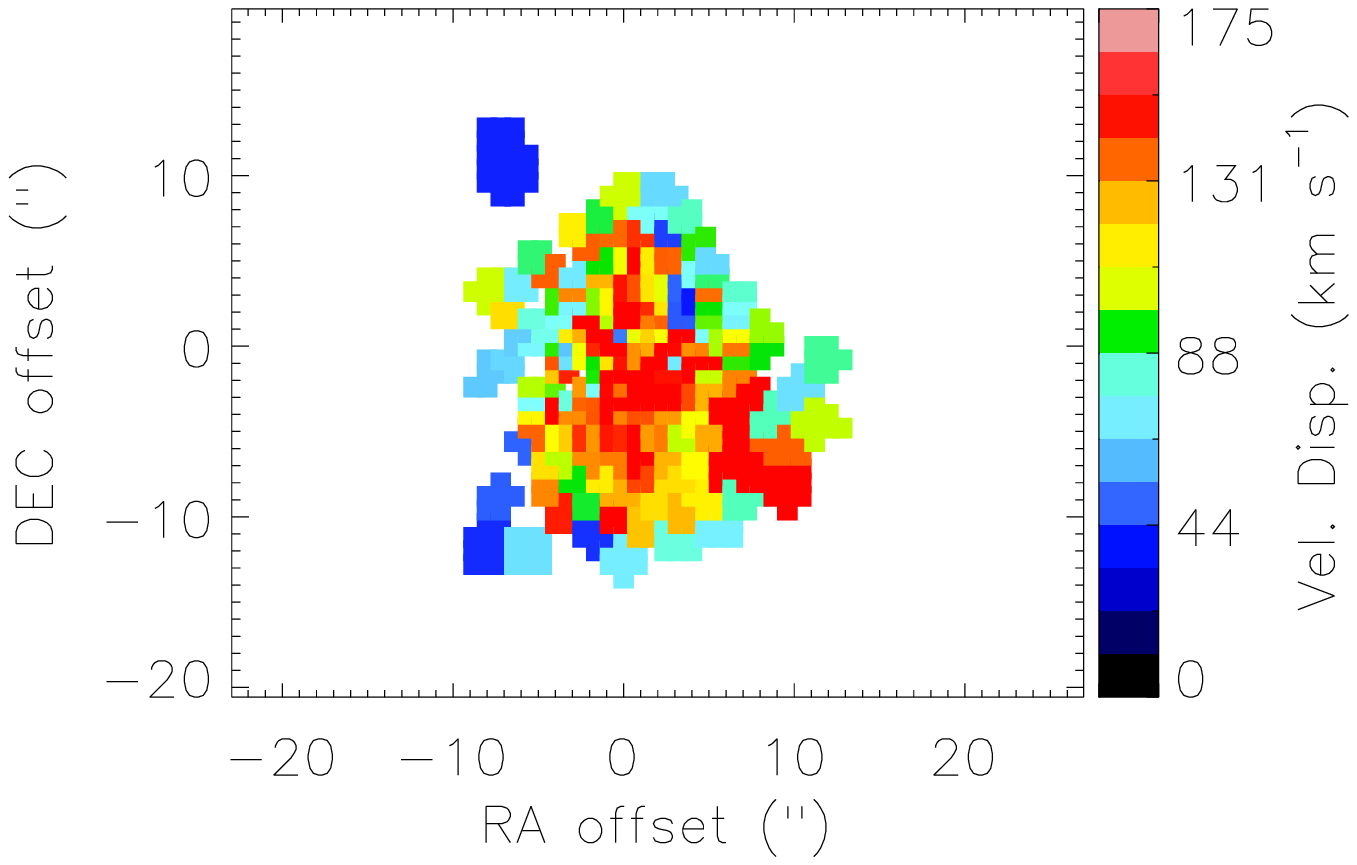}}\\
\subfigure[]{\includegraphics[width=0.42\textwidth,clip,trim=0cm 0.5cm 0cm 1cm]{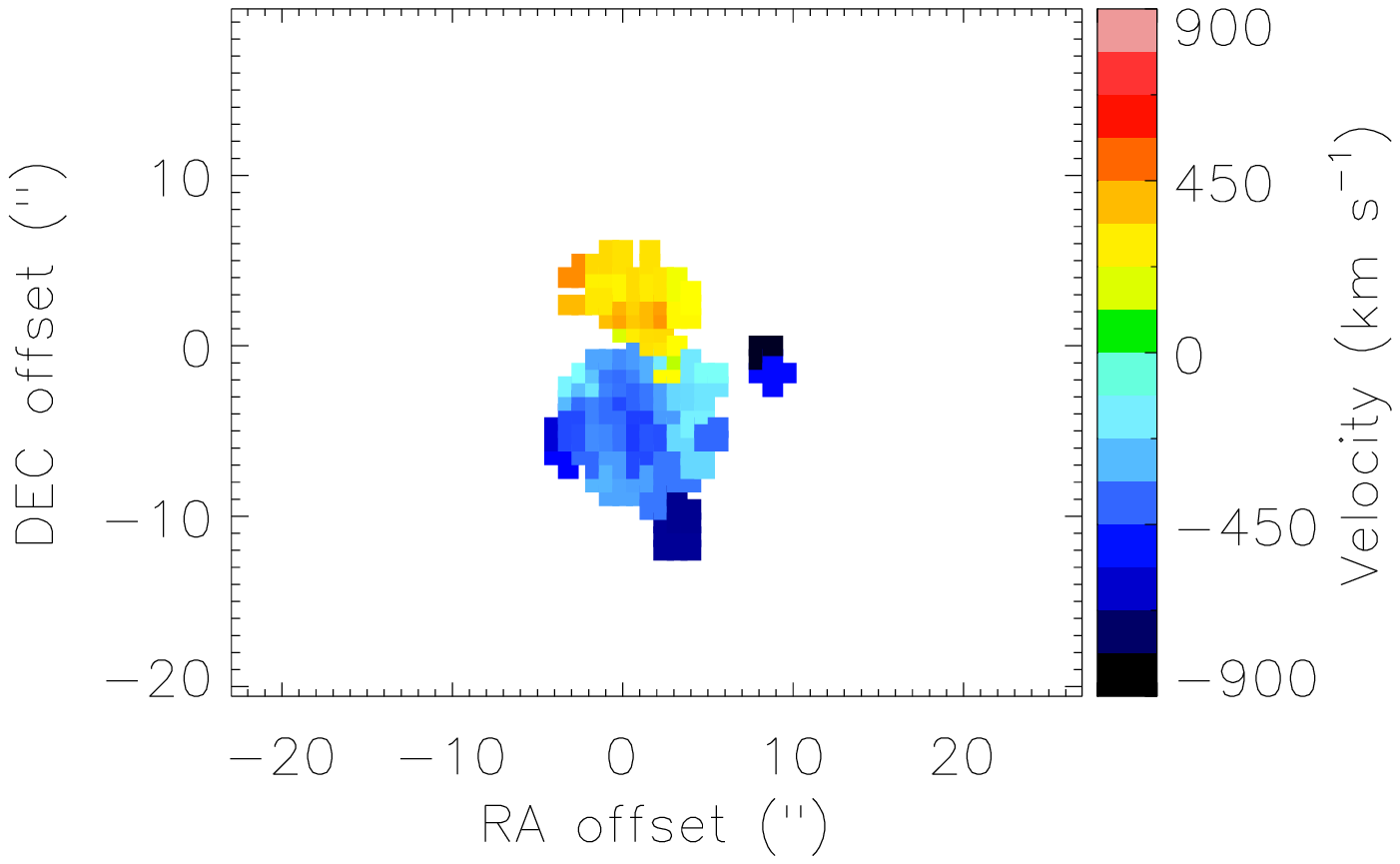}}
\subfigure[]{\includegraphics[width=0.42\textwidth,clip,trim=0cm 0.5cm 0cm 1cm]{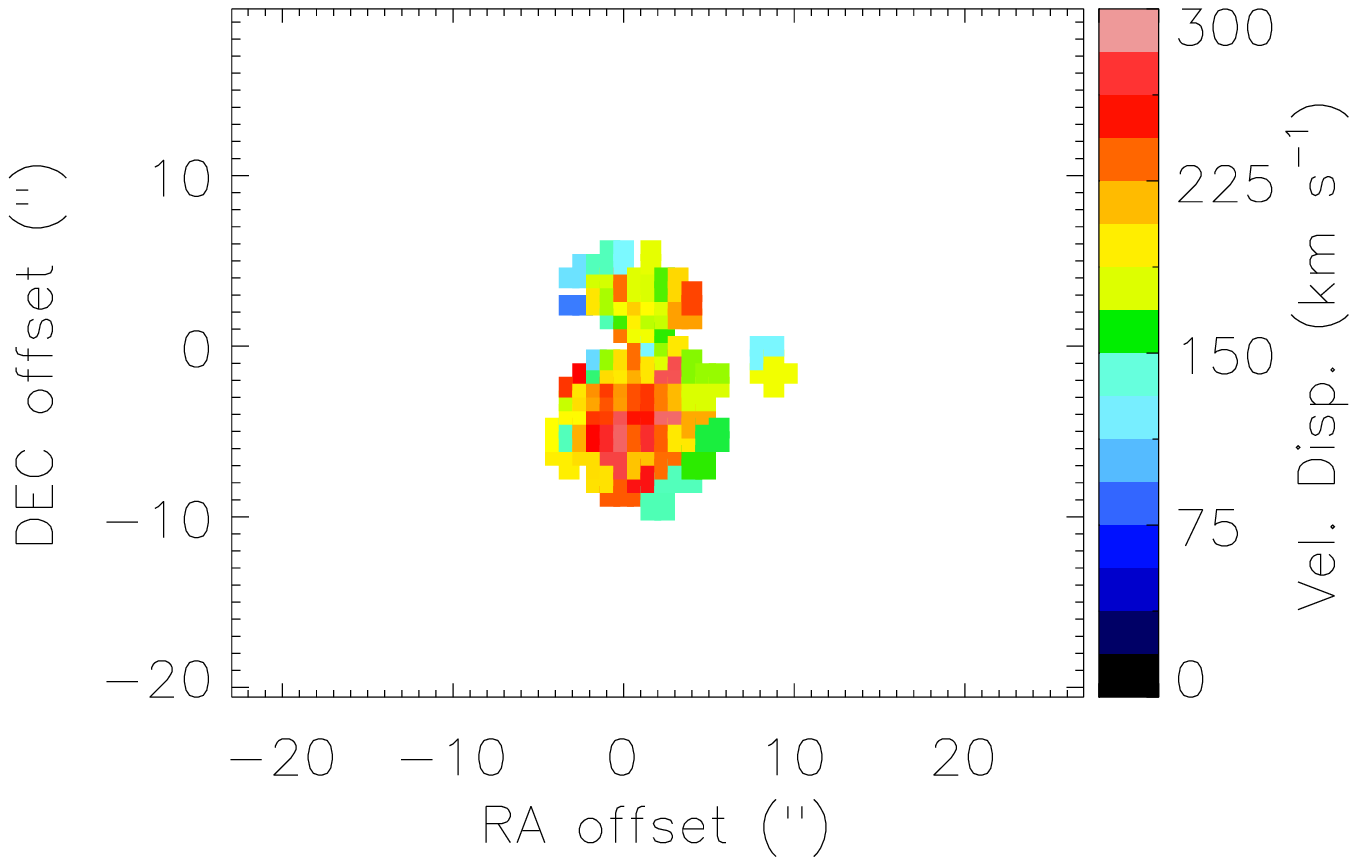}}\\
\subfigure[]{\includegraphics[width=0.42\textwidth,clip,trim=0cm 0.5cm 0cm 1cm]{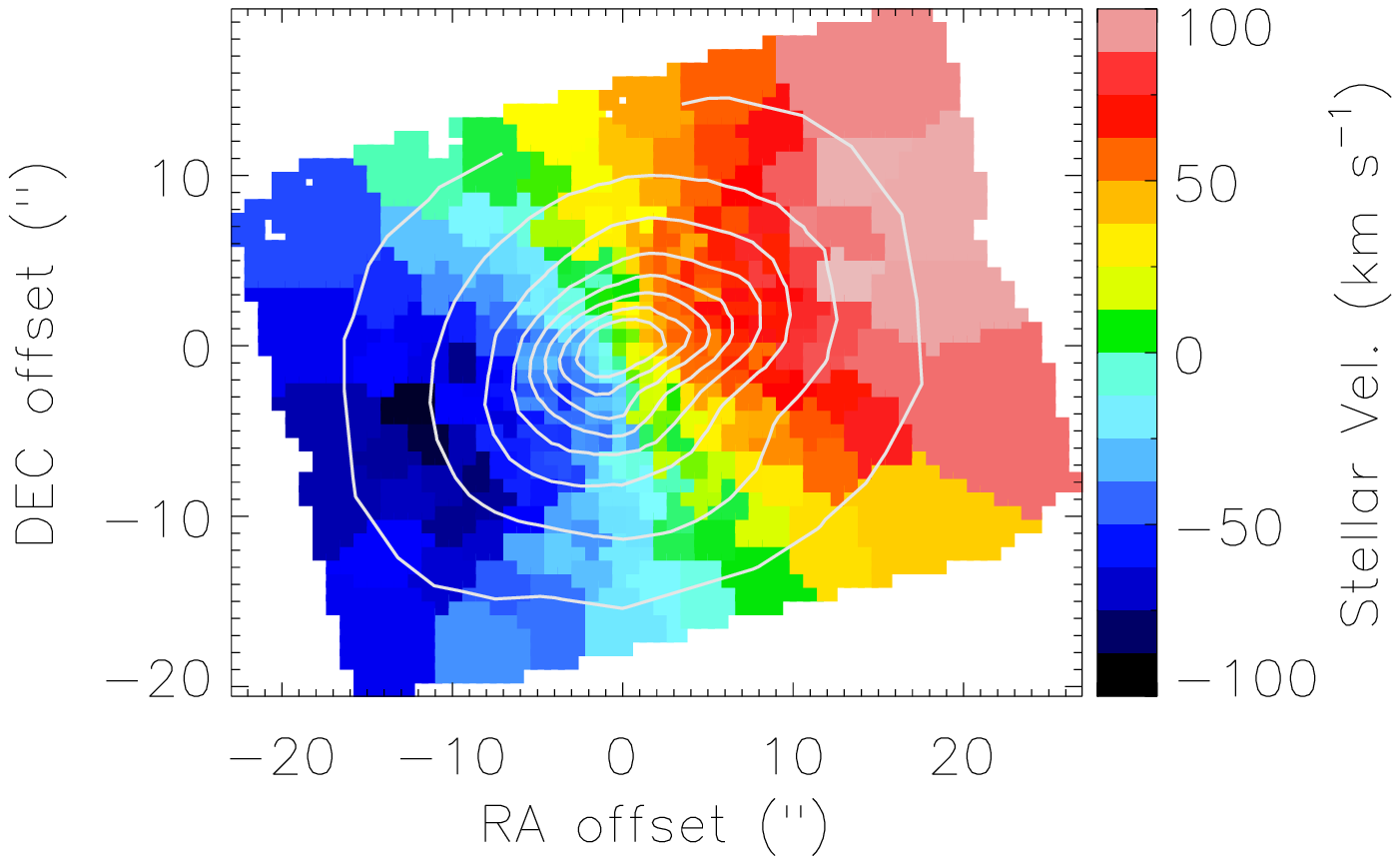}}
\end{center}
\caption{\small Ionised gas kinematics derived from the SAURON IFU data reduction process discussed in Section \ref{sauronfit}. In the top row (panels a and b) we display the kinematics of component one (confined to be closest to the galaxy systemic velocity). Bins where only one ionised gas component is required are also shown in component one. The kinematics of the faster component is shown in the middle row (panels c and d). The ionised gas velocity is displayed in the left panels (a,c), and the velocity dispersion in the right panels (b,d). The stellar velocity field of this galaxy derived from these same observation in \atlas\ Paper II is shown as a comparison in panel e, with stellar continuum flux contours overlaid. These plots are centred around the galaxy position given in \atlas\ Paper I.}
\label{sauronplots1}
\end{figure*}

\begin{figure*} 
\begin{center} 

\subfigure[]{\includegraphics[width=0.42\textwidth,clip,trim=0cm 0.5cm 0cm 1.0cm]{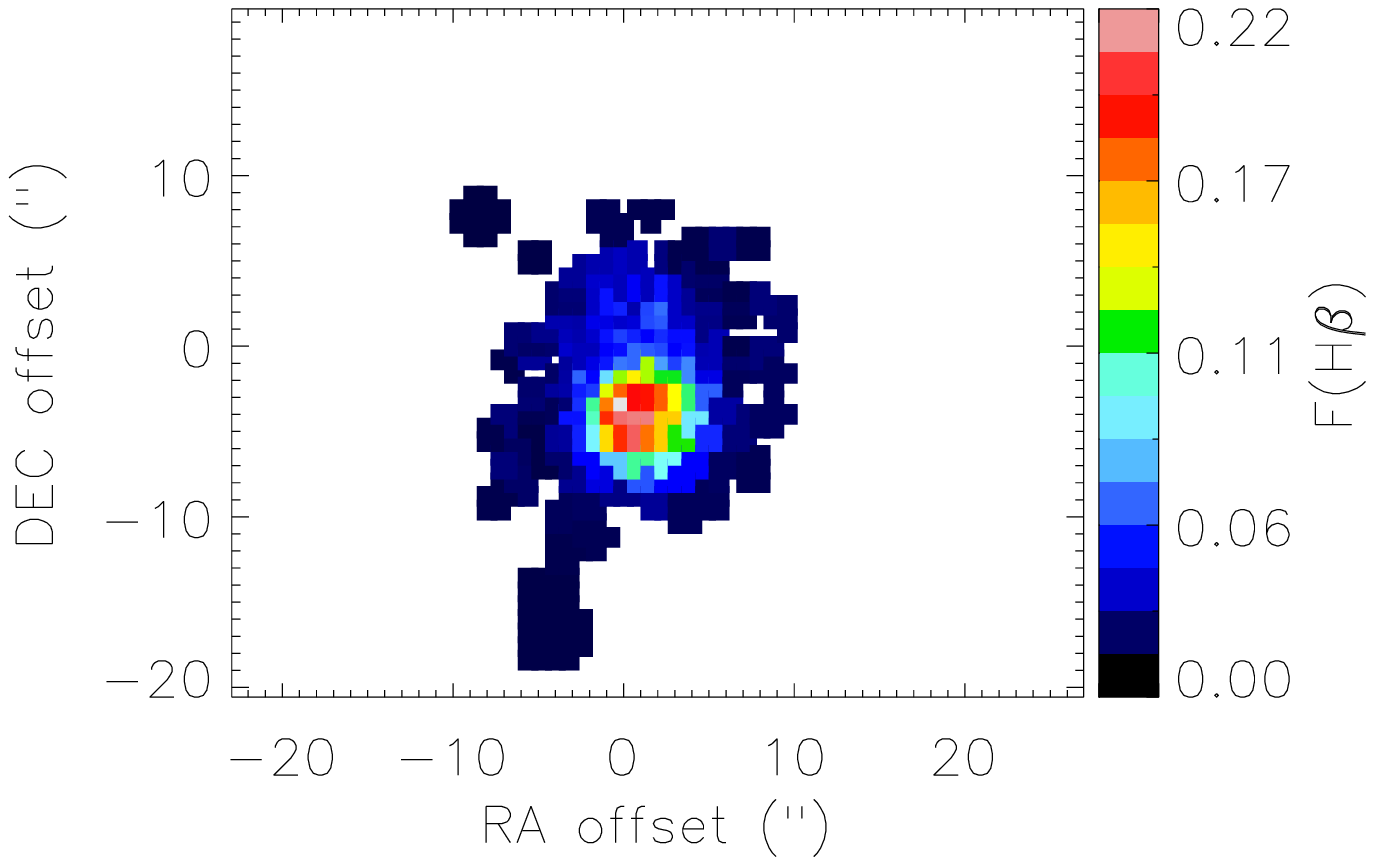}}
\subfigure[]{\includegraphics[width=0.42\textwidth,clip,trim=0cm 0.5cm 0cm 1.0cm]{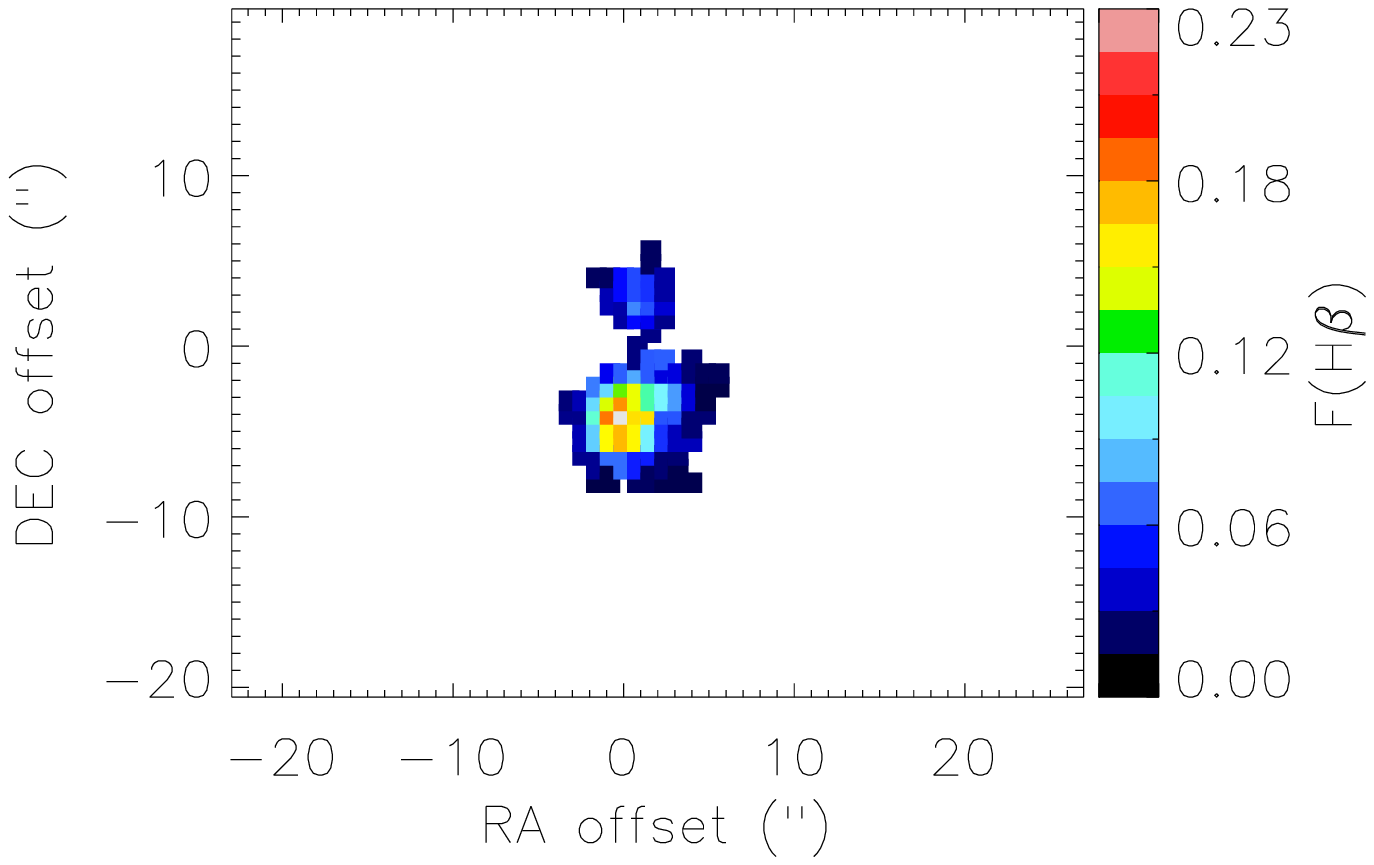}}\\
\subfigure[]{\includegraphics[width=0.42\textwidth,clip,trim=0cm 0.5cm 0cm 1.0cm]{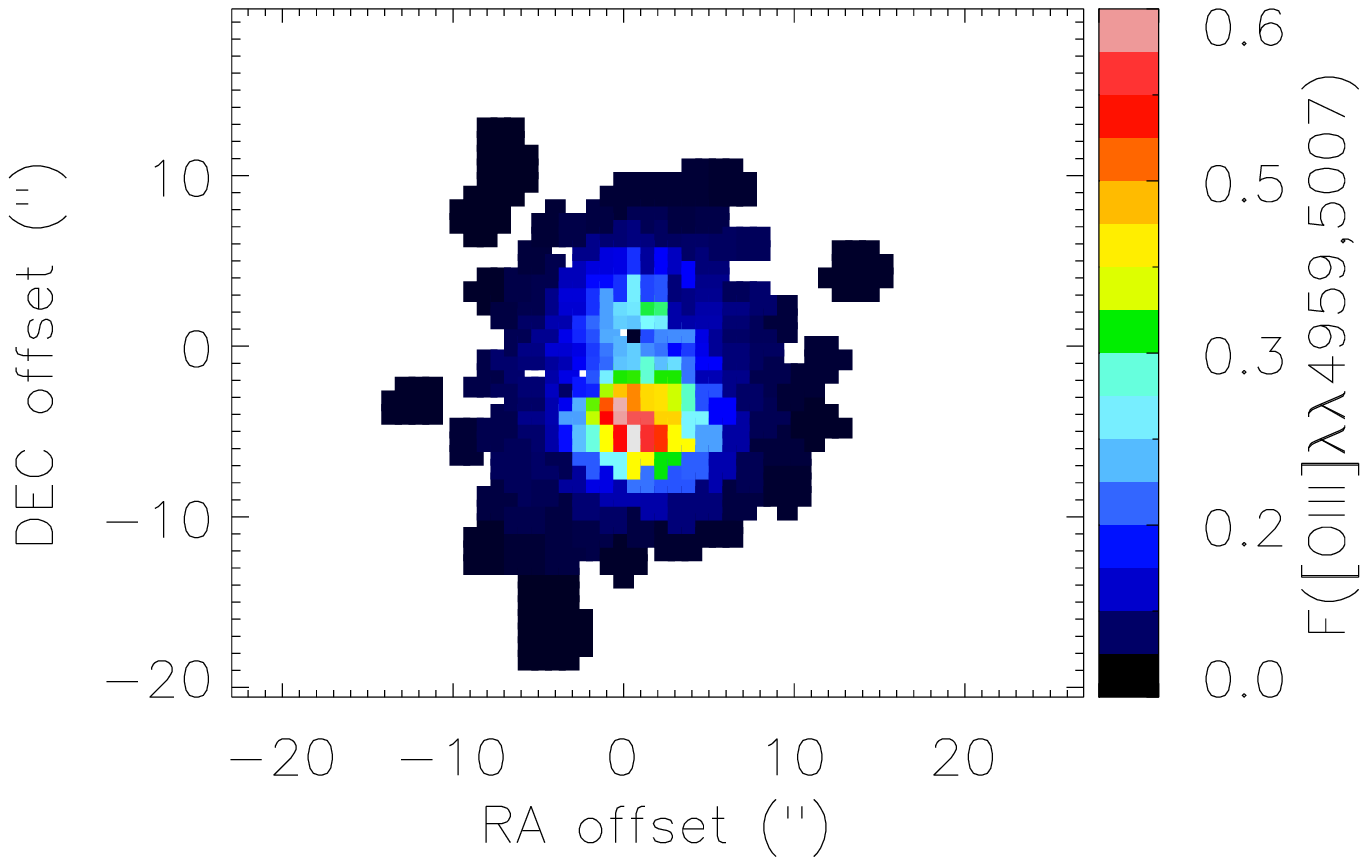}}
\subfigure[]{\includegraphics[width=0.42\textwidth,clip,trim=0cm 0.5cm 0cm 1.0cm]{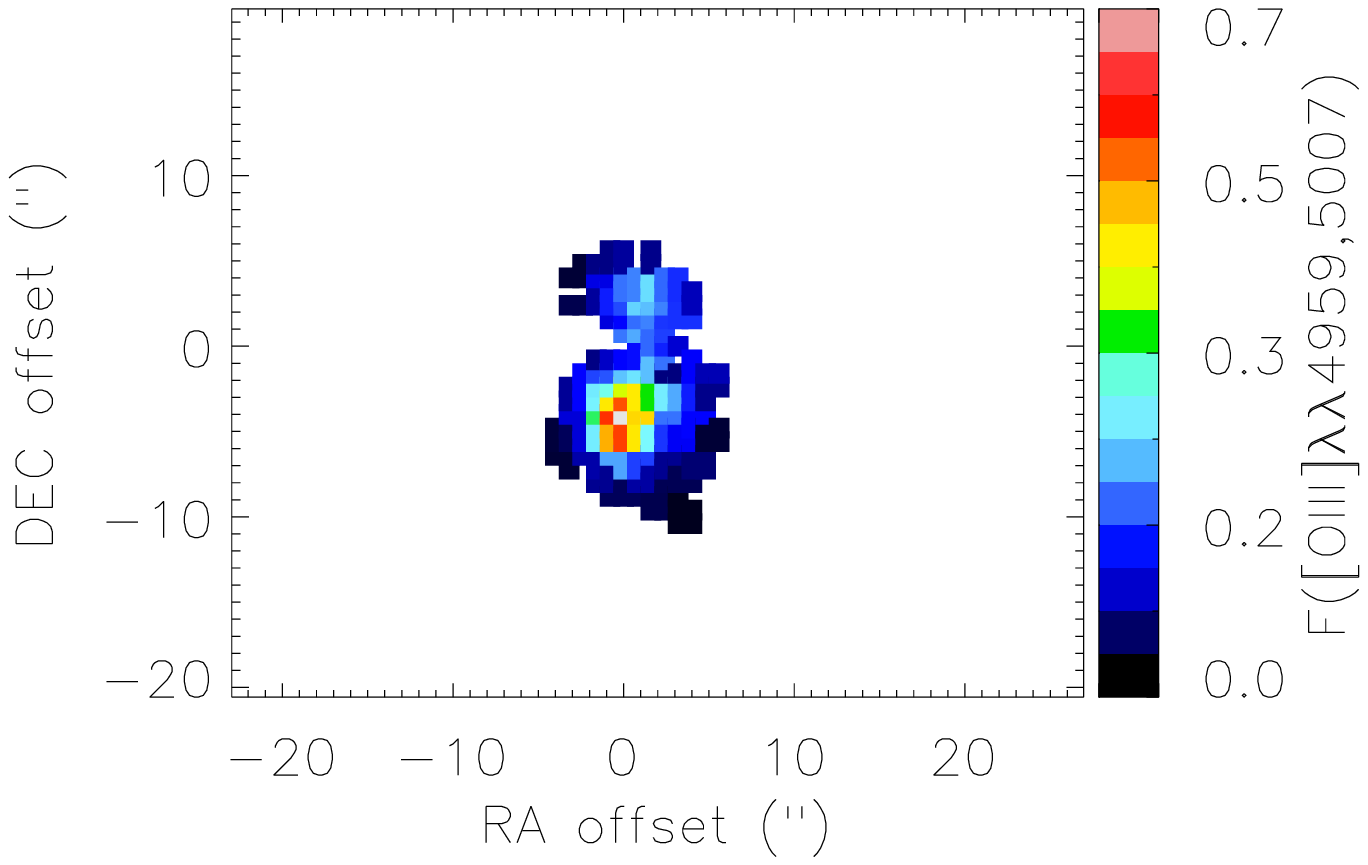}}\\
\subfigure[]{\includegraphics[width=0.42\textwidth,clip,trim=0cm 0.5cm 0cm 1.0cm]{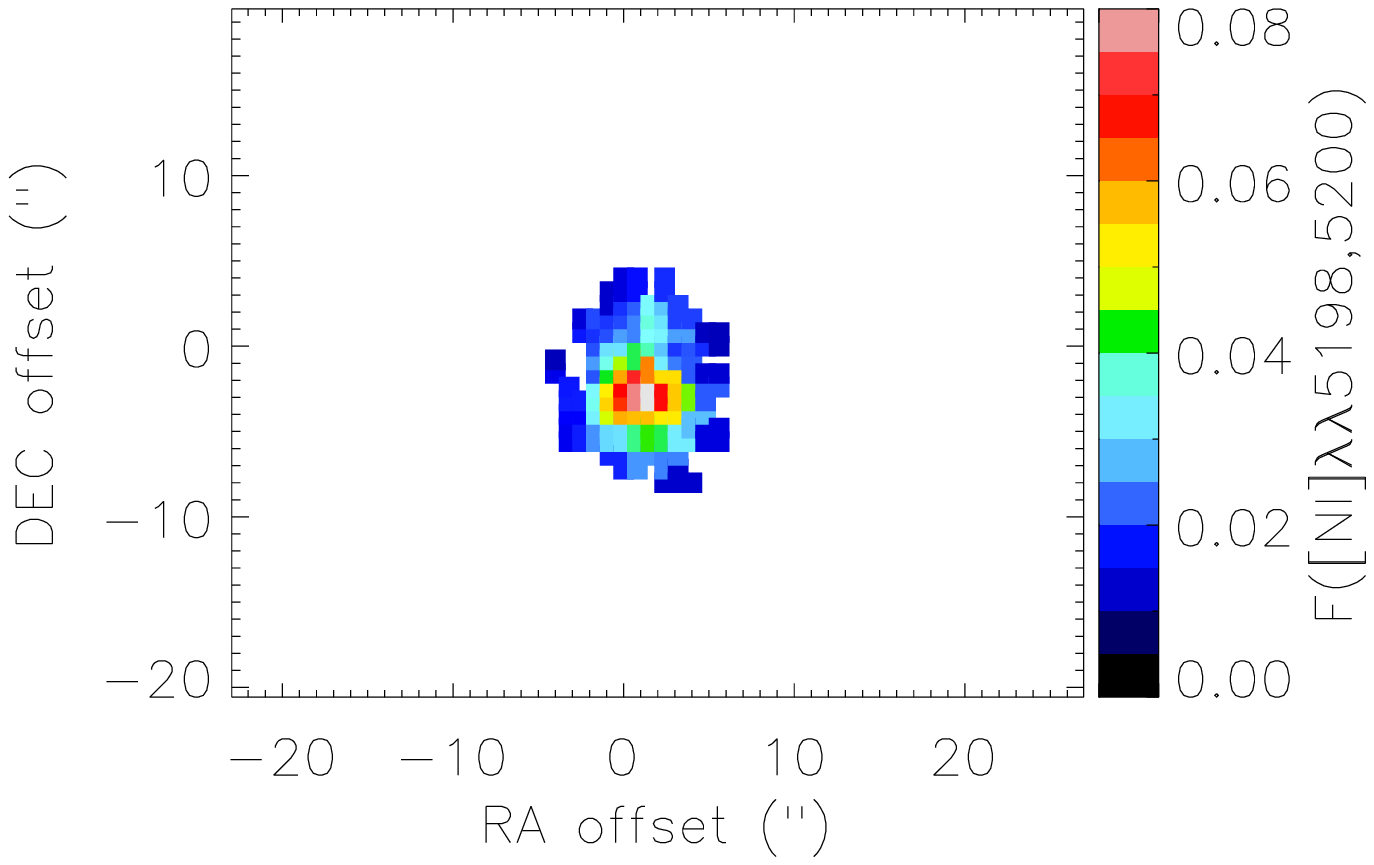}}
\subfigure[]{\includegraphics[width=0.42\textwidth,clip,trim=0cm 0.5cm 0cm 1.0cm]{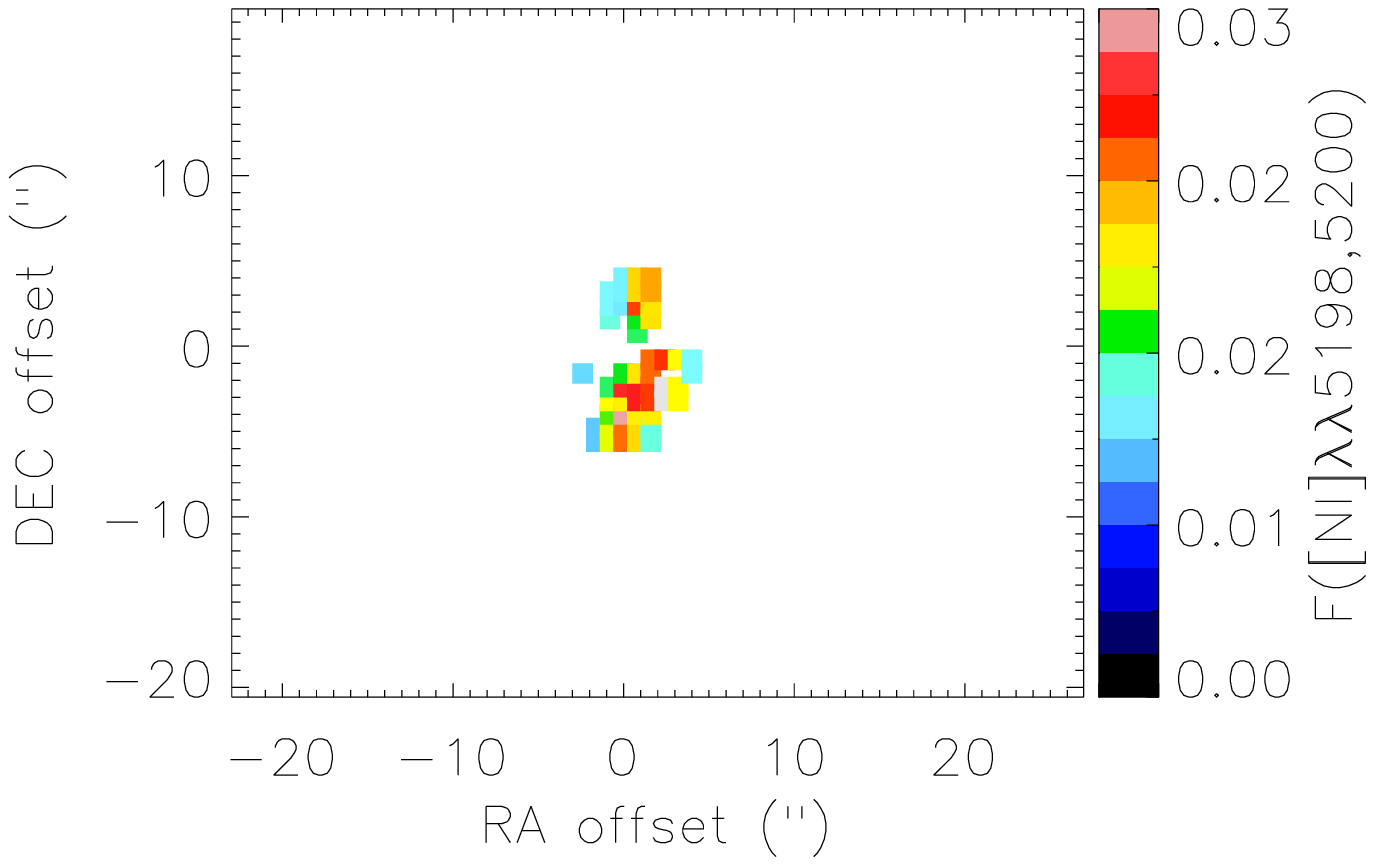}}\\

\end{center}
\caption{\small Ionised gas line fluxes derived from the SAURON IFU data reduction process discussed in Section \ref{sauronfit}. In the left column we display the flux of component one (confined to be closest to the galaxy systemic). Bins where only one ionised gas component is required are also shown in component one. The flux of the faster out-flowing component is shown in the right hand column. The top row shows the H$\beta$ line fluxes (panels a \& b), the second row the [OIII] line fluxes (panels c \& d), and the third row the [NI] fluxes (panels e \& f). Fluxes are in units of 10$^{-16}$ erg s$^{-1}$ cm$^{-2}$ arcsec$^{-2}$ in each SAURON bin.}
\label{sauronplots2}
\end{figure*}

In Figure \ref{sauronplots1} we show the observed kinematics for component one (panels a \& b) and component two (panels c \& d). The velocity dispersion maps have had the instrumental dispersion quadratically subtracted. We also show the stellar velocity field derived from these SAURON observations (panel e) for reference, as presented in Paper II . In Figure \ref{sauronplots2} we show the fitted line fluxes.

\subsection{Gemini GMOS data}
\label{gmosfit}

\begin{figure} 
\begin{center} 
\includegraphics[width=0.5\textwidth,clip,trim=0cm 0.0cm 0cm 0cm]{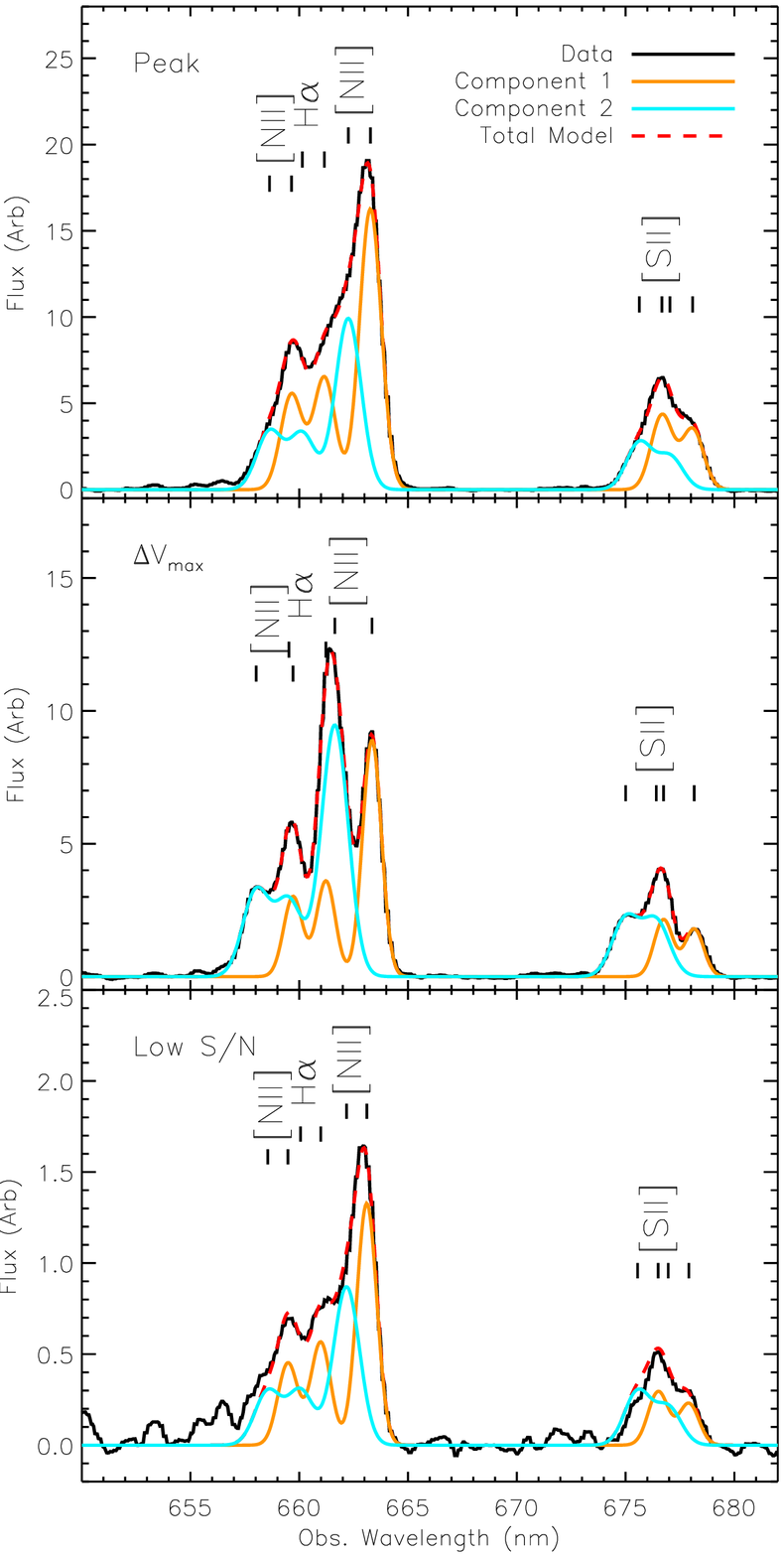}
\end{center}
\caption{\small Stellar subtracted GMOS spectrum (black solid lines)from single bins with line identifications. Shown in the top panel is the bin with the spaxel with the peak line flux (x=0\farc91, y=-2\farc08), the middle panel is the spaxel with the largest difference between the two fitted velocities (x=-0\farc2, y=-4\farc29), and the bottom panel shows the bin with the lowest total flux in which we are able to fit two components (x=-1\farc85, y=-5\farc73).  Overlaid is the two component fit produced by out fitting routine, as described in Section \ref{gmosfit}. Component one (nearest the galaxy systemic) is shown in orange, while the faster component two is shown in blue. The red dashed line shows the sum of component one and two, which closely matches the observed data. }
\label{gmosspec}
\end{figure}

In addition to the \sauron\ data, we obtained Gemini GMOS-IFU observations of the central parts of NGC\,1266, providing higher spatial resolution and a longer wavelength coverage.
 The GMOS IFU uses a lenslet array of 1500 elements to feed individual positions on the sky to optical fibres \citep{2002PASP..114..892A,2004PASP..116..425H}. The total field of view of the IFU is 5\arcsec$\times$7\arcsec, with a spatial sampling of 0\farc2. The Gemini GMOS-IFU observations of NGC\,1266 were taken over the nights of 24th, 26th and 27th of January 2009 at the Gemini North telescope (program GN-2008B-DD-1). We used a four point dither pattern to extend our coverage to a total field of view of $\approx$9\farc1$\times$12\farc5, around the optical centre of the galaxy. The resultant field of view (FOV) of our observations is shown in blue on Figure \ref{ngc1266color}. The low resolution R150 grating was used, resulting in a spectral resolution of $\approx$185 \kms (at 6500 \AA) over the wavelength range 5000 - 7300 \AA. Two different blaze wavelengths (688 and 700 nm) were used on different exposures to allow continuous spectral coverage by averaging over chip gaps/bridges. 

In order to reduce the GMOS IFU data we utilized a data reduction pipeline, as used in \cite{2010ApJ...719.1481V}. This pipeline calibrates and flat fields the data, before it is trimmed and resampled into a homogeneous data cube. This cube was binned using the Voronoi binning technique of \cite{Cappellari:2003p3284},
ensuring a signal-to-noise ratio of 40 per spatial and spectral pixel. Due to the low spectral resolution and the depth of the exposures we detect line emission to high significance over almost the entire IFU cube, but were unable to detect stellar absorption features to high significance. As we are interested in the ionised gas kinematics in this work we simply wish to remove the (relatively smooth) stellar continuum.  We do this by fitting the stellar continuum with the penalized pixel fitting routine of \cite{Cappellari:2004p3283}, as used for our SAURON data. We were able to constrain the number of stellar templates required using our best fit to the SAURON data. Once the stellar continuum was successfully removed we were left with a cube containing the ionised gas emission only. 

\begin{figure} 
\begin{center} 
\includegraphics[width=0.5\textwidth,clip,trim=0.7cm 0.5cm 0.5cm 0.8cm]{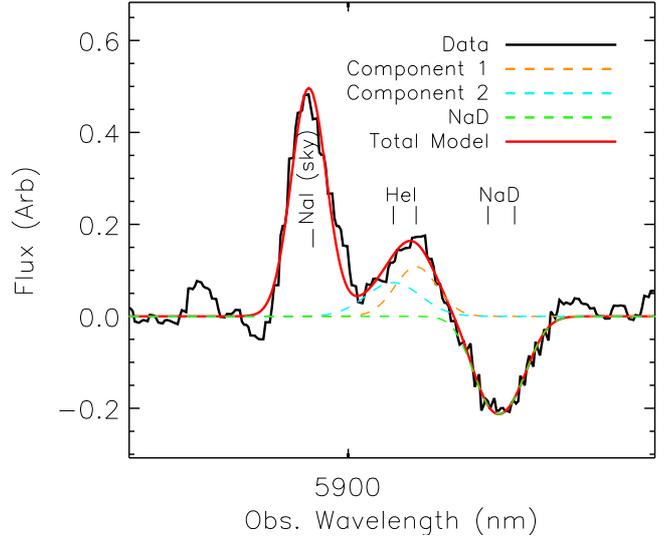}
\end{center}
\caption{\small GMOS spectrum of the NaD region from a single bin (x=-1\farc0, y=-1\farc8; black solid line). Overlaid is the two component fit produced by our fitting routine for the emission lines, as described in Section \ref{gmosfit}. Component one (nearest the galaxy systemic) is shown in orange, while the out-flowing component two is shown in blue. The red line shows the sum of component one and two (and our fit to the NaI sky line), which closely matches the observed data. Shown in green is our fit to the NaD absorption trough. }
\label{NaDspec}
\end{figure}

The spectral range of our cube includes various ionised gas emission lines. {The [OIII] and [NI] are included in our GMOS spectrum in the region which overlaps with the SAURON spectral range}. These lines however appear very weak because they are at the edge of our band pass, where the throughput is low. The main strong lines we detect are H$\alpha$, [NII]$_{6548,6583}$ and [OI]$_{6300}$, while HeI and [NII]$_{5754}$ are detected more weakly. We choose to fit the kinematics of the gas emission on the strong lines only, and impose these kinematics when measuring the fluxes of weaker lines. 
Example specta extracted from the cube are shown in Figure \ref{gmosspec}. We show here the region around the H$\alpha$, [NII] and [SII] lines only. These specta were selected to lie at the spatial position with the highest line flux (top panel), the spaxel with the largest difference between the two fitted velocities (middle panel) and at the lowest flux bin in which we can constrain two components (bottom panel). With the low spectral resolution of this data the lines are blended, however we clearly require two components to fit the line emission. We also detect Sodium D (NaD) absorption against the stellar continuum (an example spectrum is shown in Figure \ref{NaDspec}). We describe the fitting procedure used for the gas emission lines in detail in Section \ref{emission_fit}, and the procedure used to measure the parameters of the NaD absorption in Section \ref{NaD_fit}. 

\subsubsection{Emission line fitting}
\label{emission_fit}
GMOS emission line fitting was carried out as described in Section \ref{sauronfit}, with some modifications, described below. We 
fit here the H$\alpha$, [NII], [SII] and [OI] lines with single and double gaussians.
Initial guesses for the velocity and velocity dispersion were made using the derived kinematics from the SAURON cube. 
The two [NII] lines in the H$\alpha$ region of the spectrum have a fixed line ratio determined by the energy structure of the atom, and we forced the line ratios to the theoretical line ratio (F$_{6584}$=2.95$\times$F$_{6548}$: Acker 1989). The [SII] doublet is an electron density tracer, with the line ratio F$_{6731}$/F$_{6717}$ varying from 0.459 in the high density limit to 1.43 at the low density limit \citep[e.g.][]{1987JRASC..81..195D}. Here we constrain the [SII] lines ratio to lie somewhere within this region.
We here allowed the lines in the second component to have a maximum velocity dispersion of 500\kms, but again note that good fits were found in most bins with velocity dispersions $<$300\kms. 

As for the SAURON data, to ensure that the fits were robust, and spatially continuous we implemented an iterative fitting regime where the fitting processes described above were performed multiple times, using a smoothed version of the output from the previous run as the initial conditions. Here we found that the parameters usually converged within four iterations, again with very little variance between fitting attempts. Figure \ref{gmosspec} shows the GMOS spectra from a single bin (from the same spatial region as selected before), overlaid with the two component fit, as found by this procedure. 

\begin{figure*} 
\begin{center} 
\subfigure[]{\includegraphics[width=0.42\textwidth,clip,trim=0cm 0.5cm 0cm 1.0cm]{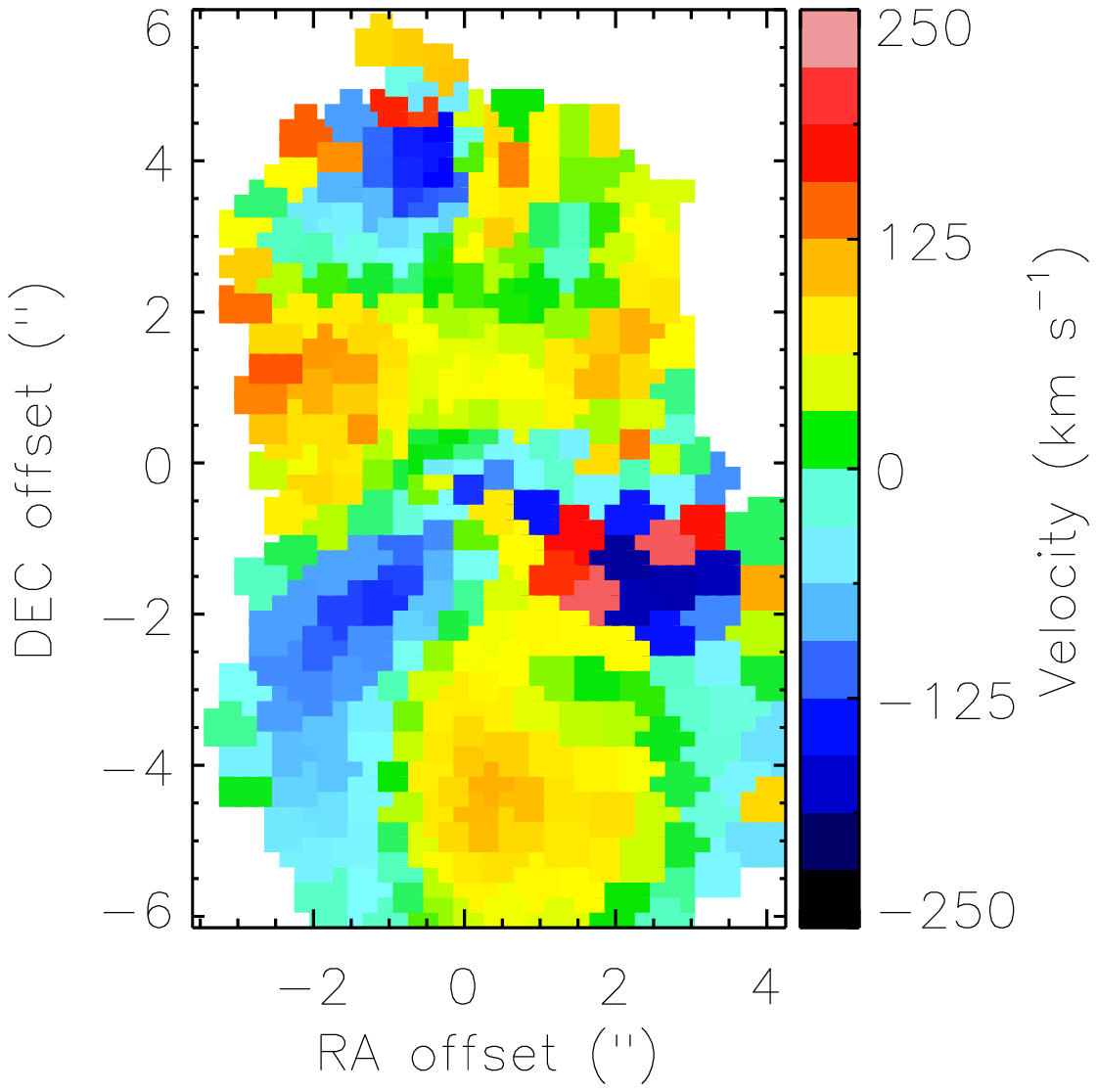}}
\subfigure[]{\includegraphics[width=0.42\textwidth,clip,trim=0cm 0.5cm 0cm 1.0cm]{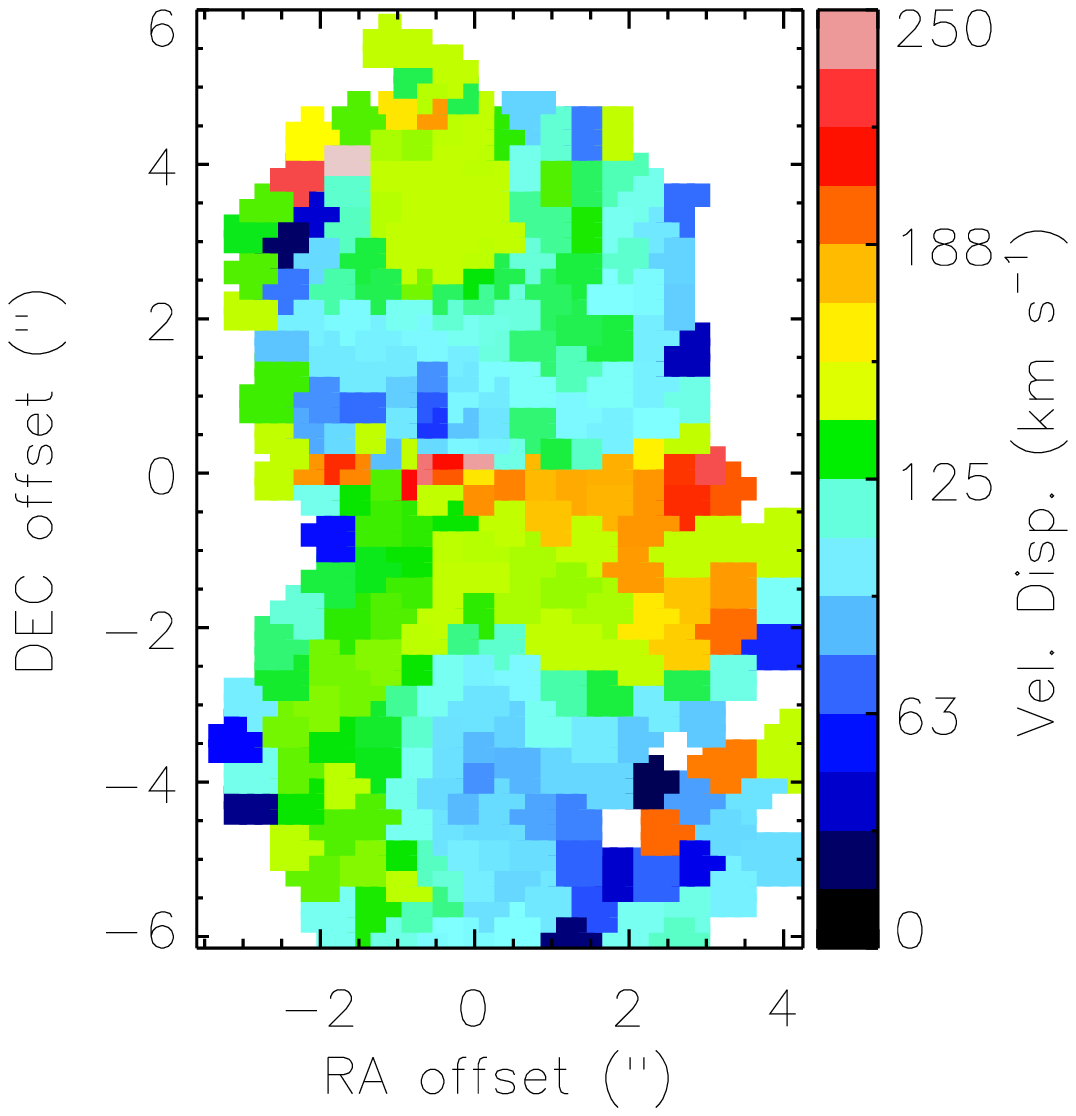}}\\
\subfigure[]{\includegraphics[width=0.42\textwidth,clip,trim=0cm 0.5cm 0cm 1.0cm]{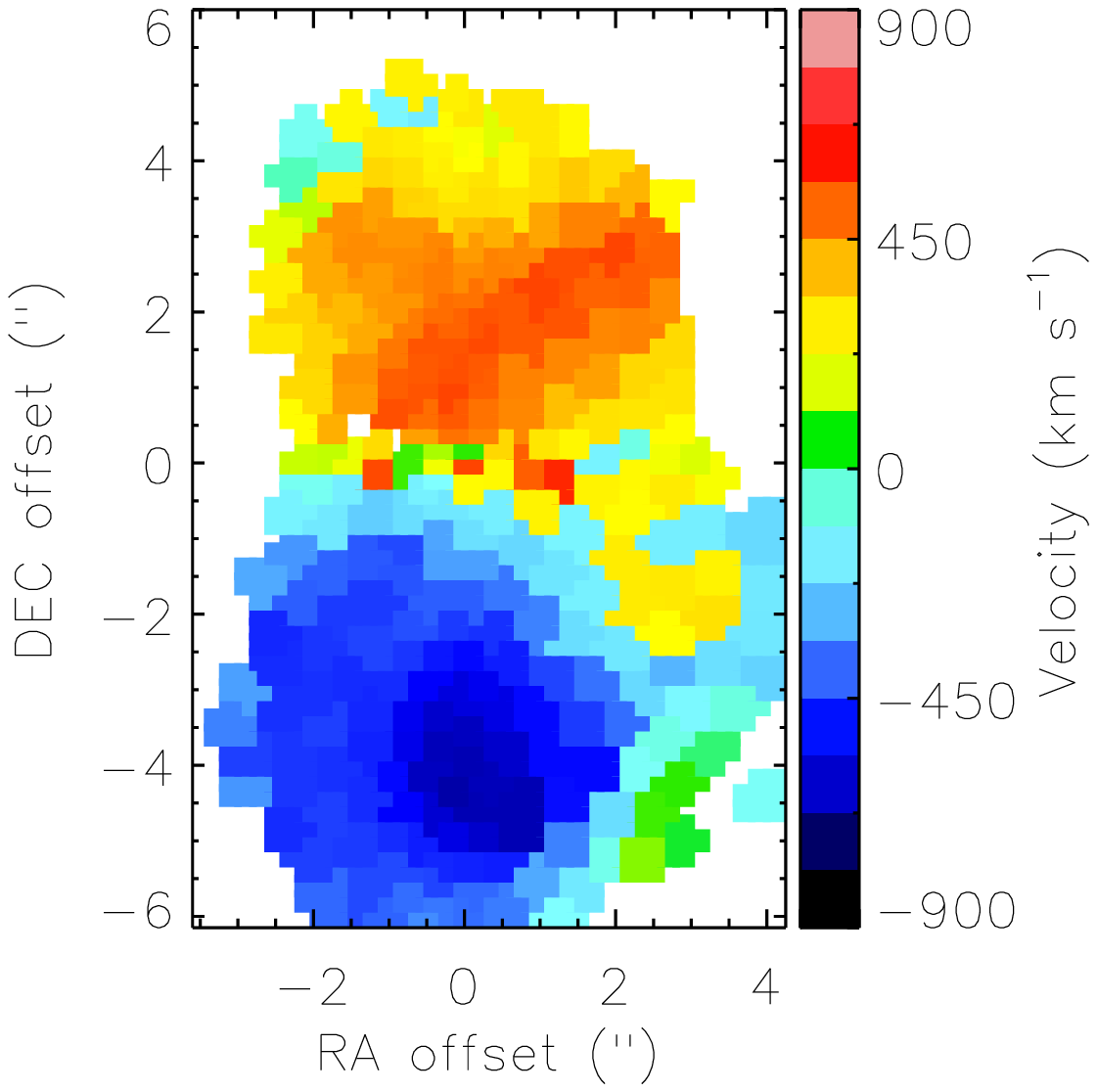}}
\subfigure[]{\includegraphics[width=0.42\textwidth,clip,trim=0cm 0.5cm 0cm 1.0cm]{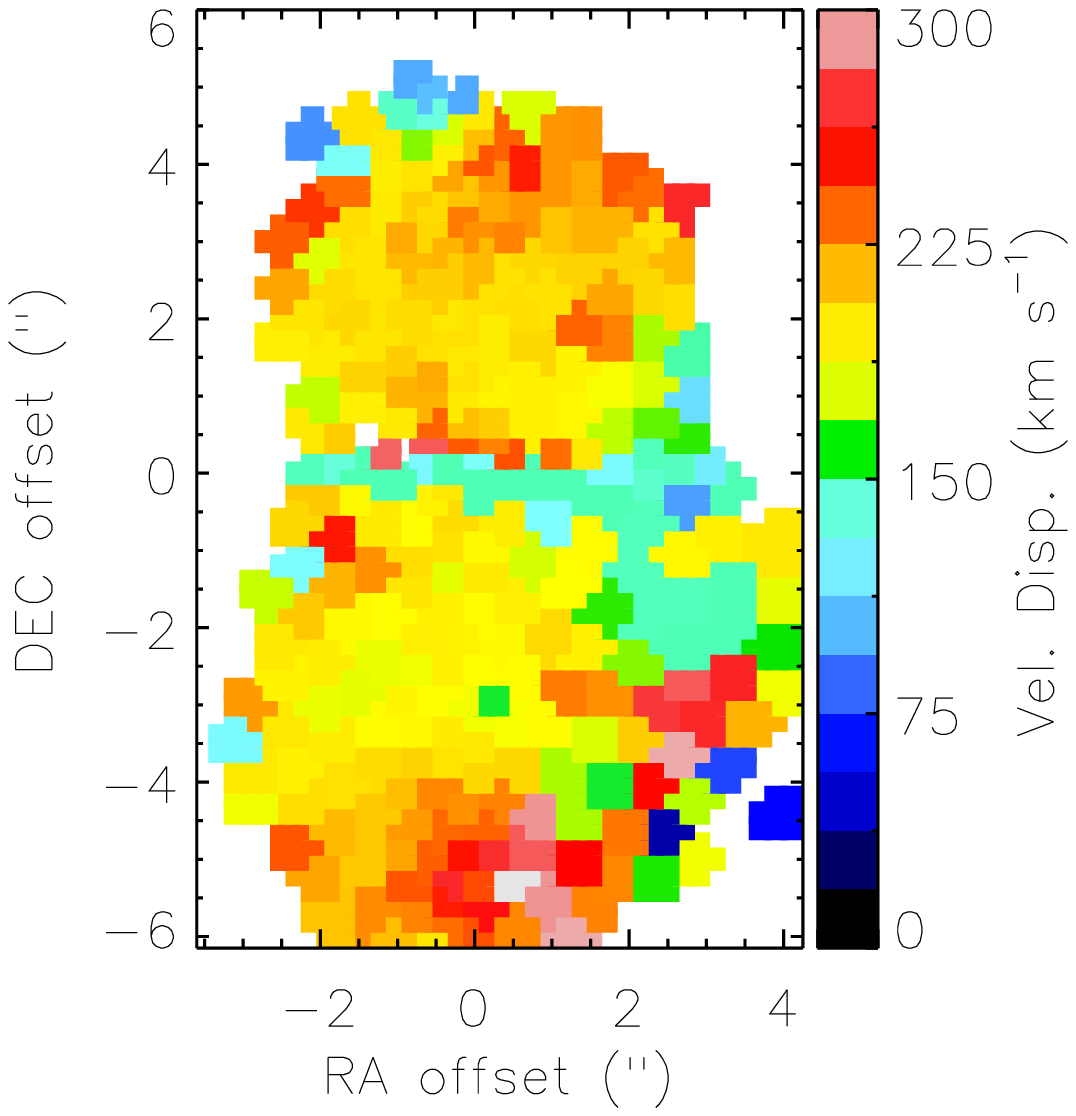}}
\end{center}
\caption{\small  Ionised gas kinematics derived from the GMOS IFU data reduction process discussed in Section \ref{gmosfit}. In the top row (panels a \& b)we display the kinematics of component one (confined to be closest to the galaxy systemic). Bins where only one ionised gas component is required are also shown in component one. The kinematics of the faster out-flowing component are shown in the bottom row (panels c \& d). The ionised gas velocity is displayed in the left panels, and the velocity dispersion in the right hand panels. }
\label{gmosplots1}
\end{figure*}

\begin{figure*} 
\begin{center} 
\subfigure[]{\includegraphics[width=0.33\textwidth,clip,trim=1cm 0.5cm 2.5cm 1.0cm]{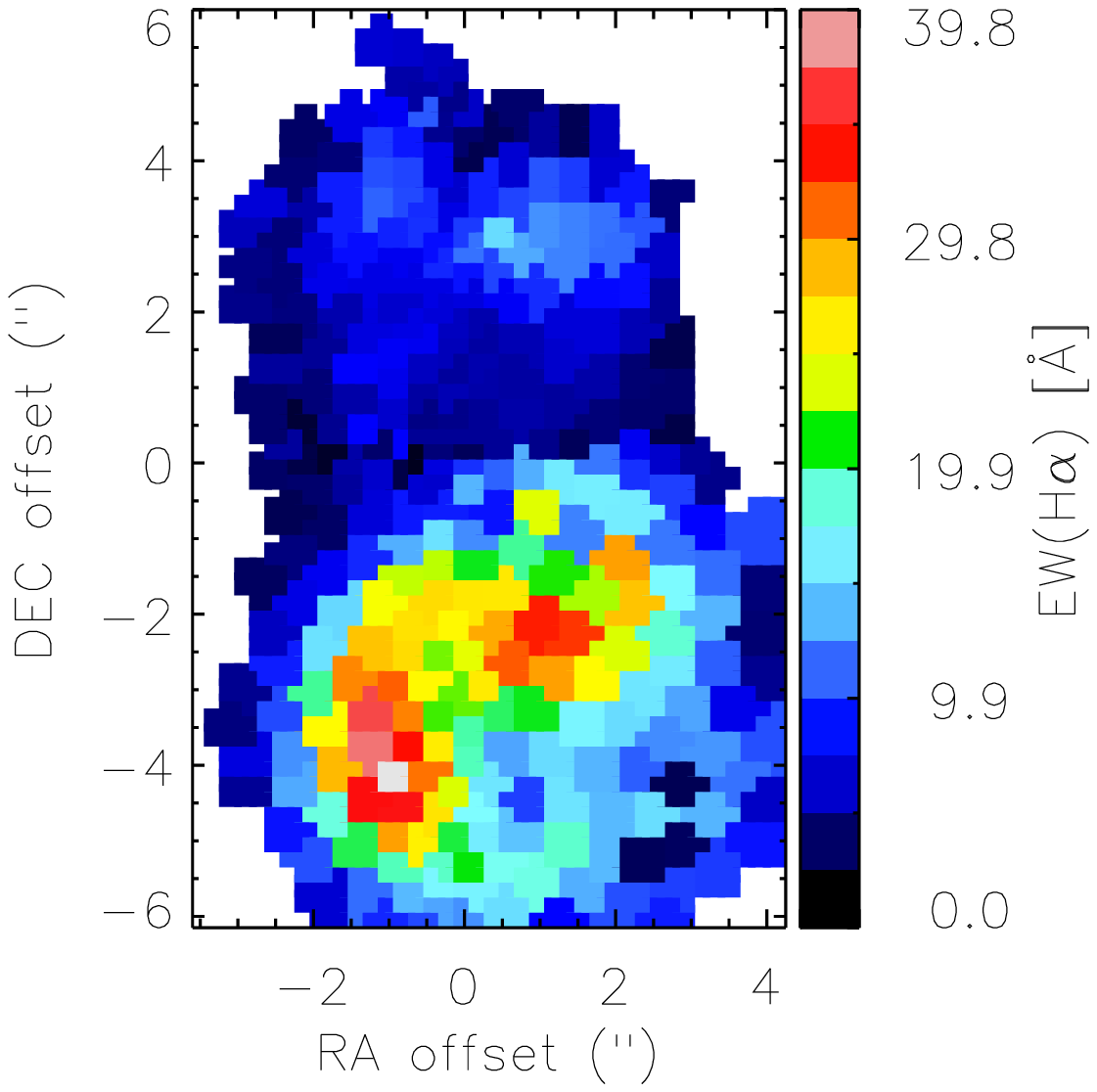}}
\subfigure[]{\includegraphics[width=0.33\textwidth,clip,trim=1cm 0.5cm 2.5cm 1.0cm]{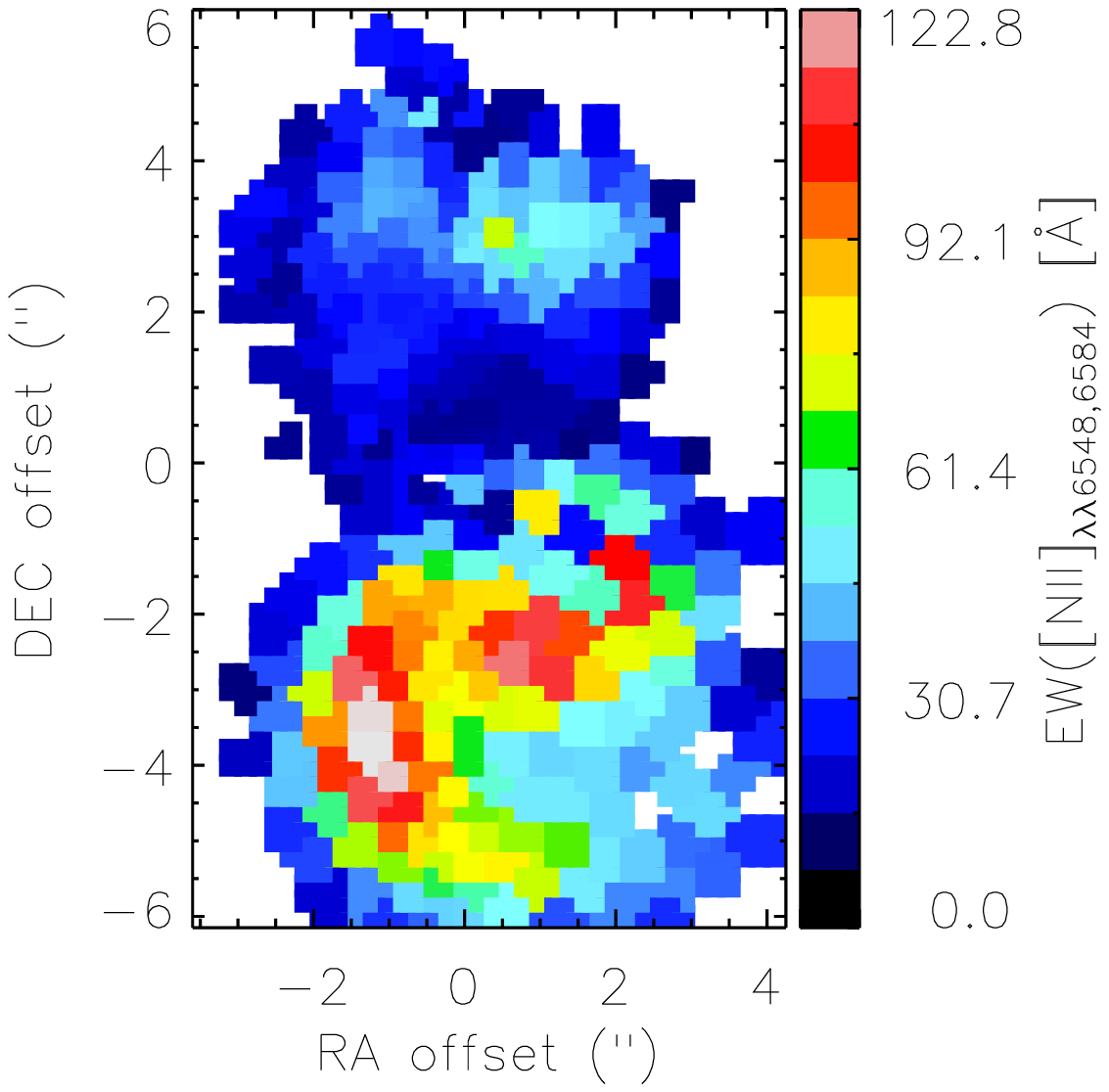}}
\subfigure[]{\includegraphics[width=0.33\textwidth,clip,trim=1cm 0.5cm 2.5cm 1.0cm]{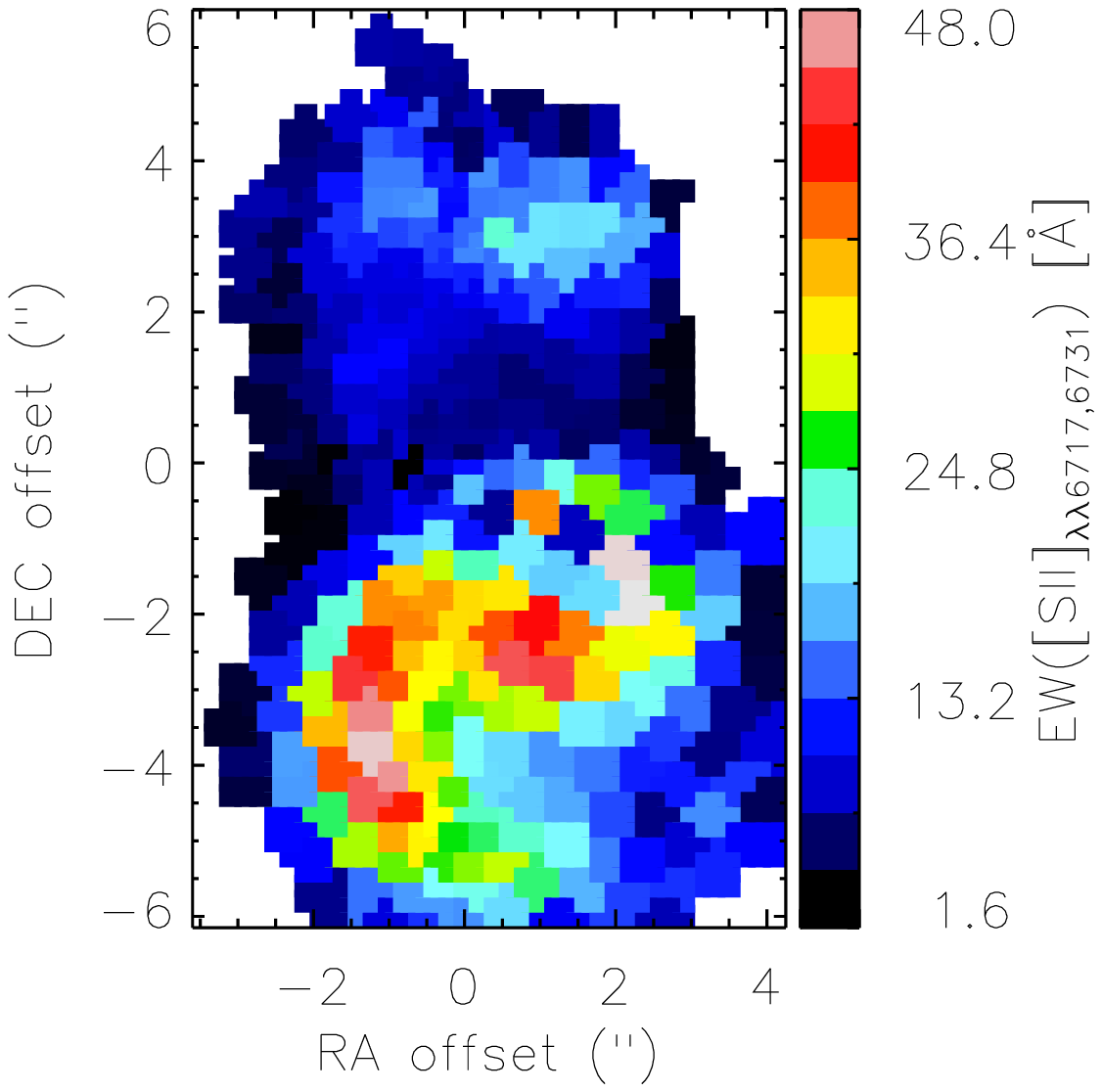}}\\
\subfigure[]{\includegraphics[width=0.33\textwidth,clip,trim=1cm 0.5cm 2.5cm 1.0cm]{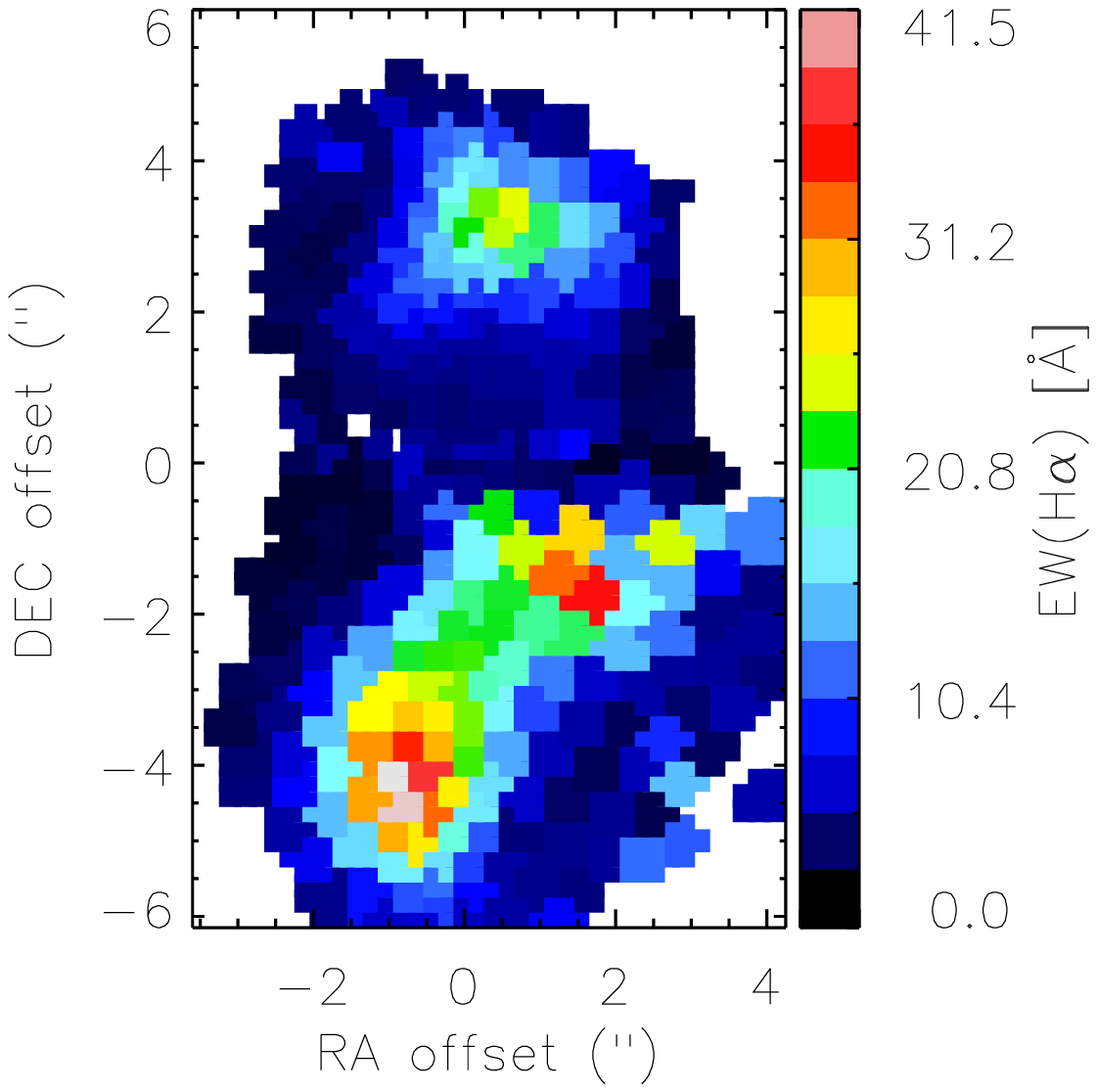}}
\subfigure[]{\includegraphics[width=0.33\textwidth,clip,trim=1cm 0.5cm 2.5cm 1.0cm]{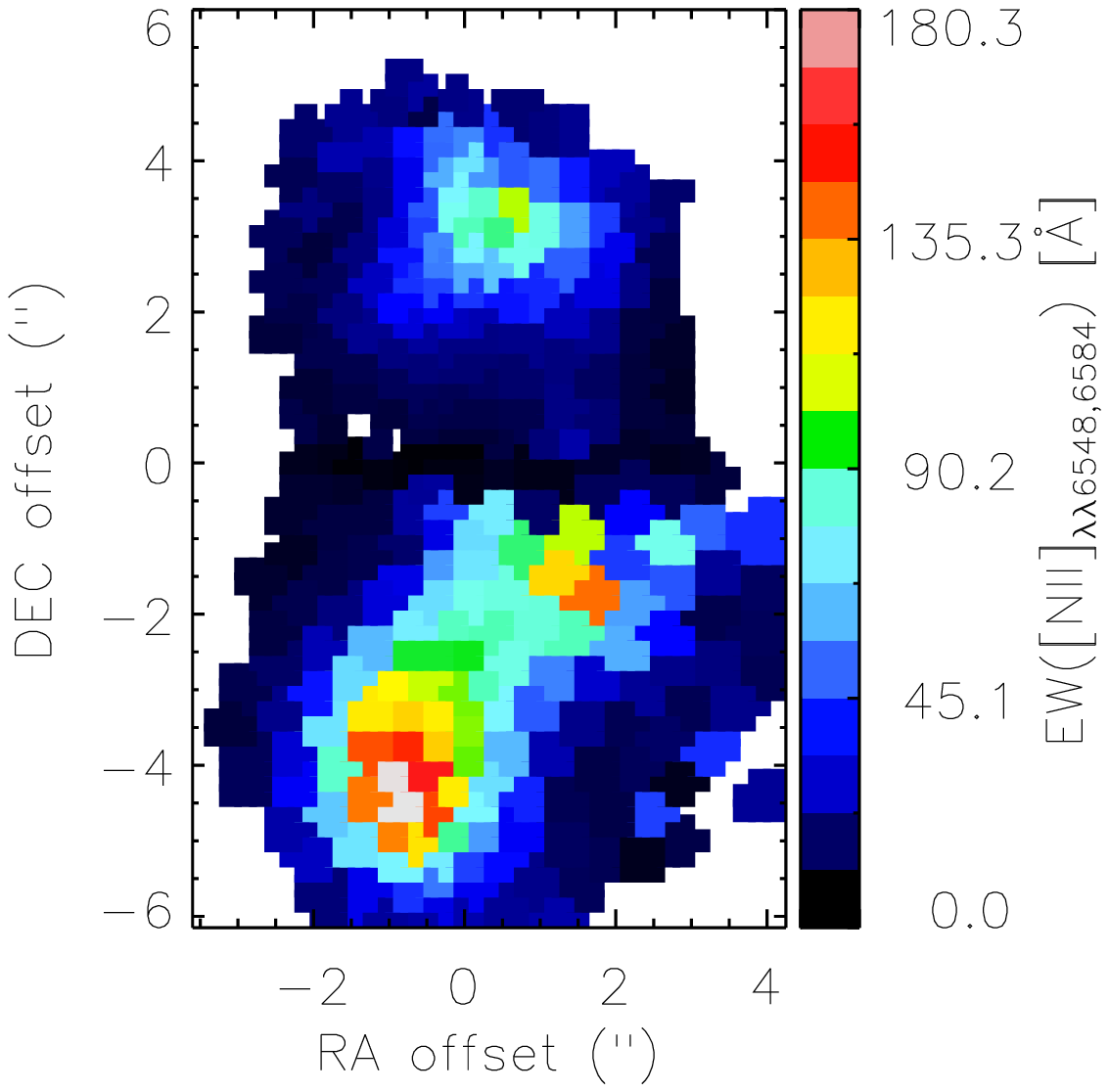}}
\subfigure[]{\includegraphics[width=0.33\textwidth,clip,trim=1cm 0.5cm 2.5cm 1.0cm]{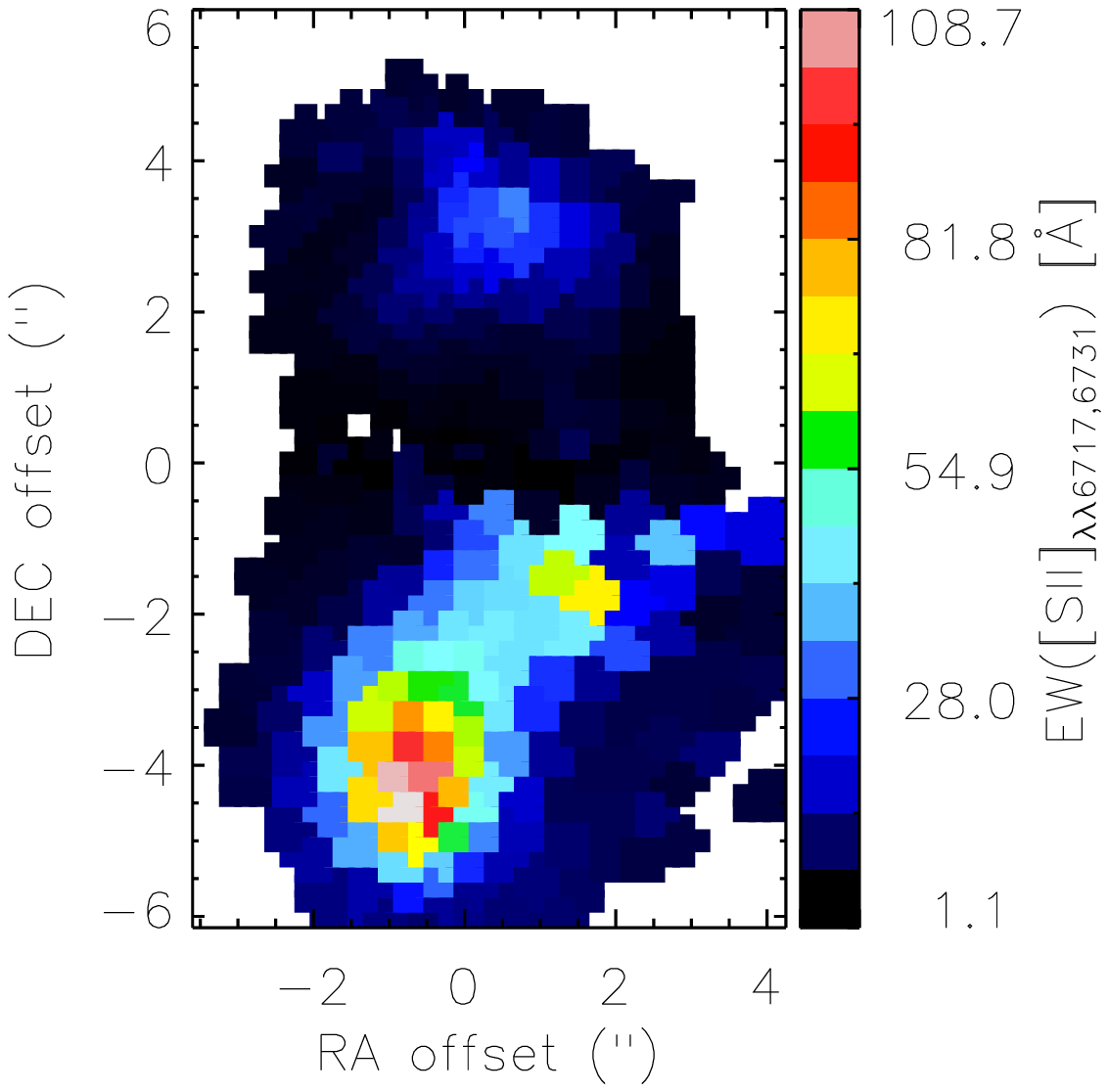}}
\end{center}
\caption{\small Ionised gas line equivalent widths derived from the GMOS IFU data reduction process discussed in Section \ref{gmosfit}. In the first row we display the EW of component one (confined to be closest to the galaxy systemic). Bins where only one ionised gas component is required are also shown in component one. The EW of the faster out-flowing component is shown in the second
row. The H$\alpha$ line EWs are shown in panels a and d, the [NII] line EWs in panels b and e, and the [SII] EWs in columns c and f. }
\label{gmosplots2}
\end{figure*}

In Figure \ref{gmosplots1} we show the observed kinematics for component one (panels a \& b) and component two (panels c \& d). The velocity dispersion maps have had the instrumental dispersion quadratically subtracted. In Figure \ref{gmosplots2} we show the fitted line equivalent widths for the strong lines, with fitted component one in the top row and component two in the bottom row. We calculate the equivalent width in the standard way by finding the width of a rectangle, with a height which is the same as the average stellar continuum flux in the region of the lines, which has the same area as the observed lines.  
As the GMOS data were taken in non-photometric conditions, we will use only ratios of the line fluxes from this point on. 

\subsubsection{Absorption line fitting}
\label{NaD_fit}

The sodium absorption lines at 5890 \AA\ and 5896 \AA\ are detected in our GMOS IFU data (Figure \ref{NaDspec}). This feature is unlikely to be due to an imperfect stellar template leaving negative residuals after subtraction from the GMOS spectrum, as the fitted velocities of the absorption feature do not match the stellar velocities assumed when fitting the template.

In order to extract the absorption depths, and determine the neutral gas kinematics we jointly fit the absorption doublet, and the neighbouring HeI emission line (and NaI skyline). We fix the  velocity and velocity dispersion of the HeI line using the best solution for each bin derived from the stronger lines (as described in Section \ref{emission_fit}). We then fit this line with two gaussian components, as before, to determine the line fluxes.

Simultaneously we fitted the NaD absorption doublet (which is blended in our data), assuming a gaussian profile for both lines. Formally an absorption line should be fitted with a Voigt profile, as the absorption has an intrinsic Lorentzian shape, which has been convolved with the instrumental gaussian response. In the low spectral resolution data we present here however we fit gaussian profiles, as the instrument response function is much broader than the intrinsic absorption. If one fits a Voigt profile to our data, the best fit profiles always tend towards a pure gaussian, with a Lorentzian width ($\Gamma$) of zero, validating such an approach. We do not fix the NaD velocity and velocity dispersion, as the absorption arises from a different gas phase, which may have different kinematics (see Section \ref{nadresults}). The velocities of the NaD hosting gas are constrained in our fit to lie within 1000 \kms of the systemic velocity, and the velocity dispersion of this component is constrained to be greater than the instrumental, but less than 500\kms. At the spectral resolution of our data we only require a single neutral gas component in all spaxels in order to fit the absorption profiles well.

\begin{figure*} 
\begin{center} 
\subfigure[]{\includegraphics[width=0.33\textwidth,clip,trim=1cm 0.5cm 1.8cm 0.8cm]{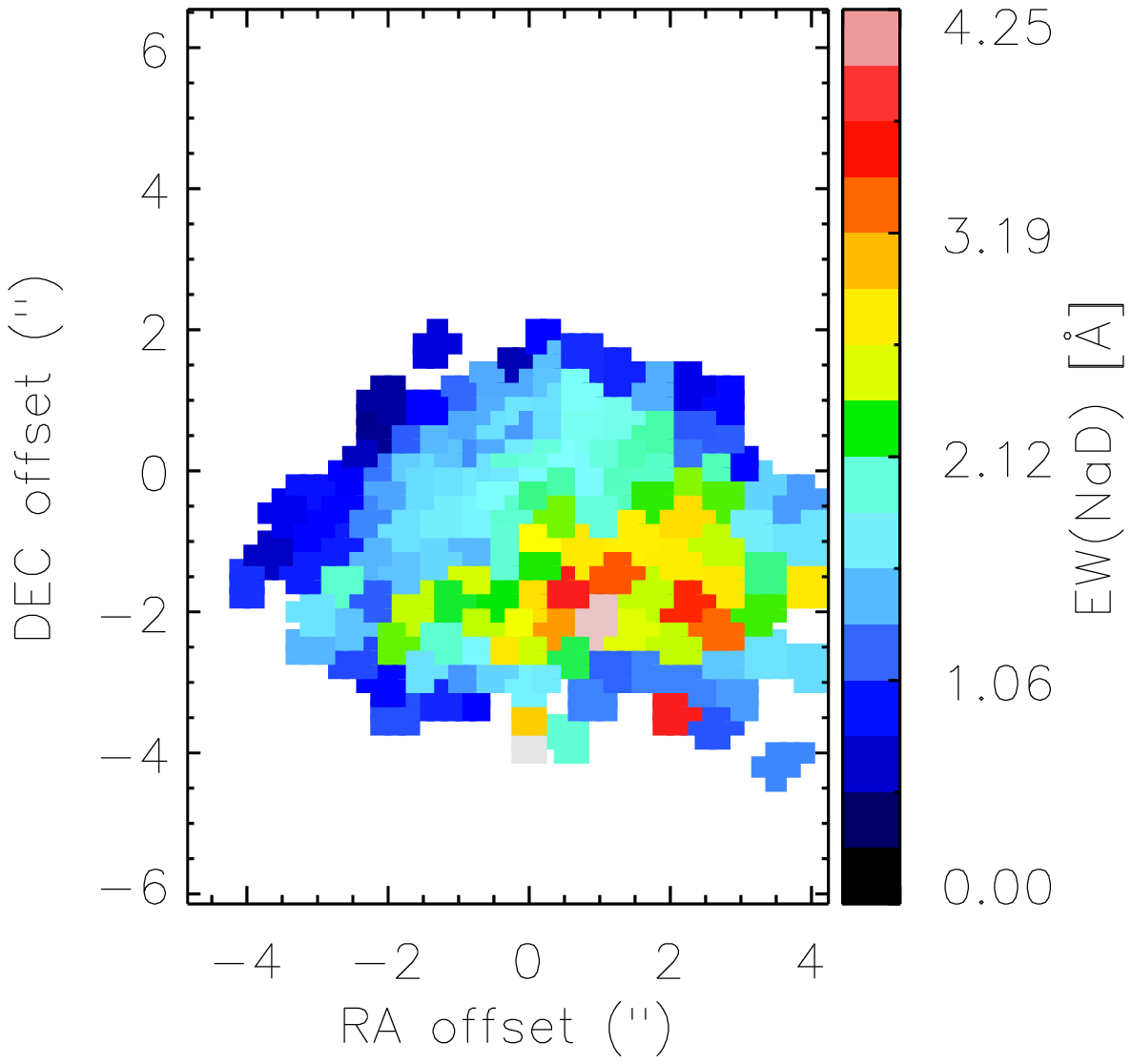}}
\subfigure[]{\includegraphics[width=0.33\textwidth,clip,trim=1cm 0.5cm 1.6cm 0.8cm]{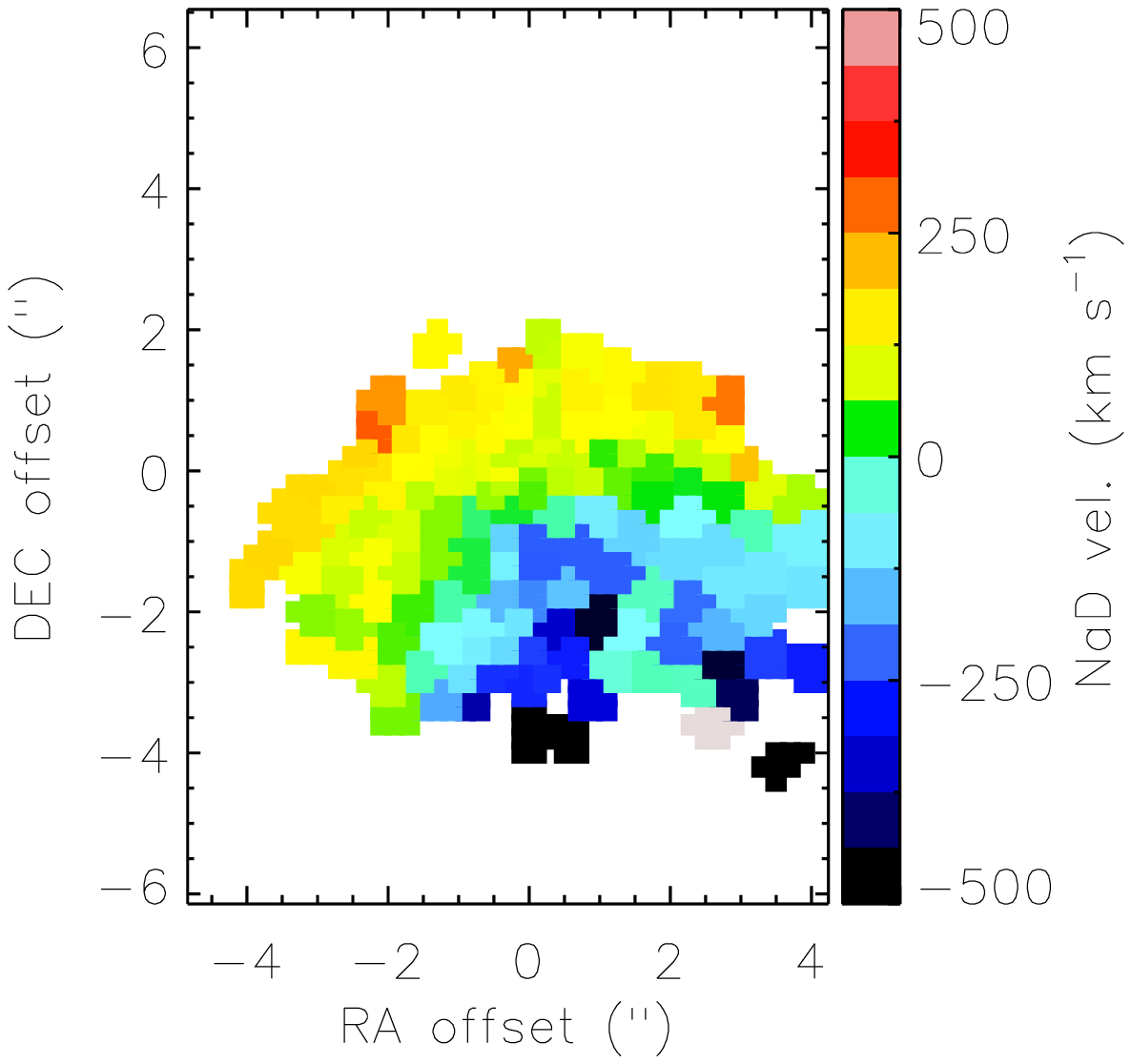}}
\subfigure[]{\includegraphics[width=0.33\textwidth,clip,trim=1cm 0.5cm 1.8cm 0.8cm]{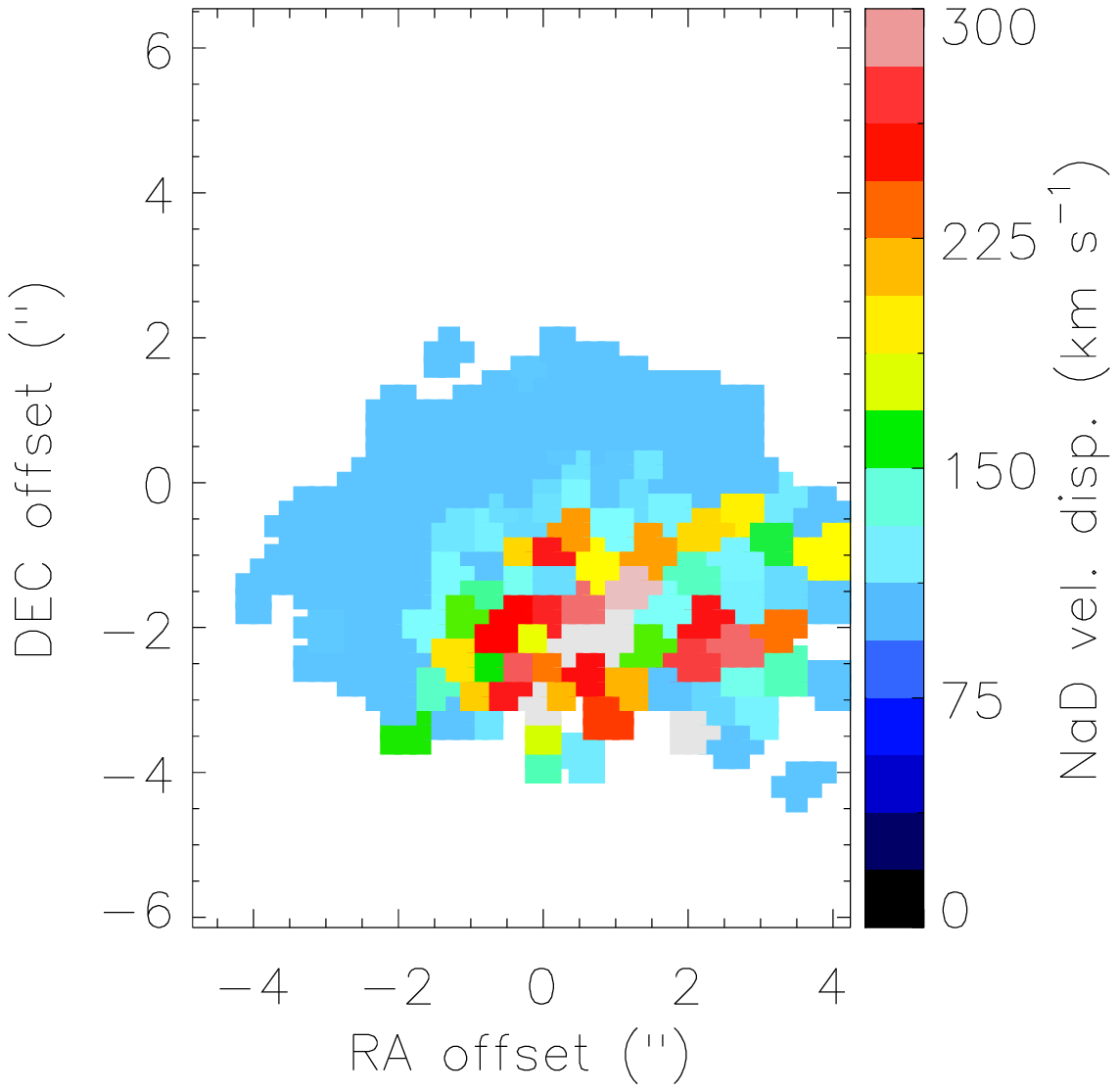}}
\end{center}
\caption{\small NaD atomic gas kinematics derived from the GMOS IFU data reduction process discussed in Section \ref{NaD_fit}. In panel a we display the equivalent width of the absorption line. Panel b shows the derived gas kinematics, and panel c the derived velocity dispersion.}
\label{NaDplots}
\end{figure*}

In Figure \ref{NaDplots} we show the observed absorption equivalent width, and the kinematics for the NaD absorbing gas. 
We calculate the equivalent width in the standard way, as above.

\begin{figure*} 
\begin{center} 
\subfigure[]{\includegraphics[height=5cm,clip,trim=1cm 0.5cm 2.5cm 1.0cm]{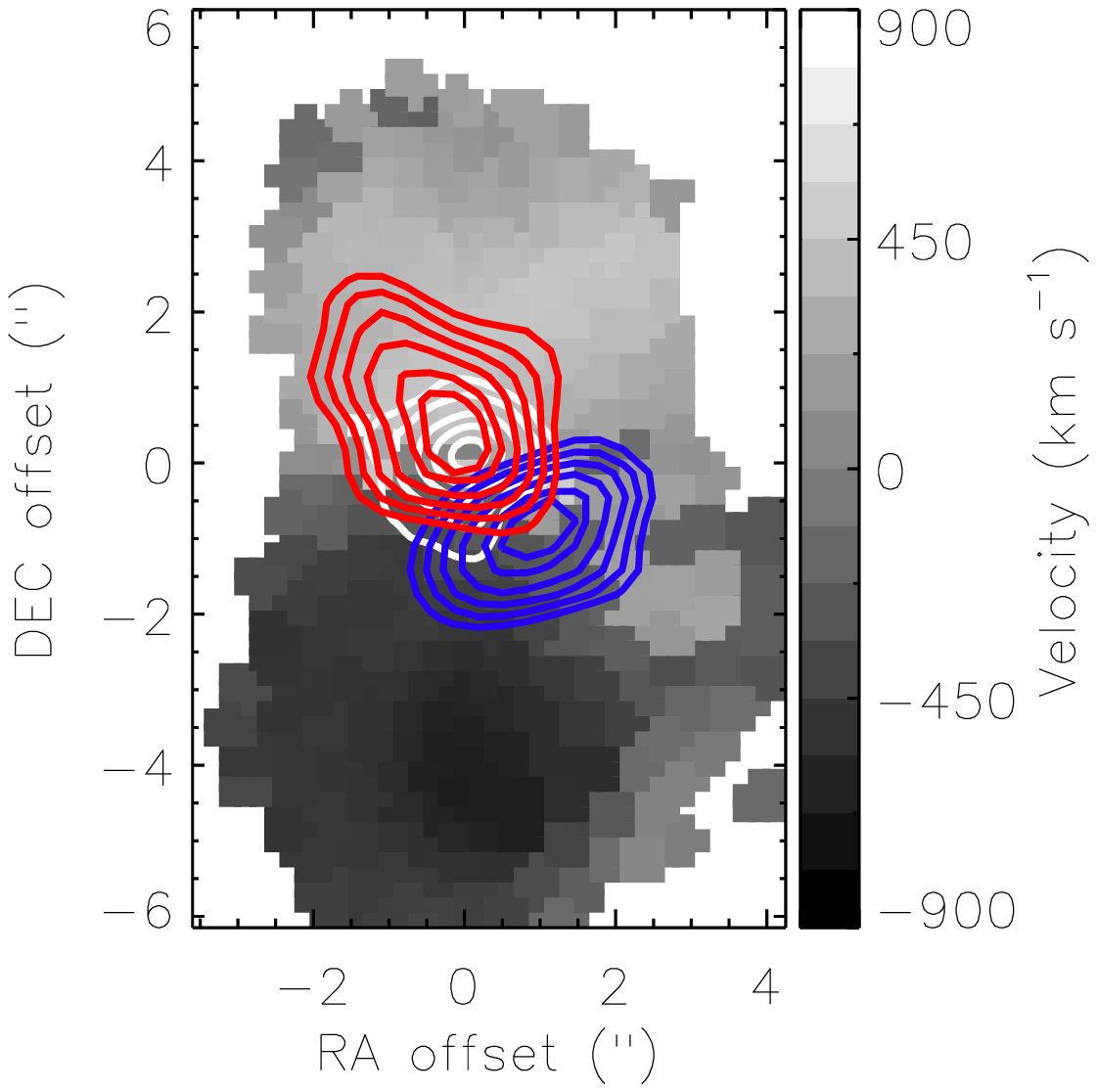}}
\subfigure[]{\includegraphics[height=5cm,clip,trim=1cm 0.5cm 2.5cm 1.0cm]{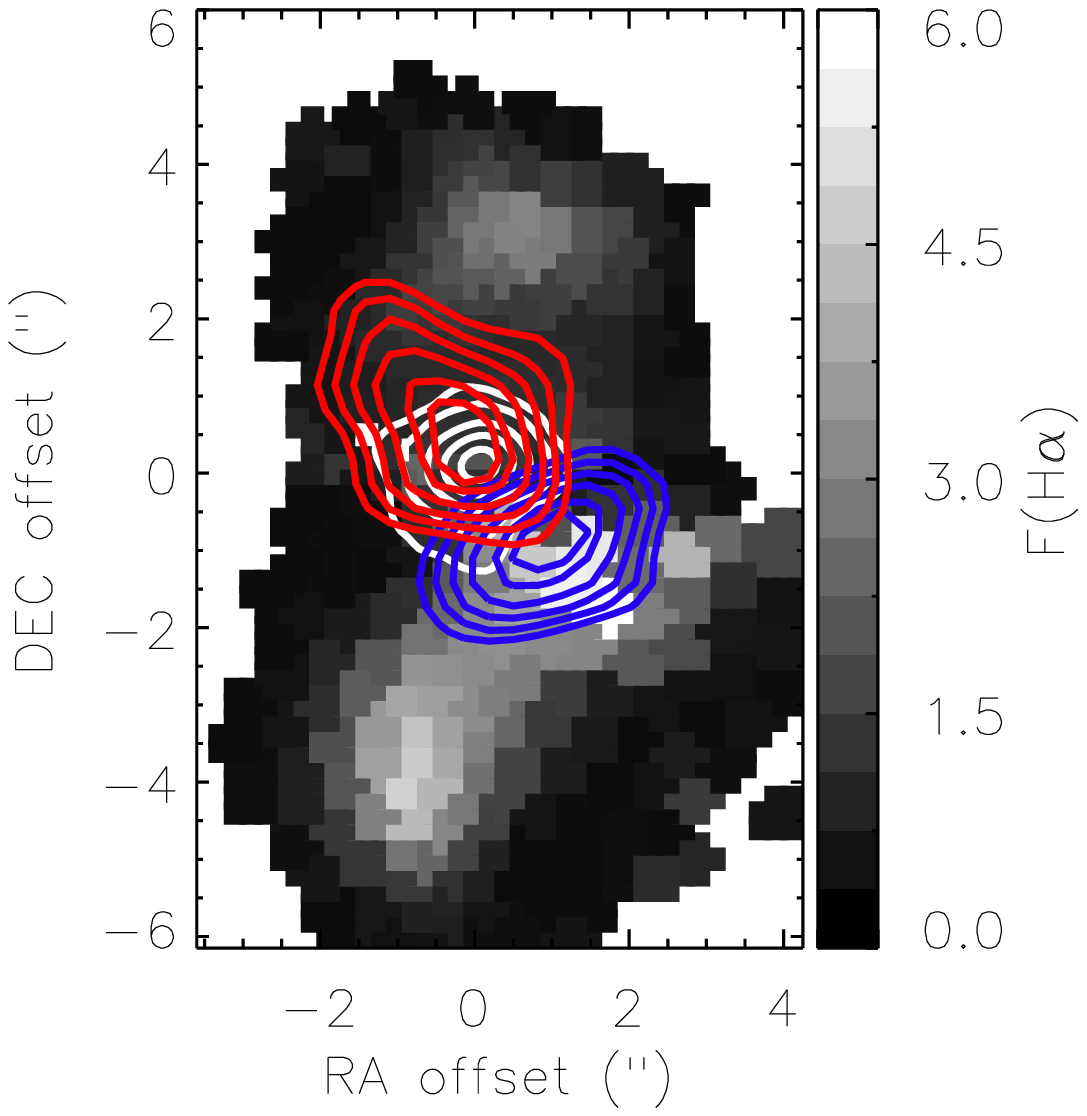}}
\subfigure[]{\includegraphics[height=5cm,clip,trim=1cm 0.5cm 1.5cm 1.0cm]{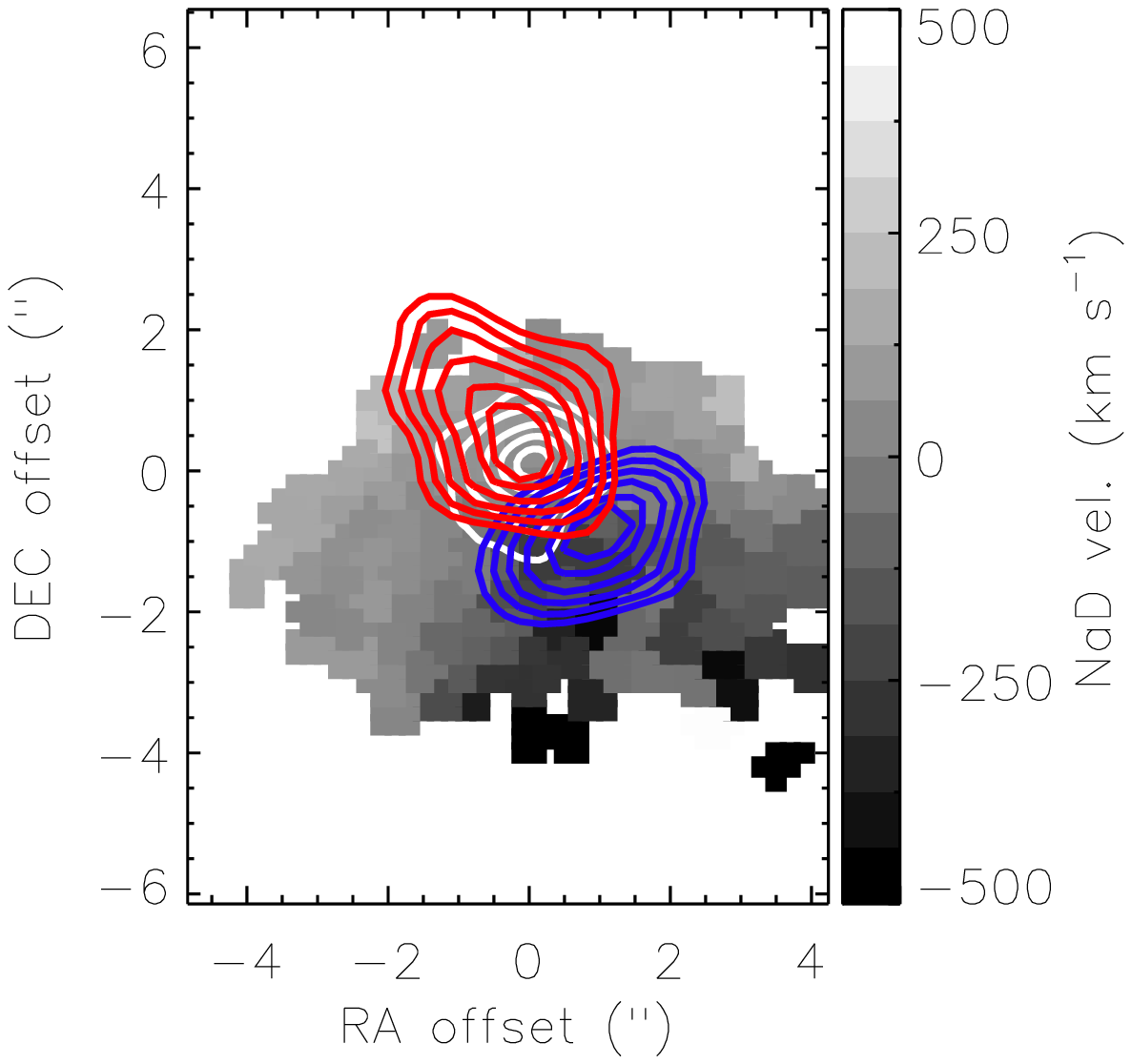}}
\end{center}
\caption{\small GMOS Ionised gas outflow kinematics (panel a), H$\alpha$ outflow EW (panel b) and NaD kinematics (panel c), as in Figures \protect \ref{gmosplots1}, \ref{gmosplots2} and \ref{NaDplots} but plotted in greyscale. These are overplotted with the CO observations of A2011. White contours show the molecular core, and blue and red contours the redshifted and blue shifted gas, respectively. }
\label{GMOSplotsCO}
\end{figure*}

\section{Discussion}
\label{results}
\subsection{Ionised gas emission line kinematics}
{
Figures \ref{sauronplots1} and \ref{gmosplots1} show the ionised gas kinematics derived from our multi-gaussian fitting procedure. The SAURON data has a much wider field of view, providing insight into the large scale kinematics of this object. The GMOS IFU data zooms in to the central portions of this object to reveal the inner regions.
}

{
Panel a of Figure \ref{sauronplots1} shows the SAURON ionised gas kinematics for the component nearest the galaxy systemic velocity (which we will hereafter call the \textit{systemic} component).  This component includes gas out to a radius of $\approx$10\arcsec\ (1.5 Kpc). {A2011} discussed the molecular gas distribution, and find evidence for a rotating molecular disk, as well as a molecular outflow. The origin of this systemic component, and its relation to the molecular disk is unclear. This component could be due to unrelated gas components at different locations along the line of sight, or it may be a coherent structure which has been disturbed.
}

{
 Some degree of symmetry appears to exist in the gas distribution around a line inclined $\approx$20$^{\circ}$ from East. This may suggest some bulk rotation, with a kinematic position angle of $\approx$30$^{\circ}$.  If this component is rotating, then comparison with the stellar rotation (shown in the bottom row of Figure \ref{sauronplots1}) shows that it is misaligned from the stellar rotation by $\approx$90$^{\circ}$,  suggesting it could be in the polar plane.  In the inner $\approx$6\arcsec however the velocity field changes sign, and is disturbed.
}

{
Figure \ref{gmosplots1} shows the GMOS zoomed in view of the centre of NGC\,1266. 
The same disturbed features present in the \sauron\ data are observed in the GMOS velocity field (Panel a of Figure \ref{gmosplots1}). These features persist whatever our choice of initial velocities for the fitting procedure, and are located at the same spatial location as the most blue- and red-shifted gas in component two. When fitting multiple components to observed velocity profiles its always possible that such reversed sign velocity structures are a result of a mis-fitting or over-fitting of the line components. Here however we find the same structure from both the \sauron\ and GMOS data independently. Our iterative fitting procedure described in Sections \ref{sauronfit} and \ref{emission_fit} is designed to avoid discontinuous fits, and thus tries to remove such disturbed structures, but is unable to find better fits in these spaxels. The middle panel of Figures \ref{sauronspec} and \ref{gmosspec} show the emission line spectrum in the bin with the largest blue-shifted velocity in each dataset (where the disturbed systemic component is also detected). Clearly the outer edges of each blended line or line-complex (which drive our determination of the relative velocities of the two components) are well fit, increasing our confidence that this structure is real.}

{
If the systemic component is a misaligned rotating structure then we speculate that the disturbed structure visible in the inner parts could be where the molecular outflow detected in A2011 has punched through, and material is flowing inward to fill the vacuum. 
Given the dynamical timescale of gas at this radius ($\approx$30-40 Myr; A2011) it is difficult to imagine that the disturbed gas in the centre of this object could exist for long without being smeared out. This suggests that either this feature is very young, or we are indeed observing unrelated clouds along the line of sight, which are not dynamically linked. A2011 estimate the age of the molecular outflow in this object as $\approx$2.6 Myr, so a recent cause of this feature is not completely ruled out.
Alternatively if these are unrelated clouds long the line of sight, the blue features south of the nucleus may be directly related to the outflow. Close correlation between some of these features and the radio jet support this conclusion (see Section \ref{driving}). Higher spatial resolution observations would allow us to disentangle these two possibilities.
}

{
Panel c of Figure \ref{sauronplots1} shows the global \sauron\ view of the gas kinematics where a second component was required. This gas appears to be in a two lobed structure, orientated approximately North-South (misaligned from the kinematic axis of the stars by $\approx$70$^{\circ}$ \citep{A3DII}).
The gas in this component has velocities up to $\approx$800 \kms\ away from the systemic velocity. We denote this component as the \textit{outflow} from this point. 
Panel c of Figure \ref{gmosplots1} shows this component in more detail. With the finer spatial sampling of the GMOS-IFU map we are able to detect gas with velocities up to $\approx$900\kms away from the systemic. 
As argued in A2011 the escape velocity in the centre of this object is at most $\approx$340\kms\, supporting the idea that the outflow in this system allows material to escape the galaxy. 
The molecular gas in the outflow of this object reaches velocities of up to $\approx$480 \kms, with a slightly smaller  spatial extent (see Figure \ref{GMOSplotsCO}). This could suggest molecular gas is destroyed as it flows out of the galaxy, or that our observations were not sensitive to detect emission from the molecular gas at the highest velocities.
}

{
The molecular gas in NGC\,1266 dominates the total mass of the ISM (with a mass of $\approx$2$\times$10$^{9}$ M$_{\odot}$), and is contained within the inner most $\approx$100 pc of the galaxy (A2011). It is very hard to explain how this gas lost its angular momentum if it was already present within the galaxy. Deep optical imaging shows some minor signatures that could be due to recent disturbances, but no signatures of a stellar bar (Duc et al., in prep). 
A minor merger could explain the compactness of the gas, if the merger happened with the correct initial parameters to leave the gas with little or no angular momentum. The dynamical mass (M$_{1/2}$) of NGC1266 within a sphere of radius r$_{1/2}$ (containing half of the galaxy light) is 1$\times$10$^{10}$ M$_{\odot}$ (Cappellari et al., in prep). 
A minor merger with a smaller galaxy could provide the gas we see, and with a stellar mass ratio of $\approx$5:1 (assuming the smaller galaxy was gas rich, with a gas fraction of one) may not leave visible traces in the luminosity weighted galaxy kinematics.
}

{
In \cite{2011MNRAS.417..882D} we analyzed the kinematic misalignment of the ionised gas in \atlas\ galaxies (extending the work of \citealt{Sarzi:2006p1474}) in order to gain clues about the origin of the gas. In that work we suggested that gas with kinematic misalignments $>$30$^{\circ}$ almost certainly have externally accreted gas.
If the systemic component is rotating (and its misalignment from the stellar rotation axis is not caused by orbit branching or similar; \citealt{1985A&A...150...97P,1985CeMec..37..387C}) then this would once again argue for a recent minor merger/accretion.
}

\subsection{NaD absorption}
\label{nadresults}

The sodium absorption doublet at 5890 \AA\ and 5896 \AA\ provides a good probe of the cold ISM in the outflow. 
The ionisation potential of NaI is lower than that of hydrogen, at only 5.1 eV.
The photons that ionize NaI are thus in the near-UV ($\lambda\approx$ 2420 \AA). These lines primarily probe the ISM in the warm atomic and cold molecular phases, where there is sufficient dust extinction to allow neutral sodium to survive \citep{1978ppim.book.....S}. For NaD lines to be observed, only relatively modest optical depths and \hi\ column densities are required, which makes these lines a sensitive probe of the outflowing (or inflowing) neutral ISM. 

As shown in Figure \ref{NaDplots} we detect NaD only in the central regions, and southern part of the galaxy. As absorption lines are viewed against the stellar continuum, if blue shifted velocities are observed it is clear that the gas is outflowing, rather than inflowing. The gas kinematics show that it is likely caught in the outflow, with typical blue shifted velocities of $\approx$-250\kms\, and extreme velocities detected out to $-500$ \kms (which is well beyond the escape velocity). The position angle of the outflow traced in NaD (and molecular gas; see Figure \ref{GMOSplotsCO}) is slightly different than that traced by the ionised gas, by $\approx$35$\pm$5$^{\circ}$. 
It is unclear if the difference between the two gas phases is significant. We discuss this further in Section \ref{driving}.

The presence of NaD in the outflow provides further evidence that NGC\,1266 hosts a multi-phase outflow, and that cold gas is being removed from the galaxy. As discussed above, the outflow extends to higher velocities when traced by ionised gas than when using a dense gas tracer. Either the gas is dissociated further out in the outflow, or we no longer have the sensitivity to detect it.

To observe the NaD lines, the extinction must be sufficient for the optical depth ($\tau$) to be $\gtsimeq$ 1 at 2420 \AA\, which corresponds to an A$_\mathrm{v}\gtsimeq$  0.43 mag in the V-band (for a \citealt{1989ApJ...345..245C} extinction law) and to a \hi\ column density of $\gtsimeq$8$\times$10$^{20}$ cm$^{-2}$.
The NaD lines we observe are likely to be associated with the \hi\ in this source, which is detected in absorption by A2011. They find a total \hi\ column density of 
N$_{\hi}$ = 2.1$\times$10$^{21}$ cm$^{-2}$ in front of the continuum source in NGC\,1266, and estimate that the \hi\ column depth in the outflow is $\approx$ 8.9$\times$10$^{20}$ cm$^{-2}$, consistent with our detection of NaD. 

Our observations provide an alternative way to estimate the reddening, and thus the atomic gas column density in each spaxel. The equivalent width (EW) of the NaD absorption lines has been shown to correlate with reddening. Here we use the relation of \cite{2003fthp.conf..200T}, derived from low resolution spectroscopic observations of supernovae:

\begin{equation}
\frac{\mathrm{E}(B-V)}{ \mathrm{mag}}=-0.01+0.16\times\mathrm{EW(NaD)},
\label{ebvnad}
\end{equation}
 
\noindent where E(B-V) is the standard colour excess. This can be combined with the relation between E(B-V) and \hi\ column density, such as that derived by \cite{1978ApJ...224..132B}:

\begin{equation}
\frac{N(\mathrm{\hi})}{\mathrm{cm}^{-2}}=\frac{\mathrm{E}(B-V)}{0.2\times10^{-21}}.
\label{ebvnhi}
\end{equation}

\noindent Using these relations we find that the average \hi\ column density in the outflow of NGC\,1266, as derived from the NaD EW (displayed in Figure \ref{NaDplots}) is (1.2$\pm$0.6)$\times$10$^{21}$ cm$^{-2}$, consistent with the \hi\ absorption spectrum presented in A2011. 

The above estimate of N(\hi) is very simplistic, and assumes that the NaD absorption comes from a `screen' in front of the galaxy. In fact, due to the small spatial scales we are probing here, some of the stellar continuum detected by GMOS will come from stars in front of the outflow gas, decreasing the measured EW. The amount of contamination will change with radius. 
Just to the North of the galaxy nucleus the gas velocity is redshifted with respect to the galaxy systemic velocity. We are thus likely to be seeing through to the receding part of the outflow. 
If we assume that the outflow is symmetric, then as we only detect redshifted material at the very centre, this implies that the outflow is not in the plane of the galaxy (as otherwise the outflow on the northern side of the galaxy would be silhouetted by the same amount of stellar continuum).

It is possible to use the above facts to introduce a new simple model which can help constrain the geometry of outflows. In this model we treat the inner regions of NGC\,1266 as a planar disk, with a vertical scale height of 700 pc. The exact value we choose for the scale height does not critically change our results, so here we adopt a value at the high end of those found for the Milky Way by \cite{2001ApJ...553..184C}. As the inner surface brightness profile of NGC\,1266 (within $\approx$5\arcsec) is reasonably flat ({Krajnovic et al., submitted}) in this simple model we neglect the change in surface brightness with radius of the system. We input an outflow with a constant surface density, and a variable length (L), which is inclined from the galaxy plane by an angle $\theta$.  From Equation \ref{ebvnad} and \ref{ebvnhi} we calculate the expected EW(NaD) produced by such an outflow. The measured EW will however be contaminated by the luminosity of the stars between the observer and the absorbing material. At each radius we calculate the amount of stellar luminosity above and below the position of the outflow, and produce a model for the observed EW(NaD) profile. 

\begin{figure} 
\begin{center} 
\includegraphics[width=0.4\textwidth,clip,trim=0.0cm 0.0cm 0.0cm 0.0cm]{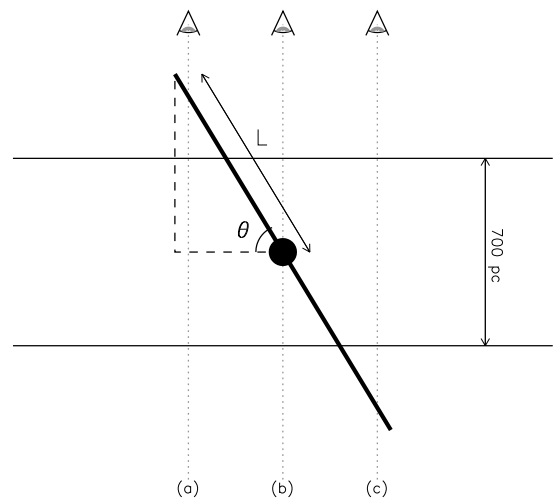}
\end{center}
\caption{\small Simple geometry assumed for the NaD absorption model described in Section \ref{nadresults}. The galaxy is represented as a flat slab with exponentially declining surface density of scale height 700 pc. The outflow has a constant surface density, a variable length (L), and an inclination from the galaxy plane of angle $\theta$. Three sight lines (referred to in the text) are labeled (a,b \& c). }
\label{nadmodel_expfig}
\end{figure}

Figure \ref{nadmodel_expfig} demonstrates the model geometry, and includes three sight-lines as examples of this procedure. In Sightline (a) the outflowing material is in front of the majority of the stars in the galaxy, and thus the expected NaD EW will be relatively uncontaminated. Sightline (b) passes through the galaxy centre, and 50\% of the stars in the galaxy are in front of the absorbing material, thus our measured EW will be 50\% smaller. Sightline (c) has almost all of the stars in the galaxy in front of the absorber, so the absorption on this sightline is likely undetectable. The exact viewing angle between the galaxy plane and the line of sight introduces a constant geometric factor, which can be neglected when one normalises the EW profile. The profile shape does not depend on the exact column density assumed for the outflow, and hence this factor is also removed when one normalises.

\begin{figure} 
\begin{center} 
\includegraphics[width=0.48\textwidth,clip,trim=0.5cm 0.0cm 0.9cm 0.0cm]{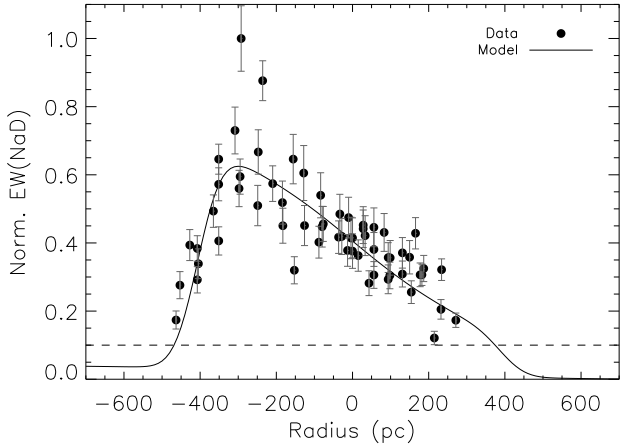}
\end{center}
\caption{\small The observed NaD EW distribution (black points) extracted from our GMOS observations in a pseudo slit of 1\arcsec\ wide, aligned in the north-south direction over the centre of the galaxy. The solid black line is the best fit model for the outflow EW produced in the way described in Section \ref{nadresults}, convolved with a gaussian to match the average seeing of our observations. The best model displayed here has an outflow inclination (with respect to the galaxy disk) of 53$^{\circ}$, and a linear outflow size of 400 pc. The dashed line shows the detection limit of our observations.}
\label{nadmodelfig}
\end{figure}

Models of the above sort can be fitted to the observed NaD EW distribution. Figure \ref{nadmodelfig} shows the observed NaD EW distribution (black points) extracted from our GMOS observations in a pseudo slit of 1\arcsec\ wide aligned in the North-South direction over the centre of the galaxy (along what seems to be the major axis of the outflow). Overplotted on the observed data is the best fit model for the outflow produced in the way described above (where additionally the model has been convolved with a gaussian to match the average seeing of our observations). Models with inclinations between 0 and 90$^{\circ}$, and with sizes between 1 and 500 pc were created, and compared to the data by calculating the reduced chi-squared. The statistical error is calculated by finding the area of parameter space around the best fit where the reduced chi-squared changes by 1.  The best fit model has an inclination (between the galaxy plane and the outflow) of 53$\pm$8$^{\circ}$, and a size of 400$\pm$50 pc. A line corresponding to such a model is shown in Figure \ref{nadmodelfig}. 

A2011 estimate the outflow to have a total extent of 465 pc, and an inclination with respect to the plane of the sky of roughly 20$^{\circ}$ (using the the average offset
between the centroids of the outflow with respect to the nucleus divided by the average extent of the lobes). If we use the inclination of 34$\pm$5$^{\circ}$ for the galaxy, as in A2011, combined with our estimate of the outflow inclination (with respect to the galaxy) this leads to an estimate of the outflow inclination with respect to the plane of the sky of 18.5$\pm$10$^{\circ}$. Both the inclination of the outflow and its size derived from our simple model are thus consistent with the estimates presented in A2011.
If we alternatively use the inclination of 26$\pm$5$^{\circ}$ estimated in \cite{2011MNRAS.414..968D}, within our error bars we still get a result consistent with the estimate of the outflow inclination in A2011.

This conclusion that the outflow in NGC\,1266 is orientated out of the plane of the galaxy is supported by the observed kinematic misalignment between the galaxy stellar kinematics, the ionised gas kinematics, and the outflow. A2011 also argue that the outflow in NGC\,1266 is orientated out of the plane of the galaxy, based on the differential reddening between the northern and southern lobes of ionised gas emission. In Seyfert galaxies no correlation between the direction of AGN jets and the galactic plane exists (e.g. \citealt{1984ApJ...285..439U}), hence the misalignment of the outflow in this galaxy is perhaps not unexpected.

Two parameter models of this type employed here are highly simplistic, and ignore a large amount of relevant physics. Despite these reservations a good fit to observational data can be obtained (as shown in Figure \ref{nadmodelfig}), and the estimated parameters seem reasonable. This highlights the power of NaD observations to constrain the geometry of outflows. 

\subsection{Gas excitation}

\begin{figure} 
\begin{center} 
\subfigure[]{\includegraphics[width=0.48\textwidth,clip,trim=0.5cm 0.0cm 0.9cm 1.0cm]{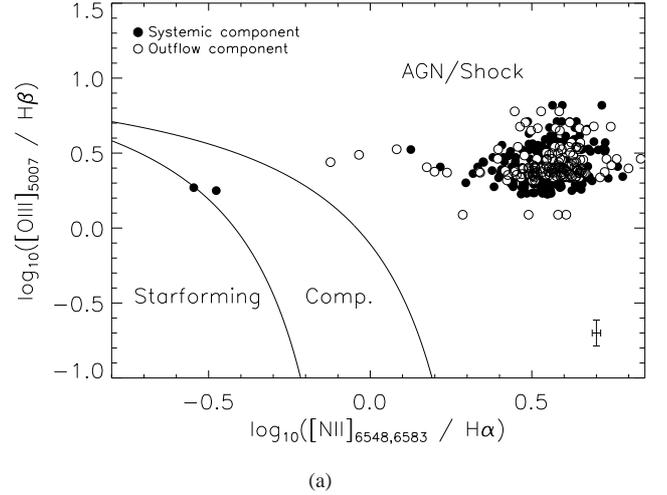}}\\
\subfigure[]{\includegraphics[width=0.48\textwidth,clip,trim=0.5cm 0.0cm 0.9cm 1.0cm]{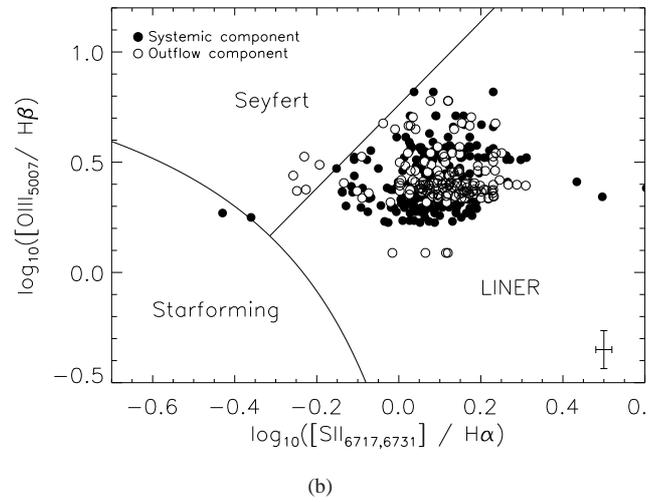}}\\
\subfigure[]{\includegraphics[width=0.48\textwidth,clip,trim=0.5cm 0.0cm 0.9cm 1.0cm]{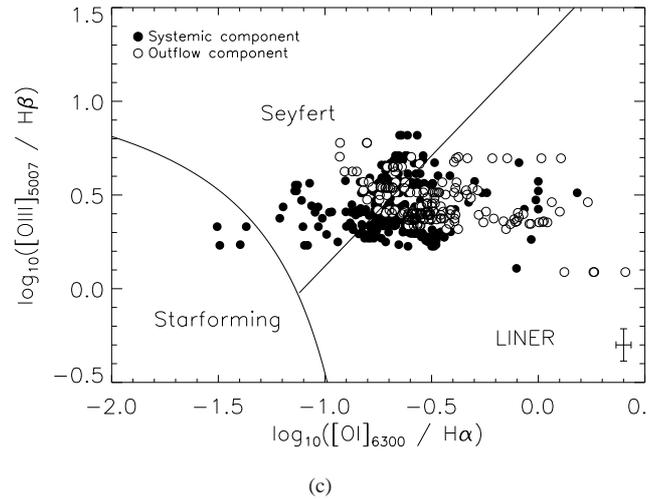}}
\end{center}
\caption{\small BPT \protect \citep{1981PASP...93....5B} type diagrams for the inner parts of NGC\,1266.  The Y-axis of each plot shows the [OIII]/H$\beta$ ratio derived from our SAURON data, and it is plotted versus the [NII]/H$_{\alpha}$ (panel a; top), [SII]/H${\alpha}$ (panel b; middle) and [OI]/H${\alpha}$ (panel c; bottom) line ratios from our GMOS observations. In the bottom right of each plot is the typical error bar associated with each point, derived from the formal fitting errors and a monte-carlo error analysis of the fluxes returned by our fitting routines. We overplot the diagnostic lines of \protect \cite{2001ApJ...556..121K,2006MNRAS.372..961K}, which indicate the dominant line excitation mechanism.  }
\label{bptdiags}
\end{figure}

\begin{figure} 
\begin{center} 
\subfigure[]{\includegraphics[width=0.48\textwidth,clip,trim=0.5cm 0.0cm 0.6cm 1.0cm]{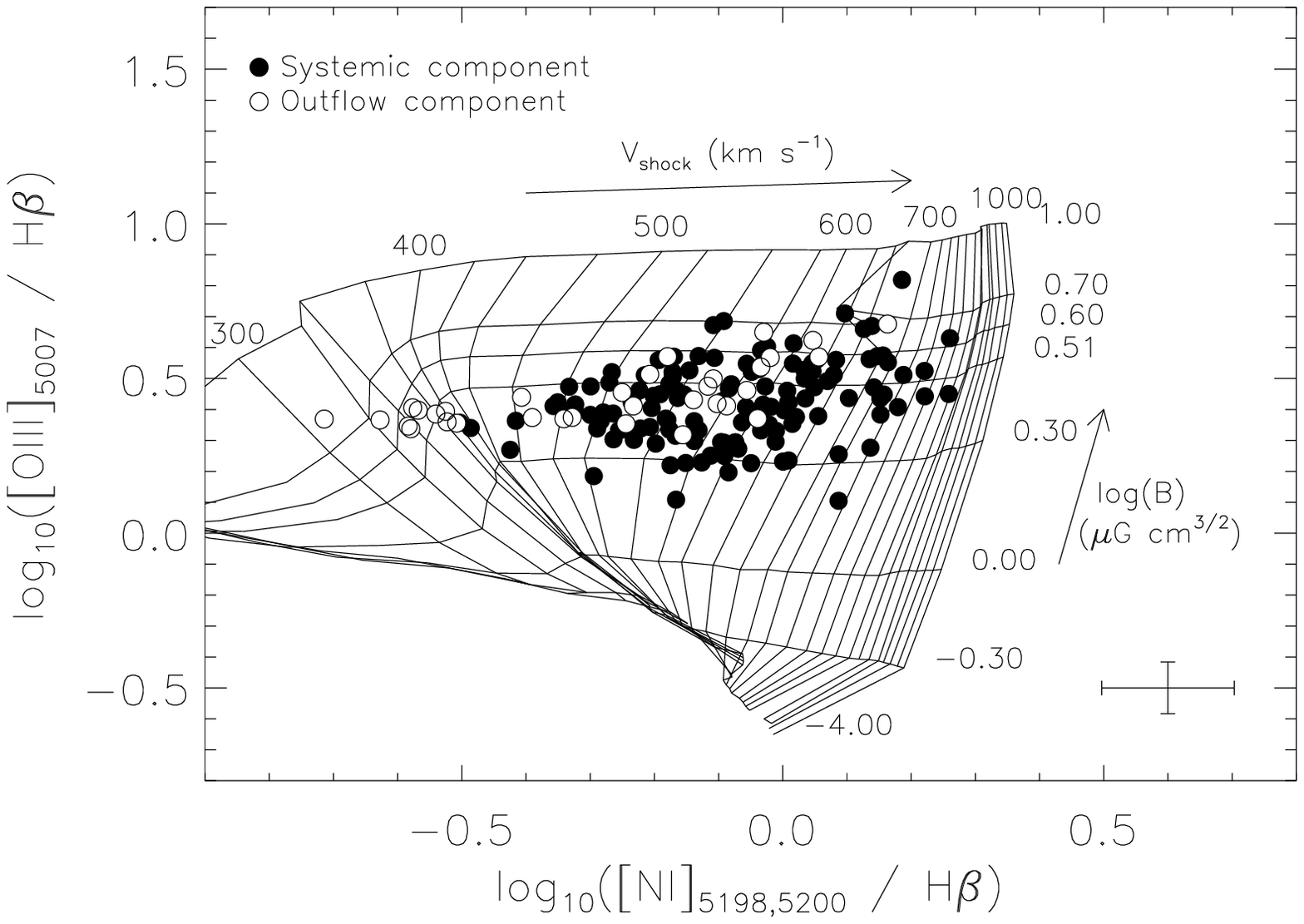}}\\
\subfigure[]{\includegraphics[width=0.48\textwidth,clip,trim=0.5cm 0.0cm 0.6cm 1.0cm]{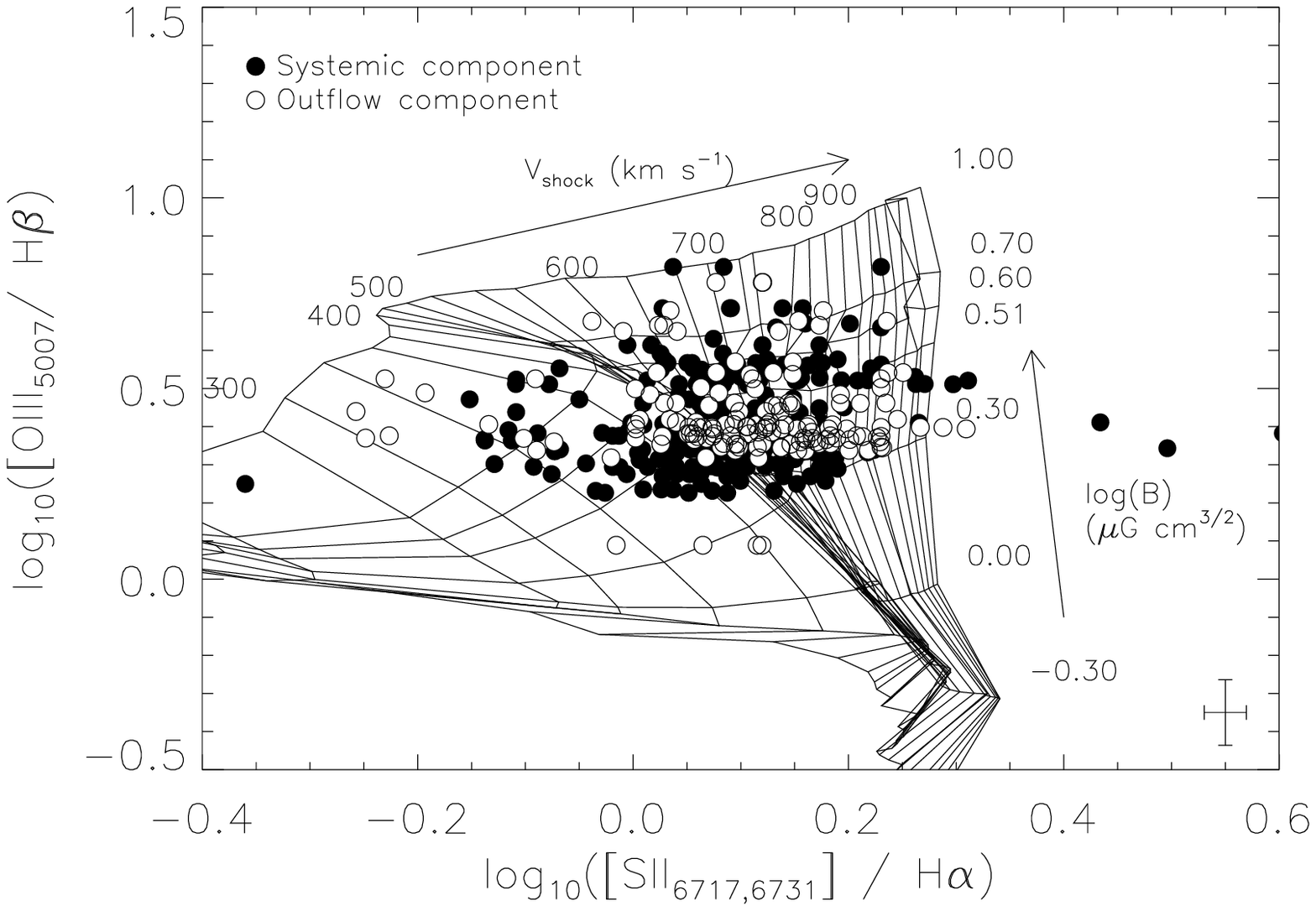}}\\
\subfigure[]{\includegraphics[width=0.48\textwidth,clip,trim=0.5cm 0.0cm 0.6cm 1.0cm]{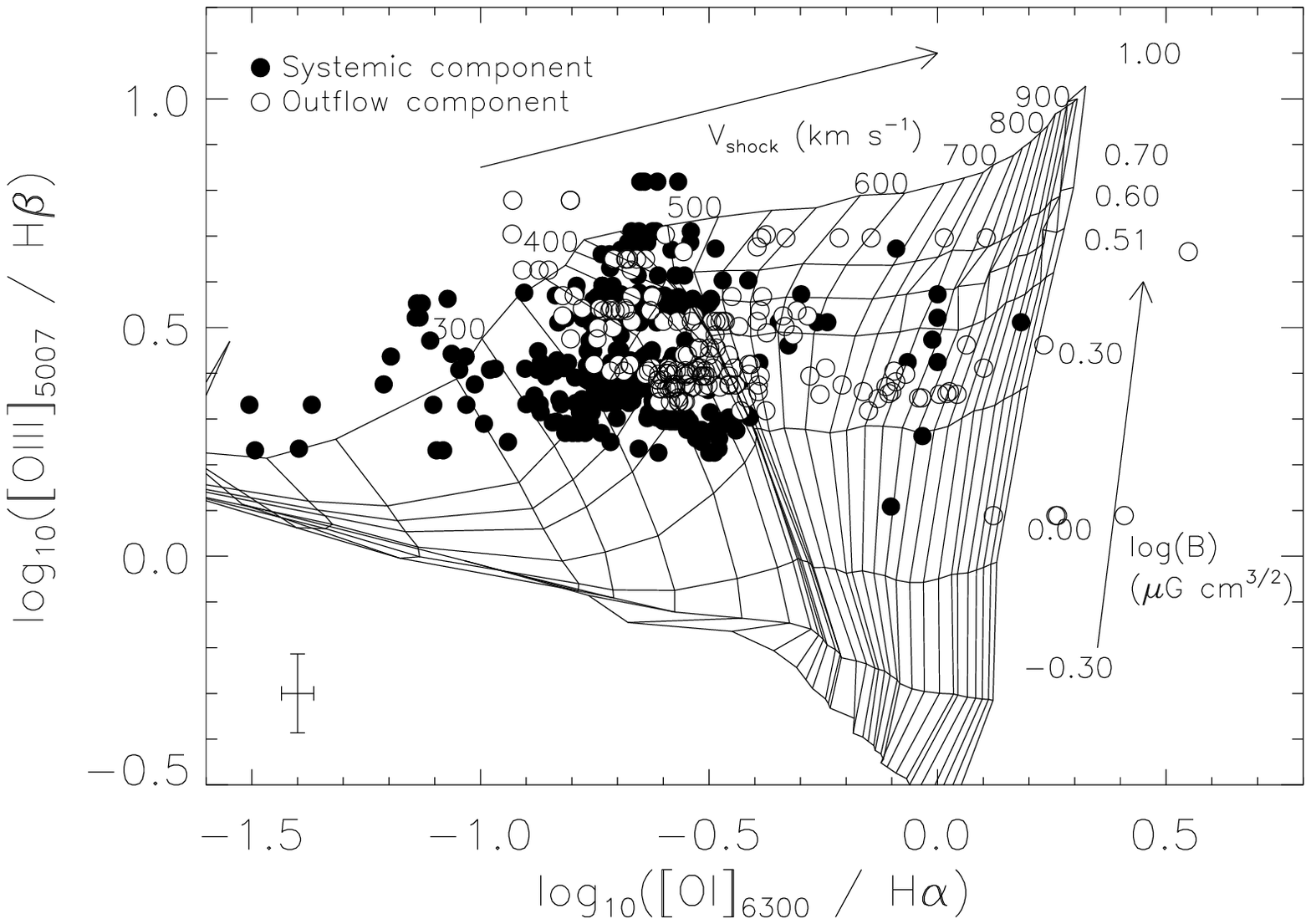}}
\end{center}
\caption{\small Observed line ratios for spaxels in the inner parts of NGC\,1266, plotted over a shock model with a super solar metallicity ($\approx$ 2Z$_{\odot}$), taken from \protect\cite{2008ApJS..178...20A}. The Y-axis of each plot shows the [OIII]/H$\beta$ ratio derived from our SAURON data, and it is plotted versus the [NII]/H${\alpha}$ (panel a; top), [SII]/H${\alpha}$ (panel b; middle) and [OI]/H${\alpha}$ (panel c; bottom) line ratios from our GMOS observations. In the bottom of each plot is the typical error bar associated with each point, derived from the formal fitting errors and a monte-carlo error analysis of the fluxes returned by our fitting routines. }
\label{shockdiags}
\end{figure}

Figures \ref{sauronplots2} and \ref{gmosplots2} show the distribution of ionised gas in NGC\,1266. Ionised gas is detected everywhere in the inner $\approx$10\arcsec\ (a projected distance of 1440 pc) of the galaxy, however the brightness distribution is not smooth. As seen in narrow band images (and previously discussed in A2011) the ionised gas emission is brighter to the south of the nucleus, and has a distinctive `kidney bowl' shape (best seen in Figure \ref{gmosplots2}). This structure correlates with the position of the outflow (and with the radio jet; Section \ref{driving}). The asymmetric brightening of the lines is present in both the systemic and outflow components, suggesting these components are linked in some way. In the rest of this section we use ratios of the line fluxes to understand the ionisation mechanism powering this emission.

Using the combination of the SAURON and GMOS IFU data is is possible to construct BPT diagrams \citep{1981PASP...93....5B} for the inner parts of NGC\,1266.  
In order to use ratios from both datasets we interpolate the \sauron\ IFU map onto the GMOS bins. As the \sauron\ bins are in general larger, some GMOS bins will have identical values of the \sauron\ line ratios.
Figure \ref{bptdiags} shows the [OIII]/H$\beta$ ratio derived from our SAURON data, plotted versus the [NII]/H$_{\alpha}$ (top panel), [SII]/H${\alpha}$ (middle panel) and [OI]/H${\alpha}$ (bottom panel) line ratios from our GMOS observations. In the bottom right of each plot is the typical error bar associated with each point, derived from the formal fitting errors and a monte-carlo error analysis of the different fluxes returned by our fitting routines. 

Onto these diagrams we overplot the diagnostic lines of \cite{2001ApJ...556..121K,2006MNRAS.372..961K}, which indicate the dominant line excitation mechanism. In all of the panels it is clear that star-formation cannot excite the lines we see in NGC\,1266. The top diagram shows that the gas is ionised either by an AGN, or shock processes. The middle and bottom panels show that the majority of the spaxels have line ratios which are consistent with low-ionisation nuclear emission region \citep[LINER:][]{1980A&A....87..152H} like activity. LINER are a controversial classification (see \citealt{2008ARA&A..46..475H} for a review), with some authors claiming that the ionisation comes from an AGN \citep[e.g.][]{1983ApJ...264..105F,1993ApJ...417...63H}, fast shocks \citep{1963ApJ...138..945B,1995ApJ...455..468D}, or photoionization by ultraviolet (UV) bright stellar sources \citep{1990ASSL..160..301D,1994A&A...292...13B,2010MNRAS.402.2187S,odea}. 

In addition to the BPT diagrams, it is also possible to plot the observed line ratios over grids that predict the line ratios of ionised gas lines in given scenarios. Both obscured and unobscured AGN models from \cite{2004ApJS..153...75G} fail to fully reproduce the observed line ratios. The closest AGN model requires a metallicity of 4Z$_{\odot}$, an electron density of 1000 cm$^{3}$, and a relatively constant radiation field. 
Even then this model cannot fully reproduce the observed line ratio distribution.  We find that the only model that can reproduce the emission line ratios we observe is a shock model from \cite{2008ApJS..178...20A}, again with a super solar metallicity ($\approx$ 2Z$_{\odot}$).  The stars in NGC\,1266 have a sub-solar metallicity (McDermid et al., in prep), and thus if the gas really is of super-solar metallicity then it must have been significantly enriched.

Figure \ref{shockdiags} shows our best fitting shock models, from the SAURON specific line ratio diagram [OIII]/H$\beta$ vs [NI]/H${\alpha}$ and the SAURON/GMOS combined [OIII]/H$\beta$ vs [SII]/H${\alpha}$ and [OI]/H${\alpha}$ diagrams. The emission observed requires shock velocities of up to $\approx$800\kms, and an average magnetic field of $\approx$3 $\mu$G cm$^{3/2}$. The shock velocities are similar to the outflow velocities, suggesting that the shocked emission is powered by the outflow.

\begin{figure} 
\begin{center} 
\includegraphics[width=0.49\textwidth,clip,trim=0cm 1.9cm 0cm 1cm]{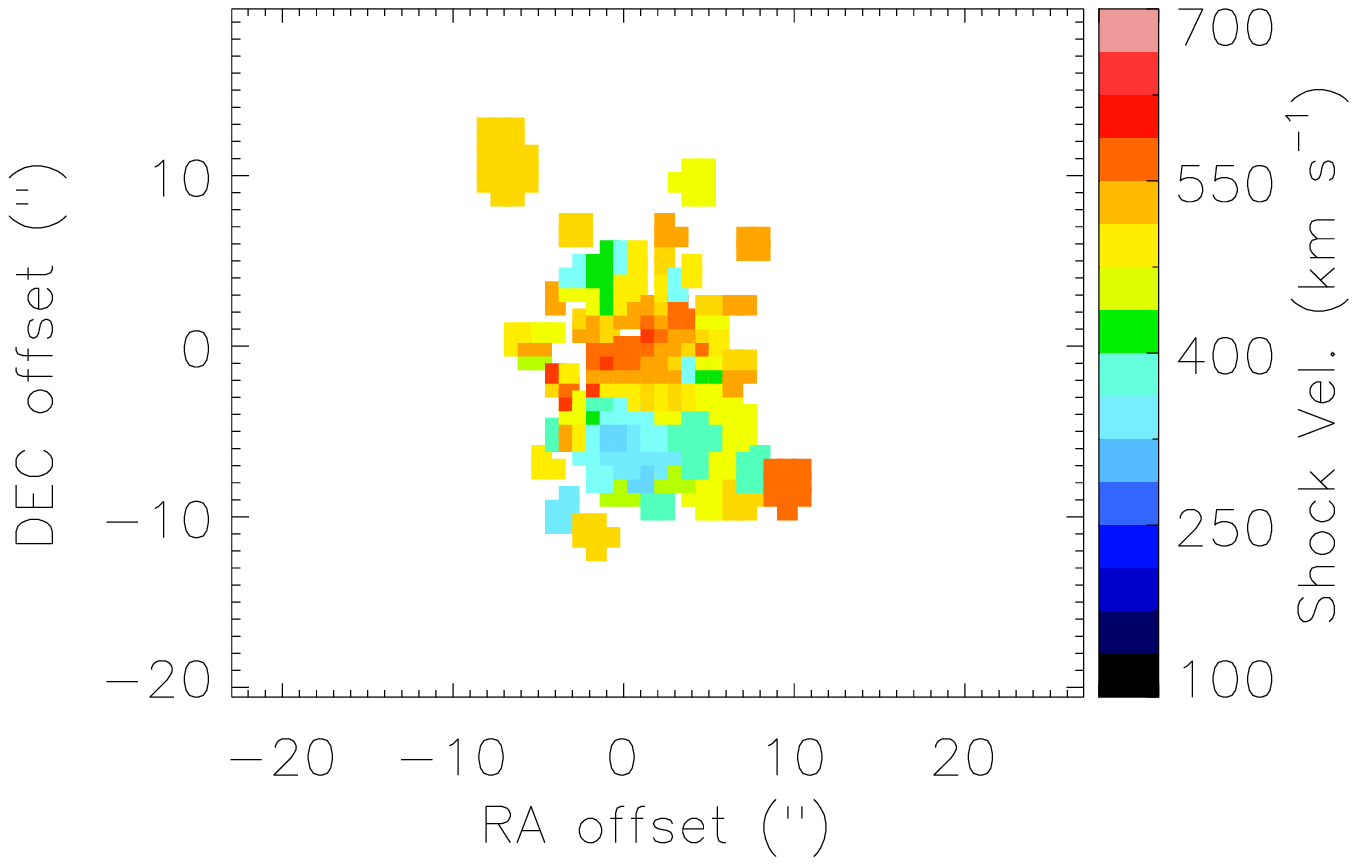}\\
\includegraphics[width=0.49\textwidth,clip,trim=0cm 0.5cm 0cm 1cm]{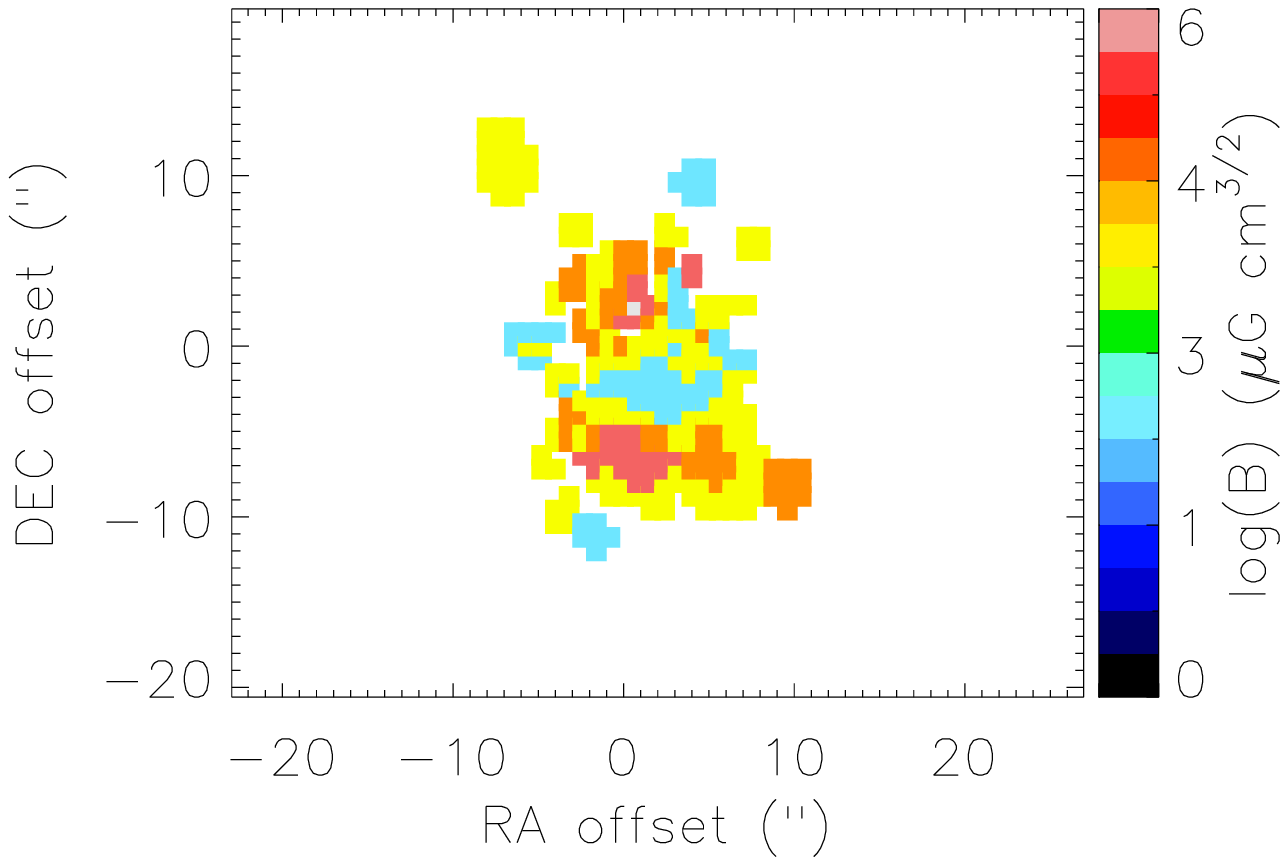}
\end{center}
\caption{\small Shock velocities (top panel) and magnetic field strengths (bottom panel) for the systemic component derived from the SAURON only [OIII]/H$\beta$ vs [NI]/H${\alpha}$ diagram by interpolating the shock grid presented in Figure \protect \ref{shockdiags}. The same trends are observed in the mixed GMOS/SAURON diagrams.}
\label{sauronshock}
\end{figure}

The shock velocities derived from these grids are highest at the centre, and towards the northern (redshifted) side of the outflow {(see Panel a of Figure \ref{sauronshock})}. This is true in all the data sets, even though the line ratio that drives the shock determination is different. The [SII]/H${\alpha}$ grid predicts higher shock velocities than the [NI]/H${\alpha}$ grid, but the average shock velocity is consistent. The [OI]/H${\alpha}$ grid predicts lower shock velocities on average, but this is driven by the bins that fall within the Seyfert region of the associated BPT diagram (Figure \ref{bptdiags}). These spaxels are mostly located in the centre of the galaxy, where the AGN contribution to the ionisation is likely to be strongest.  
It is not surprising that suggestions of the embedded AGN are most clearly seen in the [OI]/H${\alpha}$ diagram, as the presence of strong [OI]$\lambda$6300 is usually indicative of a power-law ionising spectrum.
 The [OI] line, which is collisionally excited, will only occur in a zone which has a sufficiently high electron density and temperature to excite the upper level. With a stellar input spectrum, these conditions only occur within the H$^+$ Str\"omgren sphere, where the O$^0$ abundance is negligible. However, a gas ionised by a relatively flat power-law spectrum has an extended partially ionised zone where the [OI] emission arises \citep{1997iagn.book.....P}.

The magnetic field appears to increase with increasing distance from the centre of the galaxy, perhaps due to compression of magnetic field lines as they enter the shocked gas {(Panel b; Figure \ref{sauronshock})}. 
Shock velocities are similar in both the systemic and outflow components, again arguing that these components are linked, but the magnetic field may be higher, on average in the outflowing component.

A2011 have shown that NGC\,1266 has a deeply embedded AGN, but as we show here, the ionised gas emission in this LINER galaxy is dominated by shocks, likely caused by the outflow. A similar shocked, high metallicity wind has been postulated to explain the LINER like emission from starburst-superwinds \citep[e.g.][]{RichShock}. The shocks in NGC\,1266 exist at high velocities, similar to those found in the outflow, once again demonstrating the link between the ionised gas and the material in the outflow.

\subsection{Physical Conditions}

Using the ionised gas lines we detect, it is possible to estimate the physical conditions within the ionised gas in each {spaxel}.  
From the ratio of the {6731} \AA\ and {6717} \AA\ lines in the [SII] doublet it is possible to estimate the luminosity-weighted electron density along each line of sight. 
The ratio of the [NII] doublet at $\lambda$6548,6584 \AA\ and the [NII] line at $\lambda$5755 \AA\  also provides a sensitive temperature tracer, that depends very little upon the density of the gas.
We calculated these line ratios for each spaxel where it is constrained by our GMOS data, and used the \textit{nebular} package in IRAF to estimate the electron density and temperature, based on the 5-level atom program developed by \cite{1987JRASC..81..195D}. 

\begin{figure*} 
\begin{center} 
\subfigure[]{\includegraphics[height=6cm,clip,trim=1cm 0.5cm 5.7cm 1.0cm]{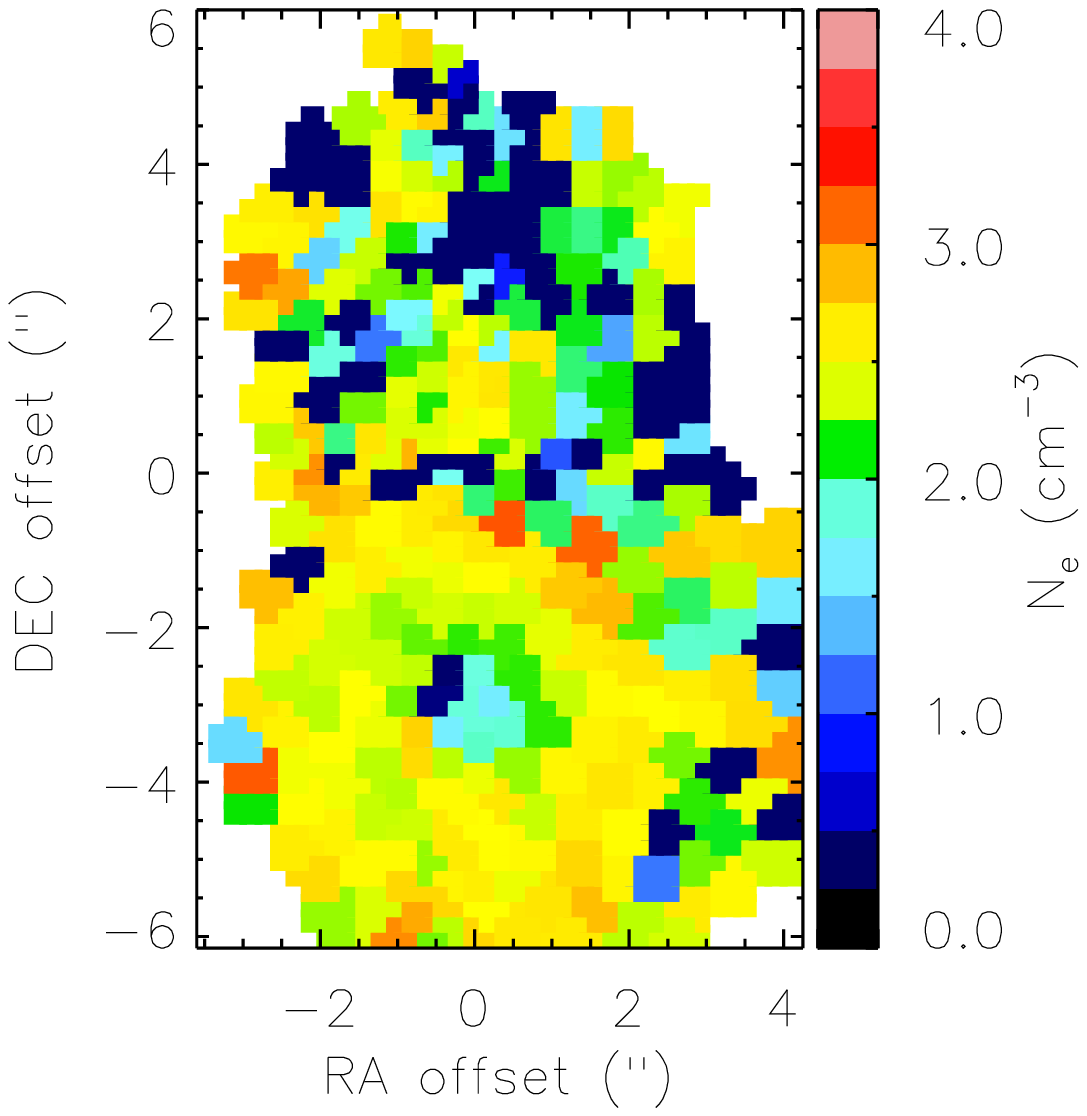}
\includegraphics[height=6cm,clip,trim=3.2cm 0.5cm 3cm 1.0cm]{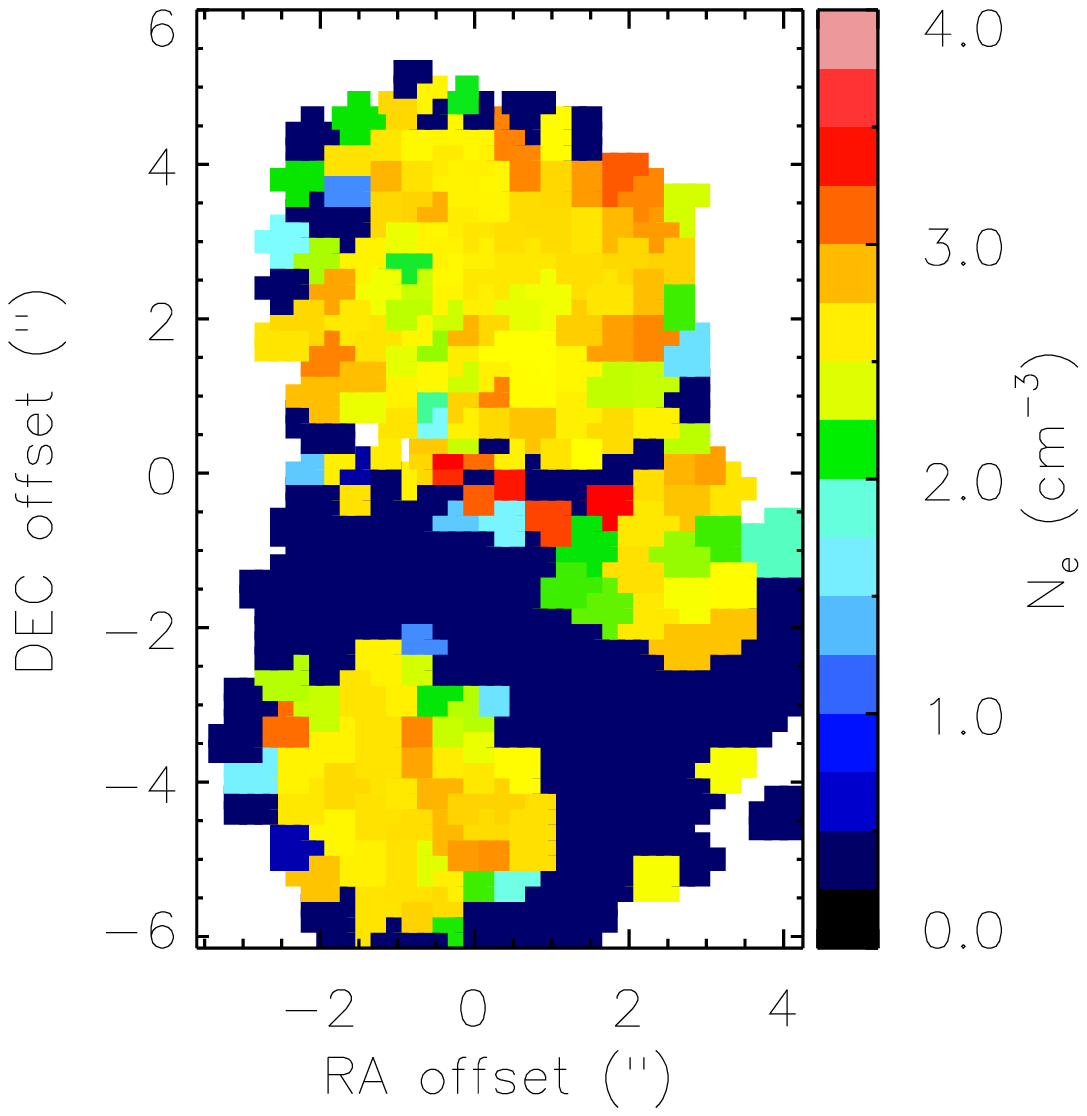}}
\subfigure[]{\includegraphics[width=0.45\textwidth,clip,trim=1.0cm 0.0cm 0.9cm 0.0cm]{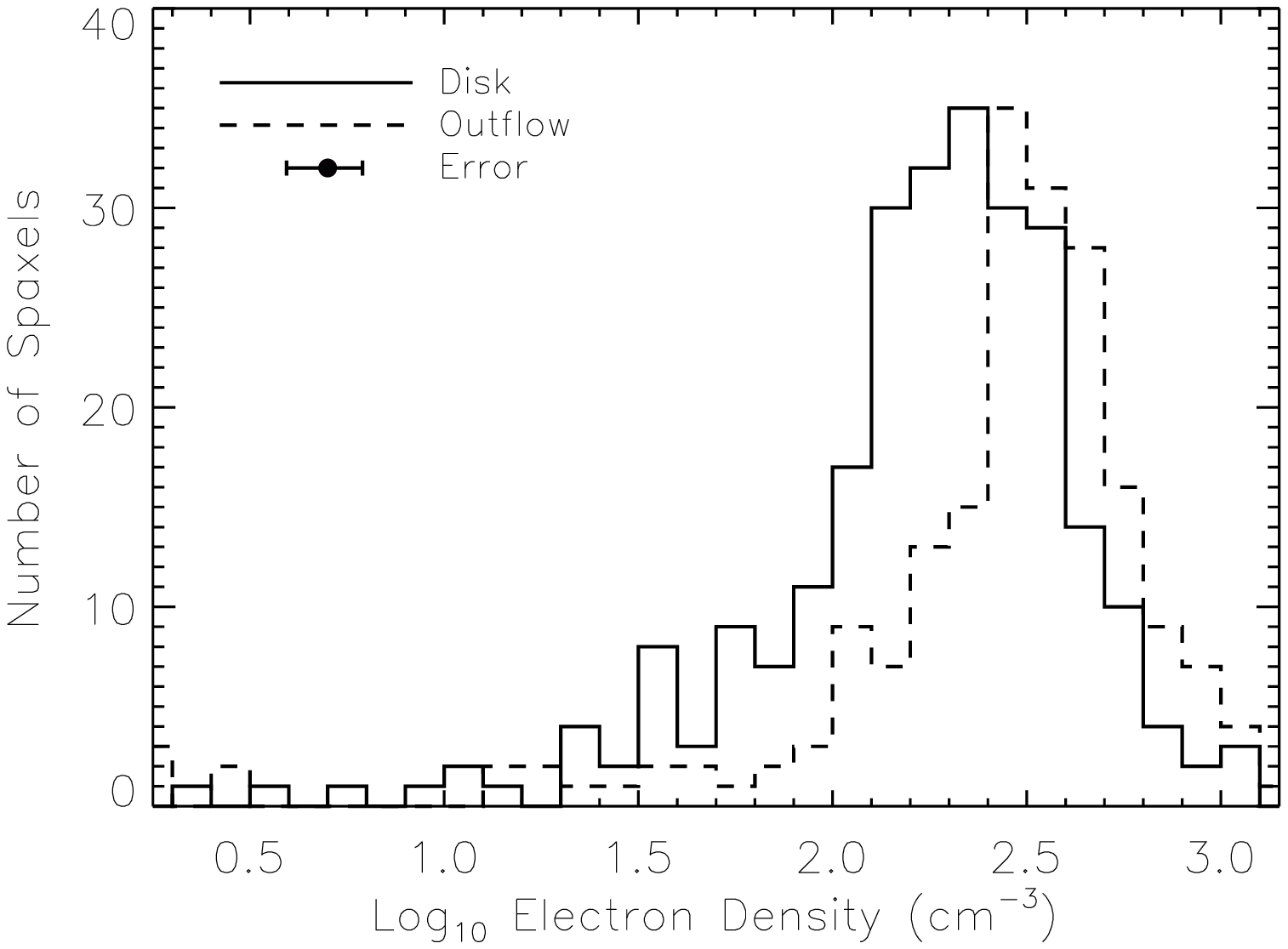}}
\end{center}
\caption{\small Panel a shows maps of the electron density (in log units), derived from the [SII] doublet, within the systemic component (left) and outflow (right panel) of NGC\,1266. Panel b shows a histogram of the electron density from the non-saturated bins, for the systemic component (solid line) and the outflow (dashed line). The median error in the electron density is shown in the legend. }
\label{electronden}
\end{figure*}

Figure \ref{electronden} shows maps of the derived electron density for the systemic and outflow component, and a histogram of the bin values in both components. In significant areas of the outflow component the [SII] ratio is saturated at the low density limit. As shown in panel b of Figure \ref{electronden}, however, in the highest velocity regions of the outflow where the line ratio is not saturated, the average density is higher than that in the systemic component.
The required lines to calculate the electron temperature were detected in fewer spaxels, and so we do not reproduce these plots here. However a reasonably consistent picture emerges, with the higher density outflow having lower average temperature, and vice versa for the lower density systemic component. 
We hypothesize that the outflow may be higher average density due to a pile up of material in-front of the ionised gas shock, and/or an increase in the number of free electrons, due to ionisation from the shock. The disturbed structure in the systemic component appears to have a lower electron density than the rest of the gas, on average, supporting the hypothesis that this could be a different gas component.

The northern half of the ionised gas systemic component has a lower electron density, on average, than the southern half. Some of the ionised gas densities in the northern regions are even at the low density limit. This could potentially offer an explanation for some of the puzzling features of the outflow. If the ISM is unevenly distributed on each side of the nucleus, then the same input energy from the central engine would dissipate faster in the denser material, producing stronger line emission in the denser southern regions, as observed. The lower density in the northern region would allow the same energy input to produce a faster shock, as observed. The magnetic field in the northern region may also be lower, and this would decrease the expected synchrotron flux from a jet, as observed in Section \ref{driving}. One might however expect that the outflow would thus be faster on the northern side of the nucleus, as the out-flowing material would encounter less resistance. This is not observed in the velocity fields presented in this paper. 

The observable tracers of the outflow that we currently posses (ionised and atomic gas in this paper, molecular gas in A2011) however have to be present at sufficient densities to be detectable, and more sensitive observations may be required to detect the highest velocity parts of the redshifted outflow. Both evidence from A2011, and from the NaD absorption presented in this paper suggest that the redshifted northern part of the outflow is seen through the galaxy, and dust obscuration could potentially prevent us from detecting low level ionised gas emission.

\subsection{Driving mechanisms}
\label{driving}

\begin{figure*} 
\begin{center} 
\includegraphics[width=1\textwidth,clip,trim=0cm 0.5cm 0cm 1.0cm]{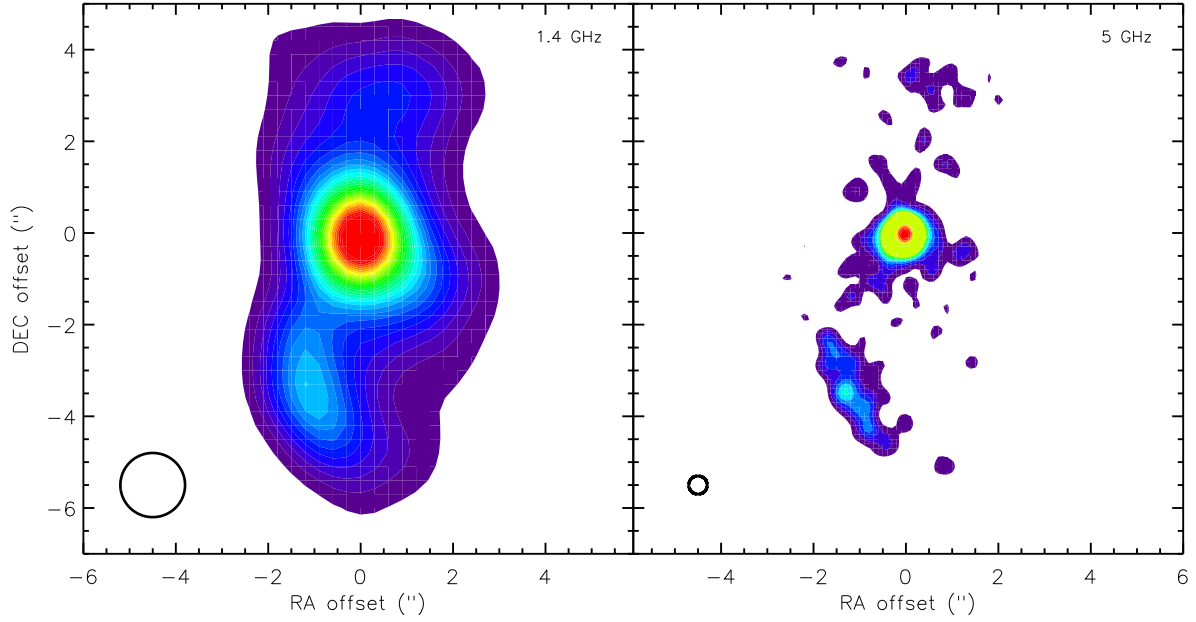}\\
\end{center}
\caption{\small 1.4 Ghz and 5 Ghz radio continuum maps of the centre of NGC\,1266 from \protect \cite{2006A&A...449..559B}. The maps are shown for all emission above 5$\sigma$ (1.4 Ghz) and 3$\sigma$ (5 GHz) with contours spaced by 1$\sigma$. The beam sizes of  $\approx$1\farc5 and $\approx$0\farc4, respectively, are indicated in the figures in the bottom left. NGC\,1266 has an integrated flux density of 90.34 $\pm$ 2.83 mJy at 1.4 Ghz (Alatalo et al., in prep).}
\label{radioimage}
\end{figure*}

The physical mechanism by which an AGN removes gas from a galaxy is still widely debated. As discussed above, radiation pressure, broad-line winds, cosmic ray heating, and radio jets have all been mooted as potential methods for driving gas out of galaxies. In NGC\,1266 the gas is clearly being accelerated and removed from the galaxy, and the cold gas phase survives reasonably far out into the outflow. A2011 argue that radiation driven winds are unable to provide enough mechanical energy to drive the outflow, and thus an AGN jet is a more likely driving mechanism. 1.4 Ghz and 5 Ghz radio continuum maps are available for NGC\,1266 \citep{2006A&A...449..559B}, and are shown here in Figure \ref{radioimage}. The 1.4 Ghz image shows a central bright point source, and a jet-like extension to the south and north.

A similar jet-like structure is visible in the 5 Ghz image, but in this higher resolution image the emission to the south and north of the nucleus is not connected to the point source. This lack of connection could be real, or because the VLA observations used here have resolved out the connecting structure. We could be seeing emission from where the jet hits the surrounding ISM (c.f. hotspots) or the jet itself may contain discrete blobs of synchrotron emitting plasma. 

The central source in this object appears to be driving the outflow, despite being a low-luminosity AGN (having a total radio power at 5 GHz of $\ltsimeq10^{22}$ W Hz$^{-1}$; Alatalo et al., in prep). This highlights that low-luminosity AGN may also be important sources of feedback.
The spectral index of the central point source (which is unlikely to be affected by $uv$-coverage issues) has a spectral index typical of an low-luminosity AGN source ($\alpha$[1.4-5Ghz]$\approx$-0.66; Alatalo et al., in prep). The spectral index when including the outer structures steepens ($\alpha$[1.4-5Ghz]$\approx$-0.79), as is seen in hotspots in local radio galaxies \citep[e.g.][]{1997A&A...319..757H}, but as some emission could have been resolved out, we cannot be sure this value is robust. 

\begin{figure*} 
\begin{center} 
\subfigure[]{\includegraphics[height=5cm,clip,trim=1cm 0.5cm 2.5cm 1.0cm]{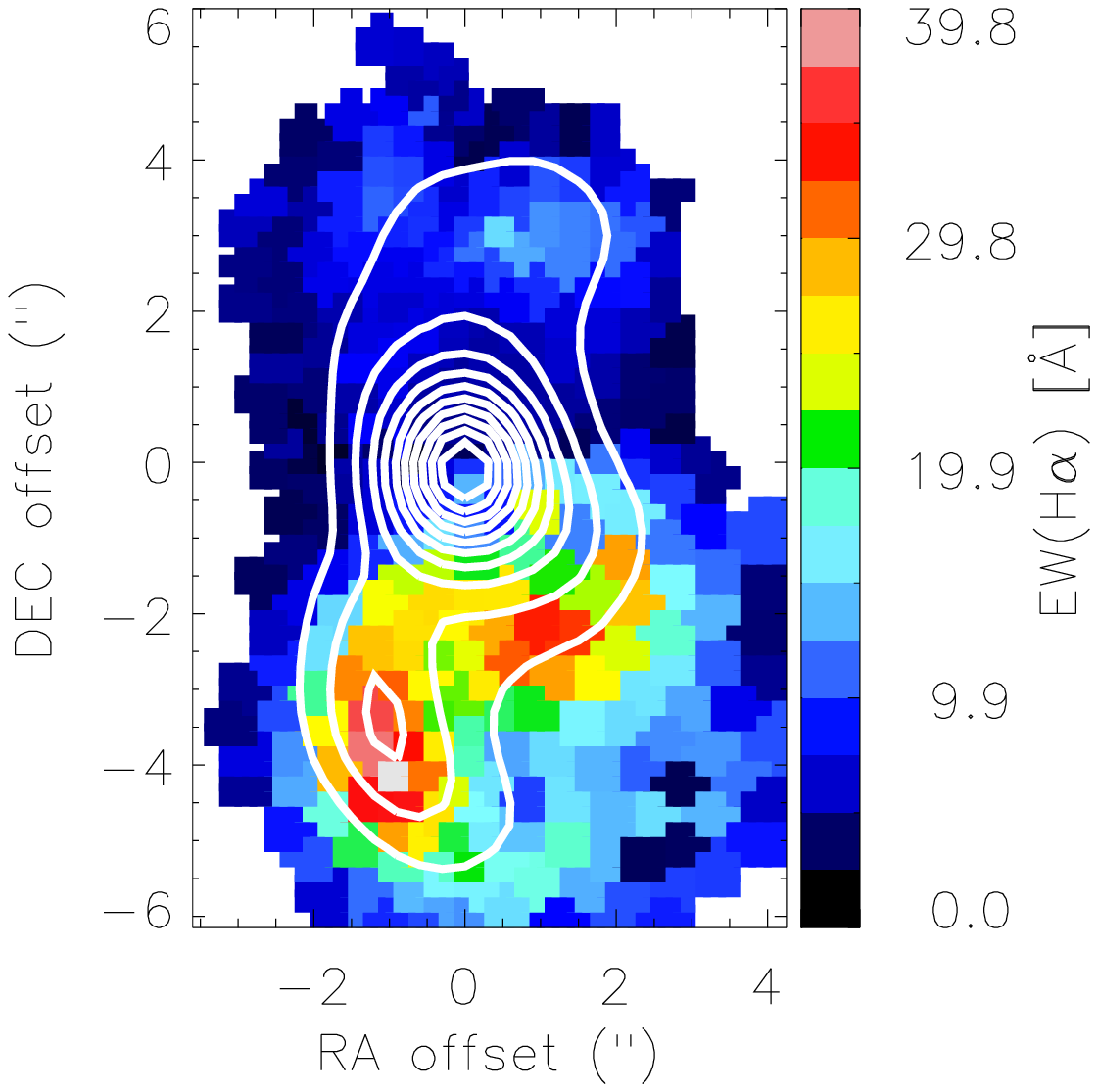}}
\subfigure[]{\includegraphics[height=5cm,clip,trim=1cm 0.5cm 2.5cm 1.0cm]{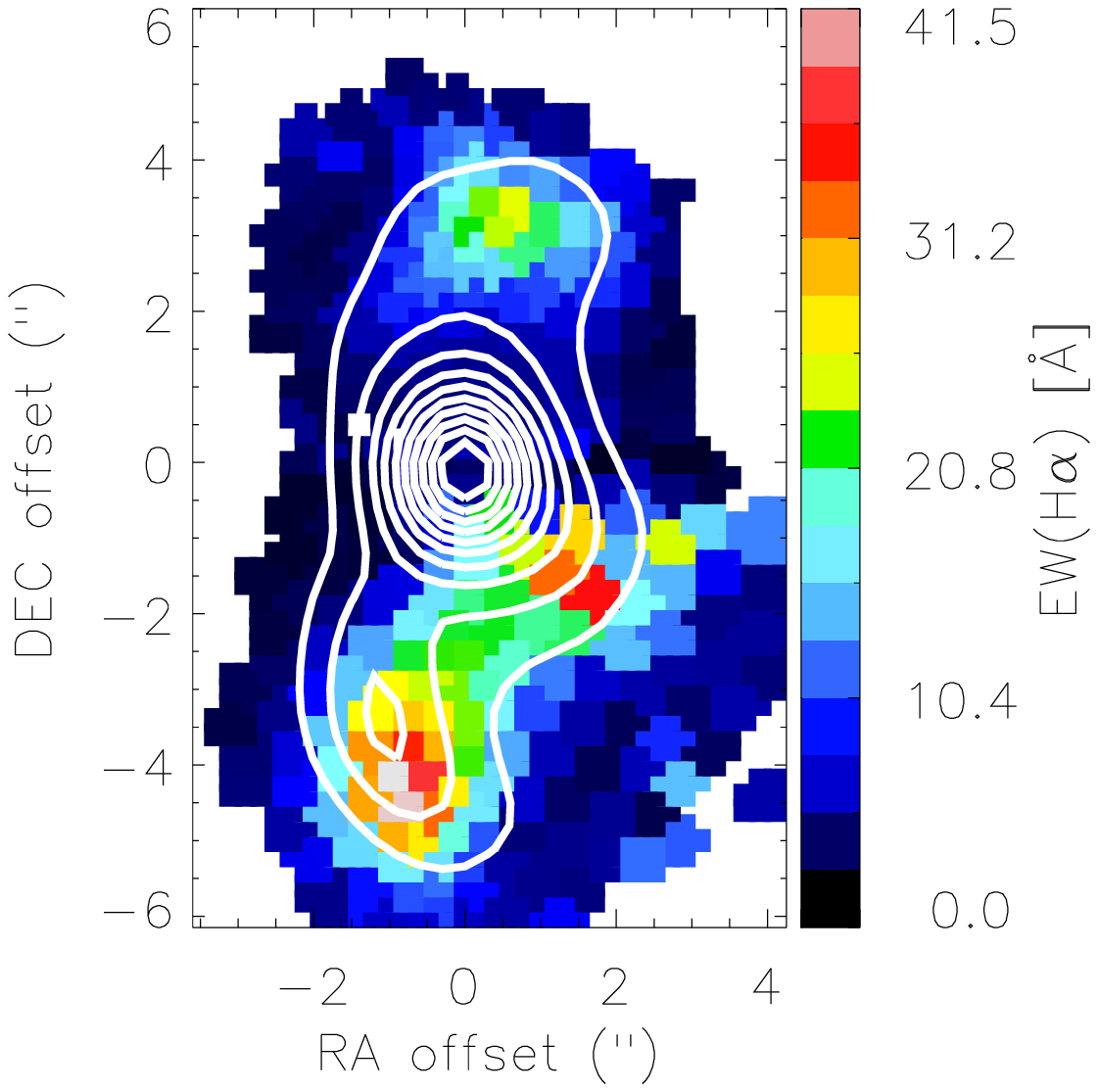}}
\subfigure[]{\includegraphics[height=5cm,clip,trim=1cm 0.5cm 1.5cm 1.0cm]{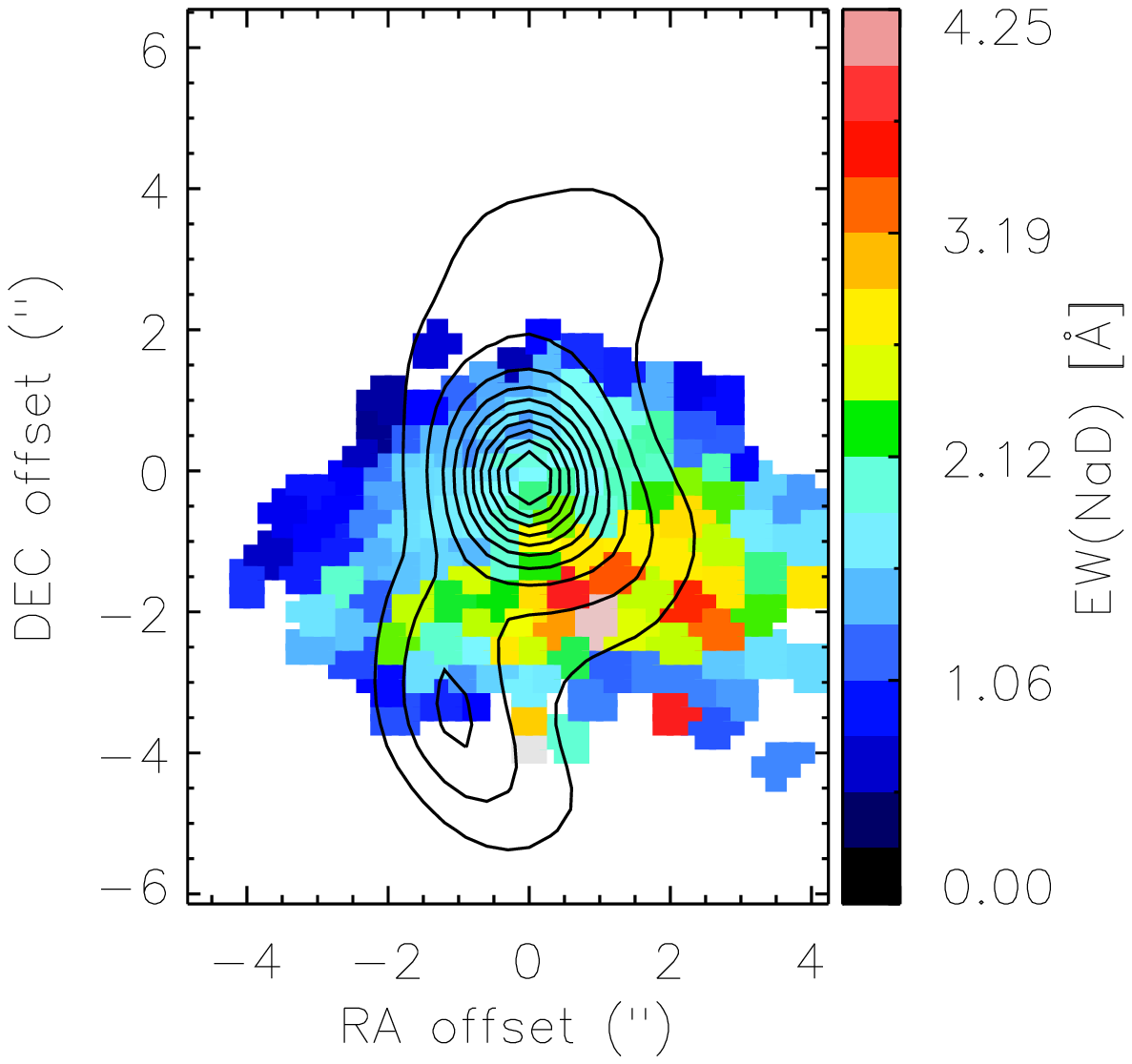}}\\
\subfigure[]{\includegraphics[height=5cm,clip,trim=1cm 0.5cm 2.5cm 1.0cm]{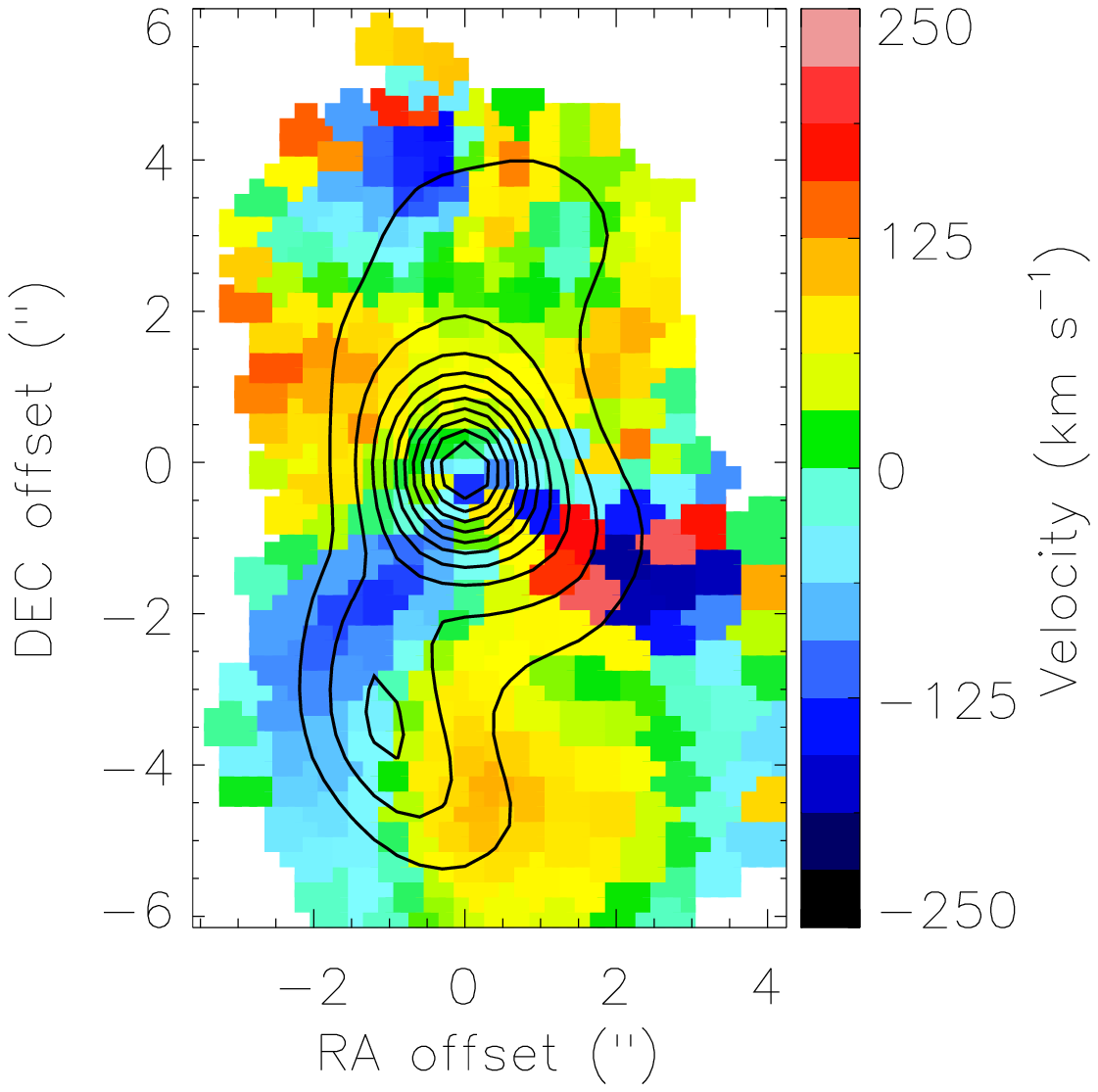}}
\subfigure[]{\includegraphics[height=5cm,clip,trim=1cm 0.5cm 2.5cm 1.0cm]{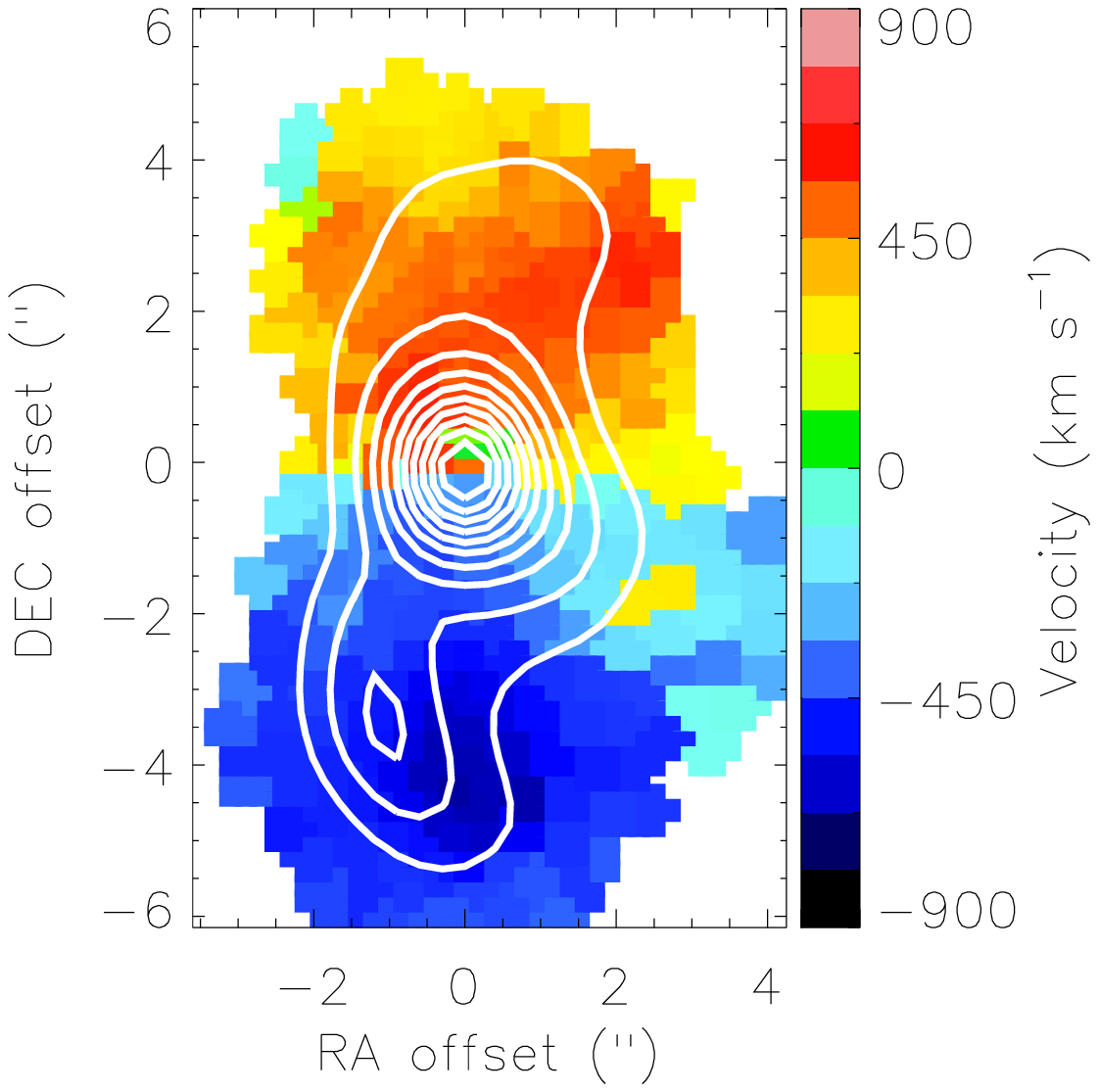}}
\subfigure[]{\includegraphics[height=5cm,clip,trim=1cm 0.5cm 1.5cm 1.0cm]{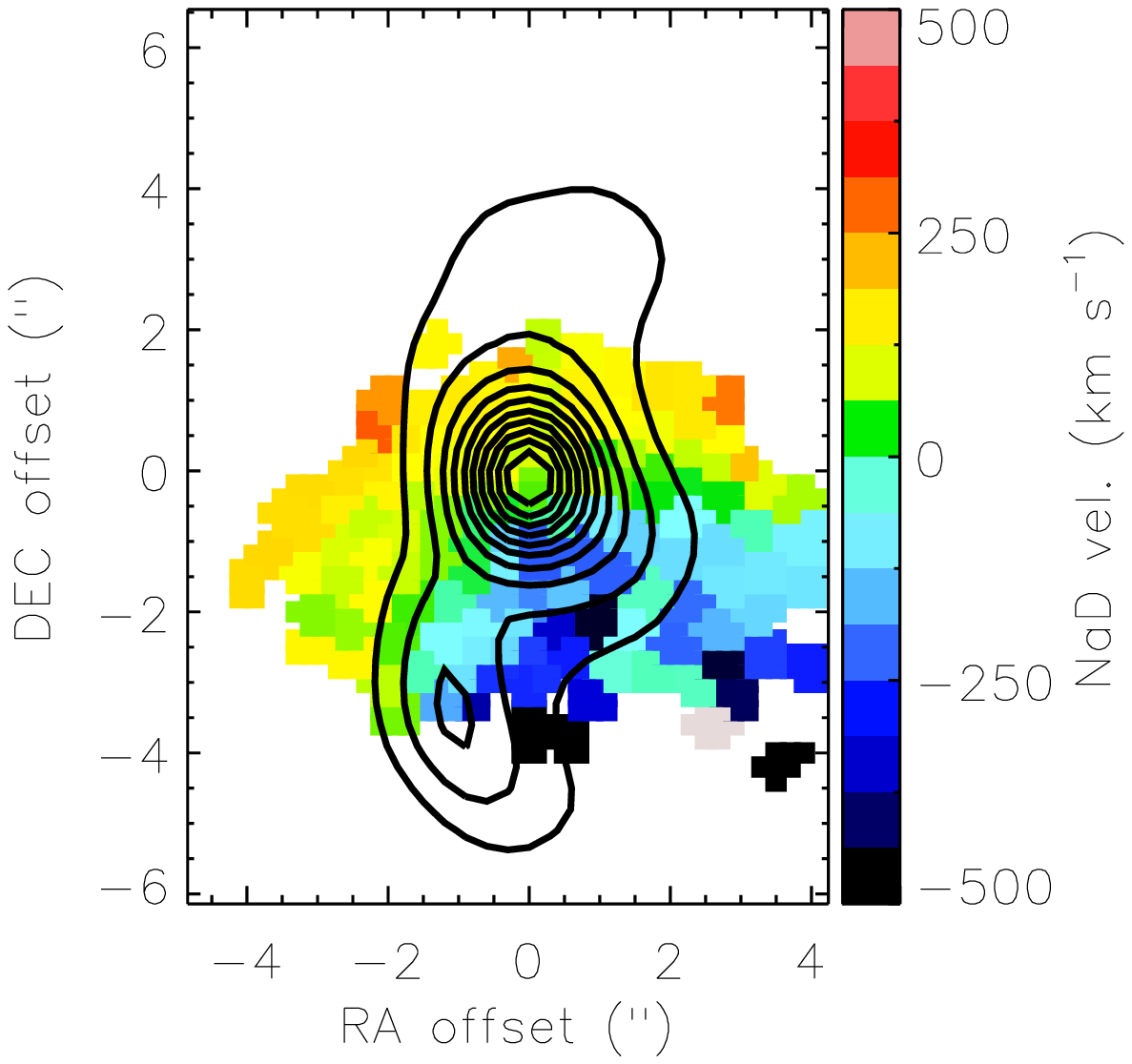}}
\end{center}
\caption{\small 1.4 Ghz radio emission contours (spaced every 8\% from 90\% to 8\% of the peak flux) over plotted on H$\alpha$ flux maps for the systemic and the outflow components (panel a \& b), velocity fields for the systemic and the outflow components (panels d \& e) and NaD neutral gas EW and kinematics (panels c \& f ).}
\label{gmosradio1}
\end{figure*}

Figure \ref{gmosradio1} shows the 1.4 Ghz radio continuum structure overplotted on our GMOS maps of the ionised gas and neutral gas outflow. The observed radio structure correlates well with the features observed in the ionised and atomic gas. 
The radio morphology is similar to that predicted by models of disrupted Fanaroff and Riley Class I (FRI) objects, as expected given the very dense environment around the jet, where it can be rapidly decelerated and disrupted by gas entrainment \citep[e.g.][]{Laing02,Laing11}.

 The `kidney-bowl' shock structure observed in the ionised gas traces well the southern edge of the jet seen in radio continuum, as is observed in many radio galaxies (i.e. the "Alignment Effect", {\citealt{1987Natur.329..604C,1987ApJ...321L..29M})}. This suggests that the shock suggested by the ionised gas line ratios occurs where the radio plasma impacts with the ionised gas. The blue shifted feature on the south-eastern side of the systemic component also seems to correlate well with the jet. This feature may well be associated with the jet, arguing that the systemic component may not represent a single structure.

The radio continuum structure hooks around the fastest moving neutral and ionised gas.  
As discussed above, the position angle of the outflow traced in NaD (and molecular gas; see Figure \ref{GMOSplotsCO}) is slightly different than that traced by the ionised gas. Projection effects could possibly explain some of the difference between the position angles, but is unclear if this can completely remove the discrepancy. If this effect is real, then it could potentially arise because interaction with the ISM has deflected the jet, or the jet could be precessing (as observed e.g. in Cen A; \citealt{1983PASAu...5..241H}). 
Of course it is also possible that the jet itself is powering the ionised gas outflow, but that another mechanism (e.g. winds or radiation pressure) could be driving the atomic and molecular outflows, naturally leading to different position angles (see e.g \citealt{RupkeMrk231}). 
From the speed of the outflow, and its measured extent we can estimate the lifetime of the phenomenon. Both the ionised gas (extending to $\approx$1.5 Kpc and travelling at upto 900 \kms) and the neutral gas (extending $\approx$400 pc and travelling at an average 250 \kms) give a similar estimate for the lifetime of the outflow ($\approx$1.6 Myr). This would seem to suggest the outflows in ionised and neutral gas have a common cause. Radiation pressure driving the outflow also seems unlikely on energetics grounds (see A2011).
 Deep high resolution observations of the radio continuum structure could help distinguish the cause of this effect.

\section{Conclusions}
\label{conclude}

In this paper we have been able to shed some light on the outflow activity in NGC\,1266. This unusual galaxy is relatively nearby, allowing us to investigate the process of AGN feedback in action. Using the SAURON and GMOS IFUs we detected strong ionised gas emission lines (H$\alpha$, H$\beta$, [OIII], [OI], [NII], [SII] and HeI), as well as  NaD absorption. We use these tracers to explore the structure of the source, derive the ionised and atomic gas kinematics and investigate the gas excitation and physical conditions.

NGC\,1266 contains two ionised gas components along each line of sight, tracing the ongoing outflow and a component closer to the galaxy systemic, the origin of which is unclear.  The gas appears to be being disturbed by a nascent AGN jet. 

We have presented further evidence the outflow in this object is truly multiphase, containing radio emitting plasma, ionised and atomic gas, as well as the molecular gas and X-ray emitting plasma (as detected in A2011). With outflow velocities up to 900 \kms, the outflow is very likely to be removing significant amounts of cold gas from the galaxy. The ionised gas morphology correlates well with the radio jets observed in 1.4 Ghz and 5 Ghz continuum emission, supporting the suggestion of A2011 that an AGN jet is the most likely driving mechanism for the ionised gas outflow. 

The line emission in NGC\,1266 causes it to be classified as a LINER in optical diagrams. We show here that although NGC\,1266 undoubtably hosts an AGN (see A2011) the line emission in this object is extended, and is most consistent with excitation from fast shocks caused by the interaction of the radio jet with the ISM.  These shocks have velocities of up to 800 \kms, which match well with the observed velocity of the outflow.

Using the observed NaD absorption we are able to set further constraints on the size and orientation of the outflow. We show that we are able to detect atomic gas entrained in the outflow out to a (deprojected) distance of $\approx$400$\pm$50 pc, and that the outflow has an inclination (between the galaxy plane and the outflow) of 53$\pm$8$^{\circ}$.
This suggests the outflow is misaligned from the stellar body. Furthermore we were able to provide an independent estimate of the column density of neutral material in the outflow N$_{\mathrm{\hi}}$=(1.2$\pm$0.6)$\times$10$^{21}$ cm$^{-2}$. This estimate is consistent with that derived from \hi\ absorption in A2011. The neutral and molecular outflows are well correlated, but appear to be outflowing along a slightly different axis to the ionised gas. The cause of this affect is unclear. 

NGC\,1266 is a highly complex object, and it is clear that further observations will be required to fully understand it. Observations of single emission lines  (either with an IFU or with a Fabry-P\'erot instrument) will be important to overcome problems with line blending inherent in this high velocity dispersion source. Higher spatial resolution would also be advantageous to obtain clear diagnostics of AGN activity, and better understand the shock structure. Sensitive interferometric radio observations at high angular resolution would also enable us to understand the morphology and orientation of the nascent radio jet. The Atacama Large Millimeter/submillimeter Array (ALMA) will allow us to study the molecular component of the outflow in greater detail, and further determine the driving mechanism. For instance if molecular shock tracers are found predominantly in the outflow then a kinetic driving mechanism would be favoured, while if photon dominated region tracing species were detected this would argue for a radiation driven component to the outflow.

This galaxy is one of the few currently known in which we can witness ongoing active feedback, where a central AGN is disrupting its star-forming reservoir. It is clear that understanding the processes removing the ISM will have widespread applications to both theoretical and observation attempts to understand AGN feedback, its affect on the ISM, and role in building up the red-sequence.

\vspace{10pt}
\noindent \textbf{Acknowledgments}
The authors thank the referee, Katherine Inskip, for comments which improved the paper.
TAD thanks Millie Maier, Niranjan Thatte, Mark Westmoquette and Grant Tremblay for useful discussions.
The research leading to these results has received funding from the European
Community's Seventh Framework Programme (/FP7/2007-2013/) under grant agreement
No 229517.

The SAURON observations were obtained at the WHT, operated by the Isaac Newton Group in the Spanish Observatorio del Roque de los Muchachos on La Palma, Canary Islands.
Also based on observations obtained at the Gemini Observatory, which is operated by the 
Association of Universities for Research in Astronomy, Inc., under a cooperative agreement 
with the NSF on behalf of the Gemini partnership: the National Science Foundation (United 
States), the Science and Technology Facilities Council (United Kingdom), the 
National Research Council (Canada), CONICYT (Chile), the Australian Research Council (Australia), 
MinistŽrio da Cincia, Tecnologia e Inova‹o (Brazil) 
and Ministerio de Ciencia, Tecnolog'a e Innovaci—n Productiva (Argentina).
RMcD is also supported by the Gemini Observatory.

This work was supported by the rolling grants `Astrophysics at Oxford' PP/E001114/1 and ST/H002456/1 and visitors grants PPA/V/S/2002/00553, PP/E001564/1 and ST/H504862/1 from the UK Research Councils. RLD acknowledges travel and computer grants from Christ Church, Oxford and support from the Royal Society in the form of a Wolfson Merit Award 502011.K502/jd. RLD also acknowledges the support of the ESO Visitor Programme which funded a 3 month stay in 2010.

MC acknowledges support from a Royal Society University Research Fellowship. SK acknowledges support from the the Royal Society Joint Projects Grant JP0869822.
TN and MBois acknowledge support from the DFG Cluster of Excellence `Origin and Structure of the Universe'.
MS acknowledges support from a STFC Advanced Fellowship ST/F009186/1.
PS is a NWO/Veni fellow.
MBois has received, during this research, funding from the European Research Council under the Advanced Grant Program Num 267399-Momentum.
The authors acknowledge financial support from ESO.

\bsp
\bibliographystyle{mn2e}
\bibliography{bibNGC1266ionout}
\bibdata{bibNGC1266ionout}
\bibstyle{mn2e}
\label{lastpage}

\end{document}